\documentclass[aps,pra,reprint,twocolumn,superscriptaddress, longbibliography, preprintnumbers]{revtex4-1}

\usepackage{url}
\usepackage{amsmath}
\usepackage{amssymb}
\usepackage{makecell}
\usepackage{comment}
\usepackage{multirow}
\usepackage{amsthm}
\usepackage{graphicx}% Include figure files
\usepackage{dcolumn}% Align table columns on decimal point
\usepackage{amssymb}
\usepackage{bm}% bold math
\usepackage[colorlinks, citecolor=blue]{hyperref}
\hypersetup{colorlinks,linkcolor=blue,citecolor=blue,filecolor=blue,urlcolor=blue}
\usepackage{physics} % Physics macros
\usepackage{qcircuit} % Quantum circuits
\usepackage{tikz} % TikZ
\usetikzlibrary{shapes.geometric}

\usepackage[normalem]{ulem}

\usepackage{subfiles}
\newtheorem{result}{Result}

\def\draft{1}
\newcommand{\xnote}[1]{\ifnum\draft=1 {\color{blue} \textbf{Xun's note:} #1}\fi}
\newcommand{\cnote}[1]{\ifnum\draft=1 {\color{brown} \textbf{Chi-Ning's note:} #1}\fi}
\newcommand{\mnote}[1]{\ifnum\draft=1 {\color{orange} \textbf{Marcin's note:} #1}\fi}
\newcommand{\bnote}[1]{\ifnum\draft=1 {\color{green!40!black} \textbf{Boaz's note:} #1}\fi}
\newcommand{\snote}[1]{\ifnum\draft=1 {\color{red} \textbf{Soonwon's note:} #1}\fi}

\newcommand{\new}[1]{\ifnum\draft=1 {\color{red} #1}\fi}

\newcommand{\innerp}[2]{\langle#1|#2\rangle}
\newcommand{\ex}[2]{\left\langle#1\right\rangle_#2}

\newcommand{\kket}[1]{|#1\rangle\rangle}
\newcommand{\bbra}[1]{\langle \langle #1|}

\newcommand{\HarvardPhysics}{Department of Physics, Harvard University, Cambridge, MA 02138, USA}

\newcommand{\HarvardSEAS}{School of Engineering and Applied Sciences, Harvard University, Cambridge, MA 02138, USA}

\newcommand{\MIT}{Center for Theoretical Physics, Massachusetts Institute of Technology, Cambridge, MA 02139, USA}

\begin{document}

%\title{Linear Cross-Entropy is a Poor Measure for Quantum Advantage}
\title{Limitations of Linear Cross-Entropy as a Measure for Quantum Advantage}
% Force line breaks with \\

\author{Xun Gao}
\affiliation{\HarvardPhysics}

\author{Marcin Kalinowski}
\affiliation{\HarvardPhysics}

\author{Chi-Ning Chou}
\affiliation{\HarvardSEAS}

\author{Mikhail D. Lukin}
\affiliation{\HarvardPhysics}

\author{Boaz Barak}
\affiliation{\HarvardSEAS}

\author{Soonwon Choi}
%\affiliation{\Berkeley}
\affiliation{\MIT}

\preprint{MIT-CTP/5321}

\begin{abstract}
Demonstrating quantum advantage
requires  experimental implementation of a computational task that is hard to achieve using state-of-the-art classical systems.
One approach is to perform sampling from a probability distribution associated with  a certain class of highly entangled many-body wavefunctions. 
It has been suggested that such a quantum advantage can be certified with the \emph{Linear Cross-Entropy Benchmark} (XEB).
We critically examine this notion.
First, we consider a ``benign'' setting, where an honest implementation of a noisy quantum circuit is assumed, and characterize the conditions under which the XEB approximates the \emph{fidelity} of quantum dynamics. 
Second, we assume an ``adversarial'' setting, where all possible classical algorithms are considered for comparisons, and show that achieving relatively high XEB values does not imply faithful simulation of quantum dynamics.
Specifically, we present an efficient classical algorithm that achieves high XEB values, namely 2-12\% of those obtained in the state-of-the-art experiments, within just a few seconds using a single GPU machine.
This is made possible by identifying and exploiting several vulnerabilities of the XEB which allows us to achieve high XEB values without simulating a full quantum circuit. Remarkably, our algorithm features \emph{better scaling} with the system size than a noisy quantum device for commonly studied random circuit ensembles in various architecture. 
We quantitatively explain the success of our algorithm and the limitations of the XEB by using a theoretical framework, in which the dynamics of the average XEB and fidelity are mapped to classical statistical mechanics models.
Using this framework, we illustrate the relation between the XEB and the fidelity for quantum circuits in various architectures, with different choices of gate sets, and in the presence of noise.
Taken together, our results demonstrate that XEB's utility as a proxy for fidelity hinges on several conditions, which should be independently checked in the benign setting, but cannot be assumed in the general adversarial setting.
Therefore, the XEB on its own has a limited utility as a benchmark for quantum advantage. We discuss potential ways to overcome these limitations.
\end{abstract} 

\pacs{Valid PACS appear here}% PACS, the Physics and Astronomy
\maketitle

\section{Introduction}\label{sec:intro}

\emph{Quantum advantage} refers to the experimental demonstration of the computational power of a quantum device far beyond that of any existing classical devices.
Such demonstration is important because it not only constitutes a milestone of quantum technology, but also challenges the so called \emph{extended Church-Turing thesis}~\cite{arora2009computational,aaronson2011computational}, which has been central to computational complexity theory. A straightforward way to demonstrate quantum advantage would be to explicitly run a quantum algorithm, such as the Shor's integer factoring~\cite{shor1994algorithms}, for problems whose size is too large (e.g. $2048$-bit integers) to be solved by any known algorithm running on classical computers.
However, this would require a quantum device with a large number of near-perfect qubits, which is well beyond the capabilities of the existing technology. State-of-the-art quantum devices consist of several dozens of \emph{imperfect} qubits~\cite{zhang2017observation,arute2019quantum,USTC,zhu2021quantum,ebadi2020quantum,Scholl.2021}.
Even the exploration of a potential scaling advantage requires larger systems, consisting of at least several hundred coherent qubits.

Instead of implementing such quantum algorithms, most of the current efforts towards demonstrating quantum advantage have focused on \emph{sampling problems}~\cite{preskill2012quantum,harrow2017quantum,lund2017quantum}, which are well suited for near-term quantum devices~\cite{arute2019quantum,USTC,zhu2021quantum,zhong2020quantum,zhong2021phase}.
In these problems, one is asked to produce a sequence of random bitstrings drawn from a certain probability distribution. A natural choice of a distribution that would be challenging for a classical computer to reproduce is one based on a highly entangled many-body wavefunction.
Indeed, it has been shown~\cite{bremner2011classical,aaronson2011computational,bremner2016average,farhi2016quantum,boixo2018characterizing,bremner2017achieving,gao2017quantum,bermejo2017architectures,terhal2002adaptive} that, for a wide class of quantum states, exact sampling by classical computers  is intractable under plausible assumptions~\cite{terhal2002adaptive,bremner2011classical,aaronson2011computational,aaronson2016complexity,bremner2016average,aaronson2016complexity,bouland2019complexity,movassagh2018efficient,bouland2021noise,kondo2021fine}.

\begin{figure}
    \centering
    \includegraphics{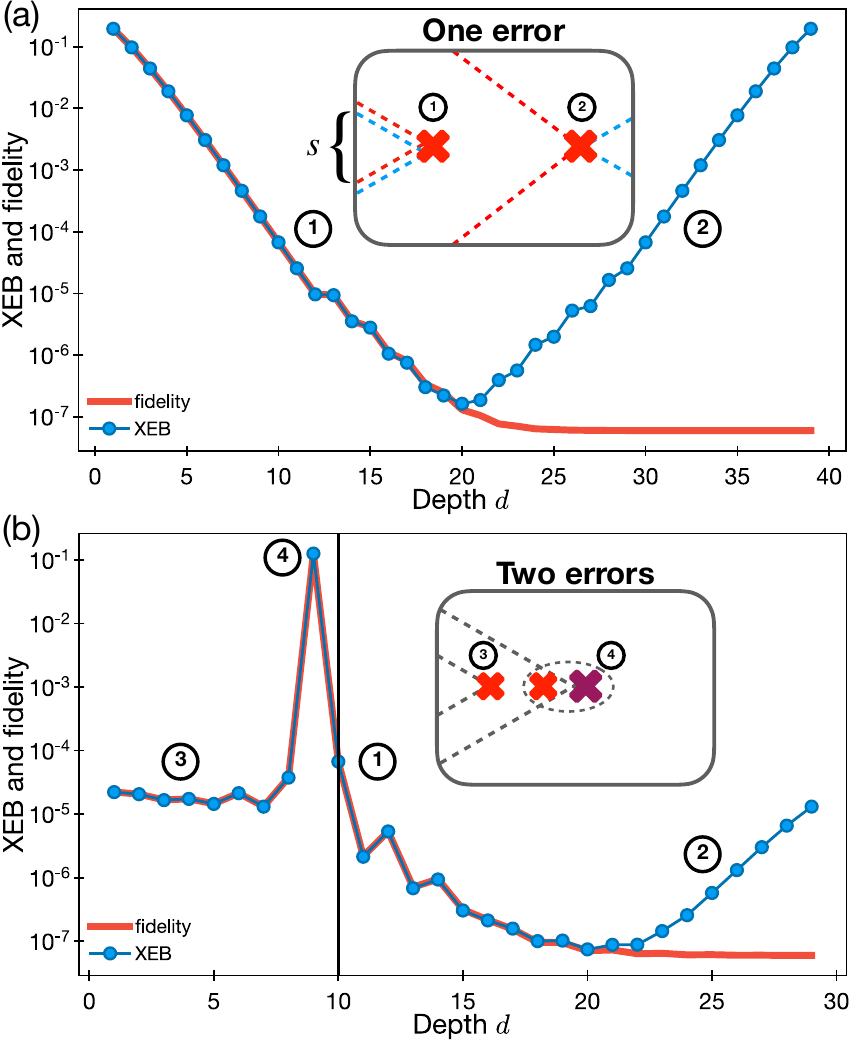}
    \caption{Effects of a single or double error at various locations on the XEB and fidelity. 
    (a) In the presence of a single error, the XEB and fidelity is reduced to an exponentially small but nonzero value that depends on the location of the error. The scaling of the XEB or fidelity can be understood in terms of the size $|s|$ of the error operator propagated to boundaries in the Heisenberg picture (inset). 
    (b) In the presence of two errors, the XEB and fidelity significantly depend on their relative location: the effect of one error can be masked (marked 3) or even cancelled (marked 4) by that of another error.}
    \label{fig:oneerror}
\end{figure}
To demonstrate quantum advantage using an actual sampling experiment, one needs to introduce a \emph{benchmark} that measures how close the sampled distribution $q(x)$ of a quantum device is to the (ideal) target distribution $p(x)$.
The idea is that on one hand, one shows that the samples from the quantum device achieve high values (indicating good correlation with the ideal distribution), while on the other hand, one presents evidence that there \emph{does not exist} an efficient classical algorithm that can produce samples achieving comparable values.
If the difference between the classical and quantum resources needed to achieve a certain value of the benchmark scales \emph{exponentially} with the system size, this demonstrates that quantum devices have an exponential computational advantage even in the regime where the gates are too noisy to allow for quantum error correction.
A prominent example of such a benchmark is the \emph{linear cross-entropy benchmark} (XEB)~\cite{arute2019quantum} defined as
\begin{equation}\label{eq:defXEB}
    \chi_p(q) = 2^N \sum_{x \in \{0,1\}^N} p(x)q(x) - 1.
\end{equation} 
Intuitively, $\chi_p(q)>0$ if $q$ places more mass on the elements $x$ whose probability is higher than the median in $p$.
A non-vanishing value of $\chi_p(q)$ is taken to mean that the sampled distribution is correlated with the ideal one.

The XEB measure has been used in recent  experiments~\cite{arute2019quantum,USTC}, where sampling from random unitary circuits was performed.
Specifically, Google~\cite{arute2019quantum}  achieved an  XEB value of $\chi_p \approx 0.002$ on a two-dimensional, 53-qubit quantum device (Sycamore) implementing circuits up to depth 20 under reasonable assumptions.
{Recently, the USTC group~\cite{USTC,zhu2021quantum} extended the number of qubits and reached the XEB value of $6.62\times 10^{-4}$ and $3.66\times 10^{-4}$, for system sizes up to 56 qubits and 60 qubits, respectively.} In both cases, it has been  conjectured that such values are challenging to achieve using state-of-the-art classical computing devices on a realistic time scale.

\begin{figure*}
\includegraphics[width=\linewidth]{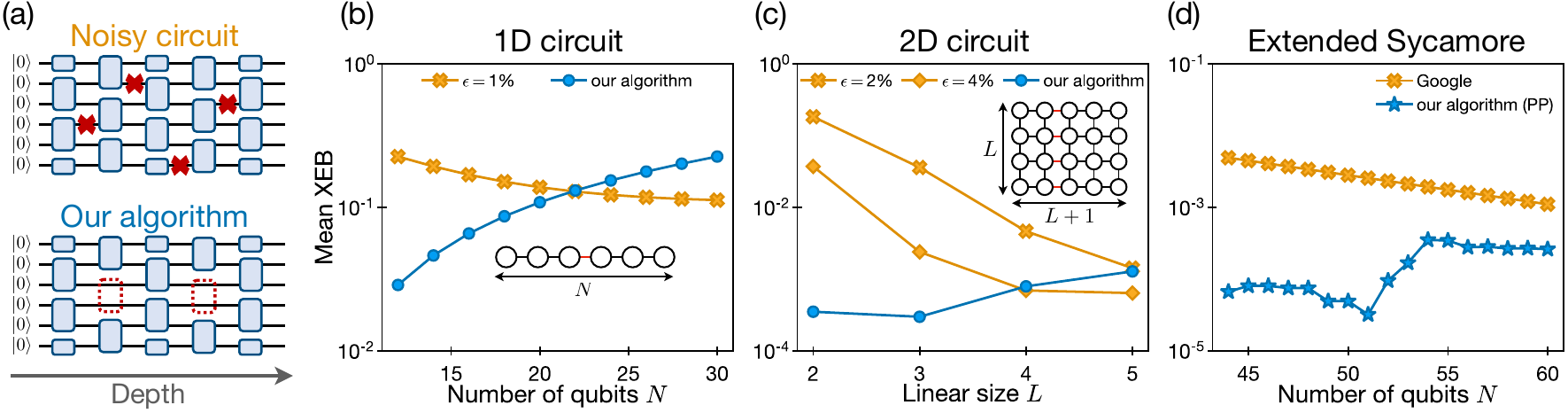}
\caption{
Classical algorithms spoofing XEB for quantum circuits in various  architectures.
(a) Schematic diagrams illustrating the key idea of our algorithm.
In noisy quantum circuits, errors (red crosses) randomly occur at a rate $\epsilon>0$, spread over the entire circuit.
In our algorithm, we introduce effective, highly localized errors by omitting or modifying a few entangling quantum gates (red dotted boxes) such that the circuit splits into smaller segments and becomes easier to simulate classically.
(b-d) Performance of our algorithm. We obtain high XEB values (blue circles and stars)  compared to noisy circuits (yellow crosses and diamonds) for 1D, 2D, and the extended Sycamore circuit architectures [see Fig.~\ref{fig:intro_partition}].
(b) 1D circuits of depth $d=16$ in the brick-work layout, with the Haar random two-qubit gate ensemble.
(c) 2D circuits of depth $d=16$ in a $L\times (L+1)$ square lattice, with the Haar random two-qubit gate ensemble.
Our algorithm outperforms noisy quantum circuits (here with error rates $\epsilon=0.02$,  $0.04$) for sufficiently large system sizes.
Insets in (b-c) show the circuit architecture and the position of omitted gates (red lines).
(d) Comparison of the mean XEB value obtained by our improved algorithm (light blue circles) to Google's Sycamore in which case we extrapolated experimental results using the ansatz ${\rm XEB}\sim \exp(-c_1 N-c_2 Nd )$. We extended the Sycamore architecture horizontally up to 60 qubits; see Fig.~\ref{fig:intro_partition} for more details.
For this simulation, we assumed a quantum circuit ensemble with random single-qubit gates similar to (but slightly modified) those used in Ref.~\cite{arute2019quantum,USTC,zhu2021quantum} [see Sec.~\ref{ssec:intro_alg_summary}].
}
\label{fig:intro_scaling}
\end{figure*}

The motivation for using the XEB as a benchmark is two-fold.
First, the XEB is relatively easy to estimate in an experiment with small number of samples.
Second, the XEB is believed to be correlated with the \emph{fidelity} of the quantum state produced by a physical noisy quantum device, against an ideal state expected from the quantum circuit without any noise or errors~\footnote{The fidelity is closely related to the \emph{Bhattacharyya coefficient} of two classical probability distributions.}.
Therefore, one may expect that achieving a high XEB value implies the demonstration of quantum advantage.
However, we emphasize that the nature of quantum advantage experiments must be inherently \emph{adversarial}: it is not sufficient to show that an experiment achieves a good value on a benchmark --- one needs to argue that \emph{every possible classical algorithm} cannot achieve the same value.
Otherwise, certain adversarial classical algorithms may take ``shortcuts'' and achieve good values on the benchmark, despite not really simulating the target quantum circuit.

To be more specific, the XEB has been used in Ref.~\cite{arute2019quantum,USTC,zhu2021quantum} to simultaneously serve two distinct purposes:
\begin{enumerate}
    \item \textbf{Proxy for fidelity:} The XEB is considered as a good approximation to the many-body fidelity $F=\bra{\psi}\rho\ket{\psi}$ for chaotic quantum systems~\cite{boixo2018characterizing,neill2018blueprint,arute2019quantum,choi2021emergent,liu2021benchmarking}, where $\ket{\psi}$ is the ideal target state and $\rho$ is the state prepared by a noisy device.

    \item \textbf{Certification of quantum advantage:} It has been suggested that obtaining bitstring samples with a significant XEB value on a classical device is computationally difficult~\cite{bouland2019complexity,aaronson2019classical}, which would allow XEB to certify  quantum advantage. 
\end{enumerate}

In this work, we critically assess these roles of XEB and present two major results. First, we characterize the relation between XEB and fidelity in the ``benign'' setting of comparing a noisy quantum device to an idealized noiseless circuit, showing how this correlation depends on the architecture and the choice of gate sets. Based on these considerations, we identify the conditions under which the XEB can be used as a proxy for the fidelity.
Second, we show that the XEB is not a good  measure of quantum advantage in the ``adversarial'' setting, by presenting a classical ``spoofing'' algorithm that achieves comparable XEB values to those demonstrated in the experiments,  using only desktop-scale computational resources within a few seconds.
This is  possible because our classical algorithm explicitly violates the aforementioned conditions, where the XEB approximates the fidelity.

Prior work challenging  quantum advantage~\cite{gray2020hyper,huang2020classical,pan,fu2021closing,refute1,refute2} obtained comparable or higher XEB values using heavy computational resources. 
While these classical methods are tailored to challenge Google's current setup (53 qubits, depth 20), up to now it was unclear if and how they could be extended to larger systems. In fact, it has been argued that by simply increasing the system size to about 60$\sim$70 qubits, one could defeat such classical spoofing algorithms~\cite{blog}.
Indeed, in  more recent experiments~\cite{zhu2021quantum} 
(60 qubits, depth 24), it has been suggested that the new device bypasses the challenge of these algorithms.
In what follows we show that the XEB has fundamental limitations as a proof of quantum advantage beyond a simple competition arising from the scaling of system sizes.

In particular, we show that XEB values produced by our algorithm feature more favorable scaling with the system size than a realistic, noisy quantum device.
As a result, our algorithm is expected to outperform such experiments on average if their architecture is  extended to involve more qubits, without a corresponding improvement  in average gate fidelities.

Before proceeding, we emphasize that efficiently measurable benchmarks such as XEB are essential for certifying quantum advantage. While fidelity could have been used directly to characterize the performance of quantum circuits, it is an inherently quantum quantity. 
As such, it can not be used to  characterize a computational task, i.e. the fidelity cannot be defined for classical algorithms. 
At the same time, the use of 
classical probability distance measures (e.g.,  Bhattacharyya or Hellinger distances, KL divergence, or total variation distance) is challenging since it is difficult to obtain empirical estimates for these quantities from experiments. 
This is because the domain of these distributions is exponentially large; all of these distances require not just an exponential computation time but also an exponential number of samples to estimate (c.f.,~\cite[Chap.~5]{Canonne20}), which is impractical. 

\subsection{Vulnerabilities of the XEB}\label{ssec:vulnerability}
In this work, we exploit three distinct properties of the XEB that make it vulnerable against adversarial attacks.
First, the XEB and fidelity may diverge from one another in the presence of errors highly correlated in their space-time locations.
Second, the XEB and fidelity exhibit distinct scaling behavior with increasing system size: when multiple systems are brought together to form a larger one, the XEB generally \emph{increases} with the number of subsystems, while the fidelity decays exponentially.
Finally, the XEB is designed to quantify the amount of correlation between an ideal probability distribution $p(z)$ and an experimentally obtained one $q(z)$, but this correlation can be dramatically amplified if one has direct access to the full description of $q(z)$ (in contrast to only having samples drawn from it).
Combining these three properties, one can devise an efficient, adversarial algorithm that achieves high XEB values for state-of-the-art quantum circuit sizes, using computational resources as minimal as a single desktop-scale GPU device.
In this work, we demonstrate this approach by introducing a simple classical algorithm.
Before presenting our main results, we elaborate on the first two properties of XEB using a heuristic and intuitive analysis; the third --- the amplification of correlations ---  is explained in Sec.~\ref{ssec:algorithm_improve} and a similar idea has been already exploited in the prior work~\cite{markov2018quantum,huang2020classical,pan,refute1}.

\medskip \noindent \textbf{Discrepancy of XEB and fidelity.}
The fact that the XEB approximates fidelity can be intuitively understood using the following simplified analysis~\cite{arute2019quantum}.
For a noisy circuit in the presence of independent, homogeneously-distributed random errors at rate $\epsilon$, the system executes the entire circuit without any error with probability $P_\textrm{no err} = (1-\epsilon)^{\# \textrm{gates}}$.
If we assume that the presence of a single or more errors leads to vanishing contributions to the XEB or fidelity, both XEB and fidelity equal $P_\textrm{no err}$.
While this argument can be made rigorous in appropriate limiting cases, the exact relation between the XEB, fidelity and $P_\textrm{no err}$ involves non-zero correction terms for finite-size systems with finite error rates.
For example, let us consider the effect of a single bit-flip error $\hat{X}$, occurring at depth $t$ in a 1D random circuit evolution, on the value of XEB and fidelity. 
This effect can be understood in the Heisenberg picture by inspecting the error operator $\hat{X}(t)$, propagated backwards in time, at $t=0$, acting on a simple initial state such as $\ket{0}^{N}$ [see Fig.~\ref{fig:oneerror}(a)].
In the case of chaotic dynamics, $\hat{X}(t)$ becomes a random linear combination of $4^{|s|}$ Pauli string operators, where the support size of the operator $|s|\approx 2c t$ grows linearly in time with an effective ``scrambling'' velocity $c$. 
Among these Pauli strings, $\sim 2^{|s|}$ operators are products of only identity $I$ or $Z$ operators, for which the initial state is an eigenstate, leading to no change in XEB or fidelity.
Consequently, even if a single error occurs, it contributes to the XEB and fidelity by a small correction $O(2^{-2ct})$.
For the case of XEB, a similar argument can be made by propagating the error operator forward to the measurement time because the combination of $I$ and $Z$ operators do not affect measurements in the computational basis, leading to a sharply distinct behavior from fidelity when an error occurs near the measurement time. 

Na\"ively, these corrections may seem small and unlikely to result in any substantial deviations of XEB and fidelity from $P_\textrm{no error}$.
However, compounding the problem is that exponentially many different events give rise to the same amount of corrections when we consider events with multiple errors.
For instance, if a second error is added to the system, such that its support is contained within that of the first error [Fig.~\ref{fig:oneerror}], their net contribution to the correction remains the same because the combined propagated error operator is still a random linear combination of Pauli operators.
In fact, one can add any number of errors within the lightcone of the first one without decreasing the net correction term.
Therefore, a substantial, non-perturbative correction may arise from a family of error events, where multiple errors are clustered at early (late) times for both XEB and fidelity (for XEB).
Even when errors occur deep inside the circuit, the effects of two consecutive errors may cancel each other with probability $\sim 1/10$ [Fig.~\ref{fig:oneerror}(b)], leading to a contribution of order unity to the fidelity and the XEB.

Based on this analysis, one can provide an approximate lower bound on total correction by summing over a few classes of error configurations~\cite{SM}.
Assuming that {\it errors are independent and homogeneously distributed over the whole system} (benign setting), we find that it is necessary and sufficient conditions for XEB, fidelity, and $P_\textrm{no error}$ to agree one another $N\epsilon f(c) \ll 1$ , where $f(c)$ is a decreasing function of order unity that depends on the microscopic details and the architecture of a quantum circuit~\cite{SM}.
Recent experiments~\cite{arute2019quantum,USTC,zhu2021quantum} approximately satisfy this condition, and we expect that XEB values would overestimate fidelities only by a few percents~\footnote{The correction is roughly $1+N\epsilon f(c)$ (see Ref.~\cite{SM}). By assuming $f(c)=1$, $N\epsilon\approx0.25\%$.}. We emphasize that this conclusion requires  the independence of errors over space and time that  needs to be explicitly checked.
In the presence of correlated errors (corresponding to adversarial setting), the corrections to XEB and fidelity may dominate their entire values, even if the total error rate remains small. This can be seen from the example in Fig.~\ref{fig:oneerror}(b): if the errors are correlated such that their position is distributed over a relatively small region, the effects of overlapping lightcones and error cancellation could be strong,
leading to potentially large (compared to $P_\text{no error}$) corrections to the XEB and fidelity.
In particular, if the errors occur in a region near the output boundary, the fidelity is suppressed due to  a large lightcone (red in Fig.~\ref{fig:oneerror}) while the XEB is affected only by much smaller overlapping lightcones (blue in Fig.~\ref{fig:oneerror}),  leading to the discrepancy between the XEB and the fidelity. Contrarily, if the errors are uncorrelated, the lightcones contributing to the XEB do not overlap, and collectively suppress the XEB value such that it is similar to the fidelity.
Based on these observation, we design an algorithm that allows for efficient classical simulation, while the discrepancy between the XEB and the fidelity is significantly amplified compared to the benign setting.

\medskip \noindent \textbf{Scaling of XEB and fidelity.}
XEB and fidelity exhibit different scaling behaviors when a system size is increased with a fixed error rate, implying that two quantities cannot agree in a certain scaling limit.
While a rigorous analysis can be made using the framework presented in Sec.~\ref{sec:DR}, here we consider a toy model illustrating the origin of the different scaling  behaviors.
Let us consider $k$ disjoint $N$-qubit systems, each undergoing noisy circuit evolution with corresponding XEB values $\chi_i = 2^N \sum_x p_i(x) q_i(x) - 1$ and fidelities $F_i$ with $i=1,2,\dots, k$.
Here $p_i(x)$ and $q_i(x)$ are bitstring probabilities for $i$-th quantum system obtained from an ideal circuit and from noisy dynamics (or any other classical algorithms), respectively.
If we consider the $k$ disjoint systems as a single composite system of $kN$ qubits, one can explicitly check that the fidelity scales multiplicatively, i.e. $F_\textrm{total} = \prod_i F_i$, while the XEB additively:
\begin{align}
    \chi_\textrm{total} &= 2^{kN} \sum_{\{x_i\}} \prod_i p_i(x_i) q_i (x_i) - 1\\
    &=\prod_i (\chi_i + 1) - 1 \approx \sum_i \chi_i,
\label{eqn:additive_nature}
\end{align}
where we assumed that $\chi_i \ll 1$ in the last line, relevant for the regime of our interest.
While this example may seem contrived as each subsystem is perfectly isolated, one can also devise an example, where all subsystems are strongly coupled by unitary gates and result in fully globally scrambled quantum states. 

This discrepancy in scaling stems fundamentally from the structure of the XEB formula in Eq.~\eqref{eq:defXEB}: as two distributions $p(x)$ and $q(x)$ become uncorrelated from one another, the first term in Eq.~\eqref{eq:defXEB} tends to a finite value, $1$, rather than approaching zero. This offset is explicitly subtracted in order to obtain a value within an interval $[0,1]$, but it also leads to distinct scaling behavior for large composite systems.

\subsection{Main results}

\begin{figure}
\includegraphics[width=\linewidth]{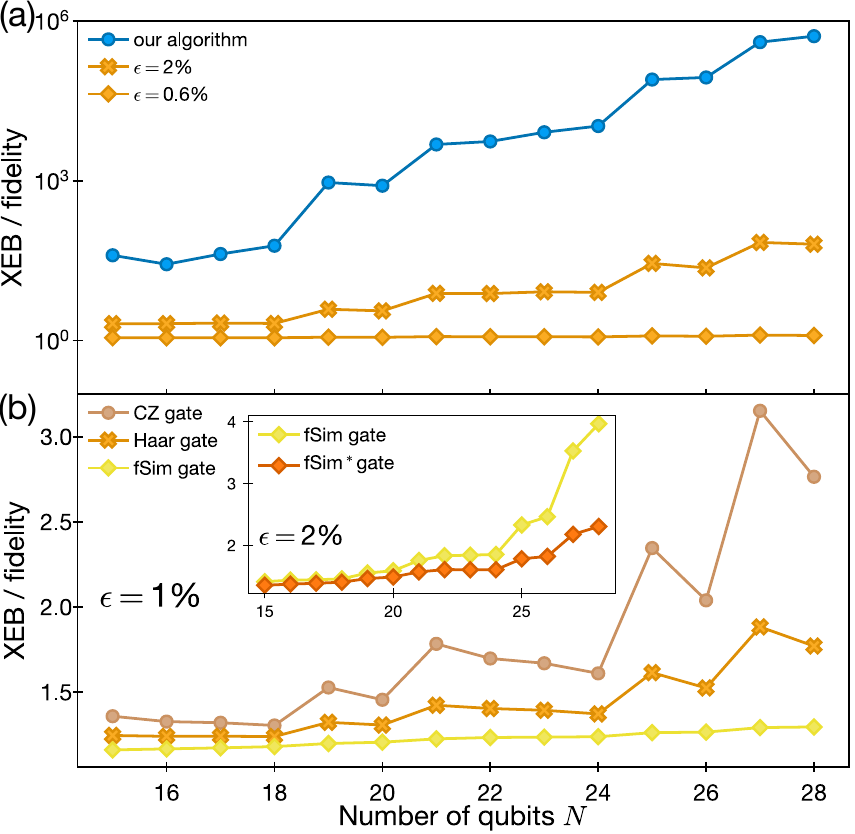}
\caption{ The ratio between 
XEB and fidelity
evaluated for quantum circuits of depth 20 in Sycamore architecture for various system sizes [according to the qubit ordering in Ref.~\cite{arute2019quantum}; see also Fig.~\ref{fig:stat_noise_fid_vs_xeb}(a)].
(a) The ratios for our algorithm (blue) are much larger than those for noisy circuits (yellow), shown with two different error rates, despite the fidelity being lower in the former case. The local gate ensemble is 2-qubit Haar.
(b) The ratios for noisy circuits with various types of gate ensembles. Out of the three standard gates (CZ, Haar, fSim with Haar random single qubit gate), the discrepancy between the XEB and fidelity is minimized in circuits with the fSim ensemble. Following on the insights from our theoretical analysis, we propose a new gate:  fSim$^*$, which corresponds to the fSim$_{\theta,\phi}$ from Eq.~\eqref{eq:fSim_formula} at $(\theta,\phi)=(90^\circ,180^\circ)$, and produces the smallest possible discrepancy between the XEB and the fidelity (see Sec.~\ref{ssec:stat_DR_mapping}). (inset) Comparison between the usual fSim gate and the new fSim$^*$ gate.}
\label{fig:intro_fid_xeb}
\end{figure}

Our key results can be summarized as follows.
First, we present a simple and efficient classical algorithm to spoof the XEB measure. 
In particular, we show that as the number of qubits increases, the performance of our algorithm scales \emph{better} on average than that of a noisy quantum simulation in a number of practical settings (see Fig.~\ref{fig:intro_scaling}).
Hence, the XEB does \emph{not} constitute a scalable measure to certify quantum advantage. 
Second, we develop a new theoretical framework to analyze and predict the XEB under various choices of quantum circuit architectures and gate ensembles.
This framework allows us to understand the relation between the XEB and the fidelity (see Fig.~\ref{fig:intro_fid_xeb}).

\medskip \noindent \textbf{Classical algorithm spoofing XEB.} Our algorithm is inspired by the observation that entanglement growth in a noisy quantum circuit is reduced by errors spread over the entire circuit in both space and time [Fig.~\ref{fig:intro_scaling}(a)]. These  effectively truncate entanglement and correlations among different subsystems.
In our algorithm, we introduce similar amount of effective errors, but they occur only at specific locations such that the quantum circuit becomes easier to simulate.
As an example, Fig.~\ref{fig:intro_scaling}(a) shows how omitting a few specific gates at certain locations (which amounts to particular types of error, i.e. \emph{gate defects}) can split a circuit into multiple disconnected sub-circuits.
Alternatively, one can apply completely depolarizing channel before and after an entangling gates.
These approaches explicitly remove correlations between subsystems.
Intuitively, when the amount of ``effective noise'' in a noisy quantum simulation is comparable to the ``effective error'' in our algorithm (proportional to the number of omitted gates), the XEB of the latter is larger due to the stronger correlation among errors [see Fig.~\ref{fig:oneerror}(b)].

Since the size of each sub-circuit is much smaller than that of the original circuit, the algorithm can be significantly faster than a direct simulation of the global circuit. In particular, when ran on 53-qubit circuits, such as Google's, it takes a few seconds using a single GPU (32GB NVIDIA Tesla V100).
 The existence of our classical algorithm has three types of implications:
\begin{enumerate}
    \item \textbf{Complexity-theoretic implications:} \emph{A linear-time classical algorithm that outperforms any noisy 1D quantum circuit.} For one-dimensional quantum circuits consisting of Haar random unitary gates~\cite{bouland2019complexity,bouland2021noise}, we present a linear-time classical algorithm which achieves higher XEB values than noisy quantum devices. Concretely, for every uncorrelated error rate $\epsilon>0$ per gate, our algorithm can spoof the XEB measure when the number of qubits is sufficiently large.
    
    \item \textbf{Experimental implications:} \emph{A highly efficient classical algorithm (1 GPU around 1s), whose performance is comparable with current experimental devices.}
    We consider a random circuit ensemble modelled after the one used in Ref.~\cite{arute2019quantum,USTC,zhu2021quantum} (see Sec.~ \ref{ssec:intro_alg_summary} and Ref.~\cite{SM} for detailed information).
    Our algorithm achieves a mean XEB value that is about 8\% of Google's experiment (53 qubits, depth 20), and 12\% and 2\% of USTC's experiments (56 qubits, depth 20 and 60 qubits, depth 24) respectively, with the running time  $\approx$1s using 1 GPU.
    
    \item \textbf{Scaling implications:} 
    Remarkably, the XEB value of our algorithm generally \emph{improves} for larger quantum circuits, whereas that of noisy quantum devices quickly deteriorates.
    Such scaling continues to hold when the number of qubits is increased while the depth of the circuit and the error-per-gate are fixed, as explicitly confirmed from numerical simulations for 1D and 2D square and the extended Sycamore architecture in Fig.~\ref{fig:intro_scaling}(b-d).
\end{enumerate}
Crucially, we show that a classical algorithm can obtain high XEB values even when the corresponding fidelity is very low. This implies that 
{\it high values of XEB cannot certify quantum advantage}. Even if one estimates the fidelity of each individual gate separately and observes good agreement between XEB and the anticipated circuit fidelity, as is the case in Ref.~\cite{arute2019quantum,USTC,zhu2021quantum}, this does not necessarily imply high many-body fidelity without additional assumptions such as the independence and the homogeneity of errors. In other words, XEB cannot be used as a ``black-box" measure for certification. 

\medskip \noindent \textbf{Understanding XEB and circuit fidelity via mapping to a statistical mechanics model.}  We present a way to analyze quantum circuit dynamics using classical statistical physics.
Specifically, for a wide class of random circuit ensembles involving single qubit Haar random gates, we show that the dynamics of
both noisy quantum circuits and our classical algorithm can be understood in terms of an effective diffusion-reaction process which was originally used to study the scrambling of circuits~\cite{mi2021information}.
In this effective description,
the application of each layer of a quantum circuit translates to particles undergoing a random walk (diffusion) for a single time step on a graph representing the  circuit architecture.
Furthermore, each particle can duplicate itself, and a pair of particles may recombine into a single particle at a certain rate (reaction).
The rates of particle diffusion and reaction are determined by the properties of two-qubit quantum gates, such as the average amount of entanglement they generate. The XEB and the fidelity of ideal circuits are given by different aspects of particle distribution at the last circuit layer, as we elaborate in Sec.~\ref{ssec:stat_DR_dynamics}.

The XEB value in a noisy circuit and our algorithm will decrease from the ideal value when a particle hits a defective (omitted or noisy) gate.
In the case of noisy quantum circuits, every gate is noisy,
so the decrease in the XEB value is proportional to the total number of particles in the diffusion-reaction process. Intuitively, when the system size grows, there are more particles hitting noisy gates and thus the XEB value becomes smaller.
In our algorithm, the XEB decreases whenever a particle hits an omitted gate at the boundaries of disconnected sub-regions.
Intuitively, when the system size grows, there is more space for particles to diffuse away from the boundary and thus, in general, the XEB value can become larger. This qualitatively explains the asymptotic scaling of XEB in Fig.~\ref{fig:intro_scaling}.

The mapping to diffusion-reaction models can also help explain the XEB's role as a proxy for the fidelity.
As we elaborate in Sec.~\ref{ssec:stat_DR_dynamics}, the XEB and the fidelity agree with each other if and only if the particle distribution at the last circuit layer reaches a certain homogeneous, steady-state profile.
Both in noisy circuits and in our algorithm, the final distribution is modified by particles hitting defective gates, 
leading to the discrepancy between the XEB and the fidelity. 

In our algorithm, the deviation from the target distribution is induced by the presence of omitted gates located along boundaries of disconnected subsystems, which leads to a strong violation of the homogeneity of the particle distribution. Therefore, XEB and fidelity are very different in this case.
On the other hand, in noisy circuits, the particles hit defective gates uniformly across the system, and thus the homogeneity is retained. This results in a small discrepancy between XEB and fidelity, especially in the weak-noise regime [see Fig.~\ref{fig:intro_fid_xeb}(a)] 

For noisy chaotic systems, it is believed that faster scrambling leads to a better agreement between XEB and fidelity~\cite{zhang2017observation,arute2019quantum,choi2021emergent,liu2021benchmarking}. In the diffusion-reaction model, the reaction rate and the diffusion rate are related to the scrambling speed and the above-mentioned intuition is reflected in a faster approach to the steady state for rapid mixing. Compared to several other commonly-studied two-qubit gates, like the control-Z and Haar-random gates,
the fSim gate used in Google's experiment has a similar reaction rate but a faster diffusion rate. Therefore, it produces  the smallest discrepancy between XEB and fidelity among these gates. However, the fSim gate is still not the optimal choice. By increasing the reaction rate further, we find the optimal gate, which we call fSim$^{*}$ due to its similar structure; see the inset of Fig.~\ref{fig:intro_fid_xeb}(b) for the comparison between fSim and fSim$^{*}$.

The choice of the single-qubit ensemble can also affect the  diffusion-reaction processes of particles.
In particular, we find empirically that Google's choice of single-qubit gates, which maps both computational basis states, e.g. $\ket{0}$ and $\ket{1}$, to their equal superposition with opposite phases, leads to significantly faster diffusion-reaction processes and makes our algorithm relatively less effective.

In the case of one-dimensional circuits with Haar random gates, a more detailed scaling analysis is possible by mapping  a quantum circuit to a two-dimensional classical Ising model~\cite{hayden2016holographic, you2018machine, hunter2019unitary,zhou2019emergent, jian2020measurement, bao2020theory, napp2019efficient}, which can be regarded as a special case of the diffusion-reaction model. 
In the case of ideal circuits, the classical model exhibits the $\mathbb{Z}_2$ Ising symmetry.
However, when noisy processes or gate defects are introduced, they appear as effective external 
magnetic fields, which break the Ising symmetry.
In this picture, the deviation of the XEB from unity characterizes the degree of symmetry violation~\cite{bao2020theory}.
Crucially, the noise and omitted gates have distinct effects, appearing as a bulk and boundary fields, respectively.
In the limit of large circuits, the bulk field has a stronger effect than the boundary field, even when the strength of the bulk field is vanishingly small. Closely related to spontaneous magnetization in the ferromagnetic phase, this phenomenon provides an intuitive explanation for the superior XEB scaling of our classical algorithm, compared to that of noisy quantum circuits. 

\subsection{Organization of the paper}
The rest of our paper is organized as follows. The next two sections include a summary of the necessary background and a detailed presentation of our results. We review the definition, properties, and applications of the XEB in Sec.~\ref{sec:intro_XEB}. In Sec.~\ref{sec:intro_alg}, we describe our algorithm and random quantum circuit ensembles in more detail, and summarize our results and their implications. 
We discuss related works on the XEB spoofing in Section~\ref{ssec:related}. 
Then, we introduce the technical aspects of mappings to statistical physics models: the diffusion-reaction model in Sec.~\ref{ssec:stat_DR_mapping}, with the detailed discussion of the relationship between the XEB and the fidelity in Sec.~\ref{ssec:stat_DR_dynamics}, and the Ising model for 1D circuits with the Haar gate ensemble in Sec.~\ref{ssec:stat_Ising}.  
Finally, we conclude in Sec.~\ref{sec:outlook}, where we discuss several potential ways to overcome the vulnerabilities of the XEB and present a few interesting future directions.
We defer the discussion of the technical details of the heuristic analysis and the diffusion-reaction model, as well as the description of our improved algorithm, to the Supplemental Material (SM)~\cite{SM}.

\section{Linear cross-entropy benchmark}\label{sec:intro_XEB}

We first review the definition of the XEB and its properties, formally introduce the XEB test, and establish our notations.

\emph{Linear cross-entropy.} The XEB corresponds to  the linearized version  of the cross-entropy (i.e., the quantity $-\sum_xq_x\log p_x$, also known as the log-likelihood), which  is commonly used to characterize the closeness between the data and target distributions~\footnote{It is known that maximizing the (standard, non-linear) cross-entropy is equivalent to minimizing the Kullback-Leibler divergence, which is a commonly used quantity for a statistical test of the closeness between the data and target distributions.}.
The motivation to adopt the linearized version is to minimize statistical fluctuation~\cite{arute2019quantum,rinott2020statistical} when estimating the XEB empirically.
Both versions can be used to estimate the fidelity under common error models~\cite{arute2019quantum} for sufficiently chaotic circuits, or (equivalently) sufficiently deep random circuits~\cite{harrow2009random}.
It is generally believed that simulating complex quantum systems with high fidelity is classically intractable, and so obtaining high XEB values is (na\"ively) expected to be hard as well.

Let $U$ be an $N$-qubit unitary, and let $p_U$ be the probability distribution induced by measuring $U\ket{0^N}$. 
If $q$ is a probability distribution over $\{0,1\}^N$ then the XEB value of $q$ with respect to $U$ is 
\[
\chi_U(q) = \chi_{p_U}(q) = 2^N \sum_{x\in \{0,1\}^N} q(x)p_U(x) - 1 \;.
\]
If $q$ is uncorrelated with $p_U$, $\chi_U(q)=0$ \cite{arute2019quantum,boixo2018characterizing}.
On the other hand, $\chi_U(q) \approx 1$ if $q$ is similar to $p_U$ for random-enough deep circuits, where we expect $p_U$ to be characterized by the Porter-Thomas distribution. Therefore, $\chi_{U}(q)$ serves as a proxy to estimate how $p$ and $q$ are correlated with one another.

We will often consider the unitary $U$ to be a random variable sampled from  a \emph{distribution} over $N$-qubit unitary transformations that correspond to choosing a circuit with random gates from a prescribed architecture. In this case, the quantity $\chi_U(q)$ is a random variable, and we denote its expectation value over different $U$
by $\ex{\chi_{U}(q)}{U}$.
\paragraph*{Empirical vs.~expected XEB value.}
The XEB value for a given circuit $U$  can be
empirically estimated with a relatively small number of samples, compared to its non-linear counterpart, using an unbiased estimator $\tilde{\chi}_{p_U}(q) = \frac{2^N}{m} \sum_{i=1}^m p_U(x_i) -1 $, where $x_1,x_2,\dots x_m$ are $m$ independent samples obtained from $q(x)$, which can be sampled in practice. In particular, the error $|\tilde{\chi}_p - \chi_p|$ scales as $\sim 1/\sqrt{m}$.
Since $U$ is sampled from an ensemble, $\chi_U(q)$ is effectively also random, when $m$ tends to infinity. Therefore, in practice, it is common to take another empirical average $\ex{\chi_U(q)}{U}$ over $K$ independent circuits $U_1,\ldots,U_K$. For example, in Google's experiment~\cite{arute2019quantum}, they choose $m\approx7\times10^6$ and $K=10$; in USTC's two experiments~\cite{USTC}, they choose $m\approx1.9\times 10^7,K=10$ and $m\approx7\times 10^7,K=12$ respectively. To get a high confidence in the estimation of $\ex{\chi_U(q)}{U}$, the number of repetitions $K$ should be chosen proportionally to the square of the inverse of the standard deviation (STD) of $\chi_U(q)$.

\paragraph*{Standard deviation.} In the body of this work, we focus on the average value of the XEB, while the statistical fluctuation of empirical estimation is ignored. However, because we are dealing with random circuit ensembles, it is important to control the STD of the classical algorithm's output. 
More discussion of the STD in various settings is presented in the SM~\cite{SM}.

\paragraph*{Classical and noisy quantum simulations.} Let $C$ be a classical randomized algorithm that takes as an input a classical description of an $N$-qubit unitary $U$ and allows us to sample an $N$-bit string $x\in \{0,1\}^N$ as outputs with probability distribution $q_{C(U)}$. We
define the XEB value of $C$ with respect to $U$ as $\chi_U(C):=\chi_U(q_{C(U)})$. We will use $\mathcal{N}_\epsilon$ to be a \emph{noise operator} such that $\mathcal{N}_\epsilon(U)$ corresponds to applying an $\epsilon$-noisy simulation of $U$.  We will model the noise as independent single-qubit noise, which can be depolarizing or amplitude-damping, see Sec.~\ref{ssec:stat_DR_mapping}. We denote by $\chi_U(\mathcal{N}_\epsilon)$ the XEB value of the distribution of the noisy circuit $\mathcal{N}_\epsilon(U)$, applied to $\ket{0^N}$, with respect to the ideal distribution induced by measuring $U\ket{0^N}$.

\paragraph*{Quantum advantage via XEB.}
The demonstration of quantum advantage consists of designing a task that can be performed on a physical quantum device, while being intractable for all polynomial-time classical algorithms.
One such task, which has been proposed recently, is to achieve a high XEB value~\cite{arute2019quantum,aaronson2019classical}. In this scenario, in order to demonstrate quantum advantage using an $\epsilon$-noisy quantum simulator, we need to come up with a probability distribution $U$ over quantum circuits such that for every efficient classical algorithm $C$, $\chi_U(\mathcal{N}_\epsilon)\gg\chi_U(C)$ with high probability over the randomness of $U$. To make the comparison between $\chi_U(\mathcal{N}_\epsilon)$ and $\chi_U(C)$ rigorous, we need to specify the parameters of the quantum device, such as the number of qubits $N$ and the noise strength $\epsilon$. In the theoretical/asymptotic setting, we pick an arbitrarily small constant $\epsilon>0$ and consider the relation between $\chi_U(\mathcal{N}_\epsilon)$ and $\chi_U(C)$ when $N$ tends to infinity and $C$ ranges over all polynomial-time classical algorithms. In practice, we use values $N$ and $\epsilon$ that are experimentally achievable. As an example, Google's quantum simulator~\cite{arute2019quantum} uses $N=53$ and the value of $\epsilon$ is empirically estimated to be less than $0.5\%$~\footnote{In realistic settings, one needs to distinguish the error rates associated with quantum gates and readout processes.}.

\paragraph*{Computational hardness of achieving high XEB values.}
As mentioned previously, when the circuit architecture and gate ensemble are chaotic enough without any noise or error, $\chi_U(U)=\chi_U(\mathcal{N}_0(U)) \approx 1$ for almost all $U$. In the presence of noise ($\epsilon>0$), the distribution $\mathcal{N}_\epsilon(U)$ is expected to approach the uniform distribution exponentially in the depth of the circuit; thus, $\chi_U(\mathcal{N}_\epsilon)$ goes to $0$ exponentially in the depth of the circuit as well. Nevertheless, when the quantum circuit size is finite and the strength of noise is sufficiently small, a noisy quantum simulation could achieve non-vanishing XEB value that implies statistical correlation between sampled and ideal distributions. For example, an XEB value of $2.24\times10^{-3}$ in $53$-qubit and depth-$20$ 2D circuits was achieved in Ref.~\cite{arute2019quantum}. In Ref.~\cite{USTC,zhu2021quantum}, XEB values of $ 6.62\times10^{-4}$ and $ 3.66\times10^{-4}$ were achieved in  56-qubit circuits of depth $20$ and 60-qubit circuits of depth $24$, respectively.

There are two types of arguments for the difficulty of achieving an XEB value $\chi_U(C)$, using a classical algorithm $C$, that is bounded away from zero.
The first argument, put forward in Ref.~\cite{arute2019quantum}, was based on the conjecture that brute-force simulation is the optimal classical approach. This conjecture was recently refuted~\cite{gray2020hyper,huang2020classical,pan,fu2021closing,refute1,refute2}.
The other, more subtle argument, relies on conjectures in computational complexity. 

Aaronson and Gunn~\cite{aaronson2019classical} reduced the classical hardness of spoofing the XEB measure to the Linear Cross-Entropy Quantum Threshold Assumption (XQUATH), which is a stronger version of the Quantum Threshold Assumption (QUATH)~\cite{aaronson2016complexity}. Our results appear to refute XQUATH, at least in some instances. 
See more details in Sec.~\ref{ssec:related} and the SM~\cite{SM}. 
\begin{figure}
\includegraphics[width=\linewidth]{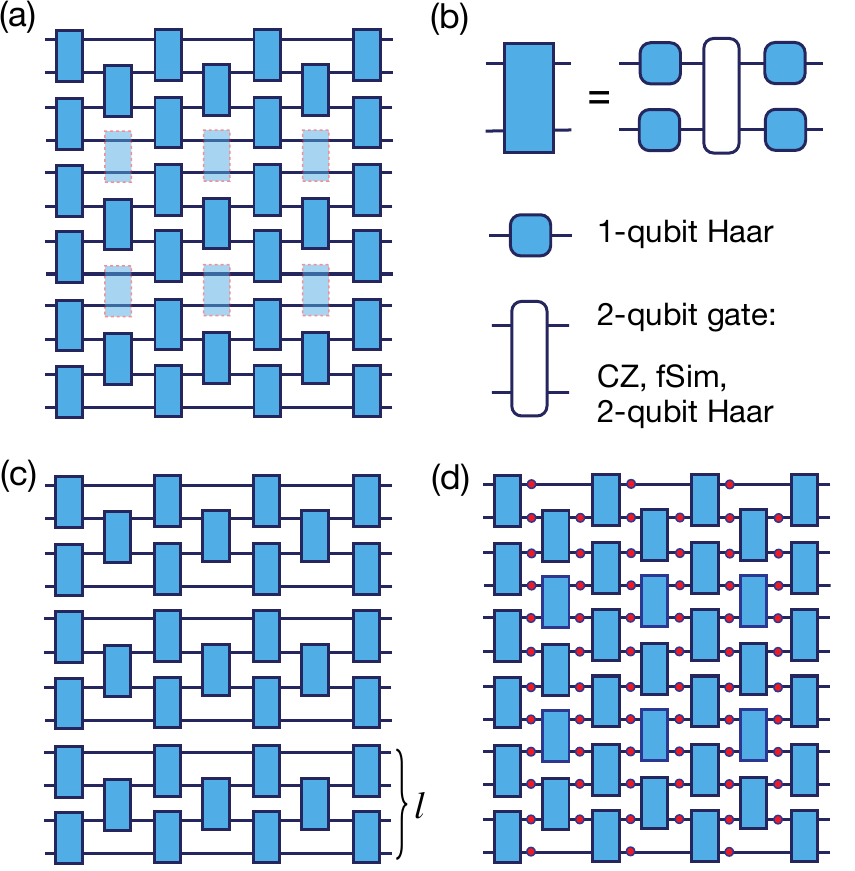}
\caption{Illustration of our algorithms.
(a) The target (ideal) circuit to simulate. The light blue gates correspond to the ones omitted in (c).
(b) Each random two-qubit gate in our circuit consists of any (potentially fixed) two-qubit gate surrounded by 4 single-qubit Haar random gates. When compared to experimental data, the single-qubit random gates are chosen to be a slight modification of those used in Ref.~\cite{arute2019quantum,USTC,zhu2021quantum},
(c) Our algorithm: one can approximately simulate the ideal circuit by simply omitting a certain subset of gates (in light blue color with red dashed boxes) in the ideal circuit (a). Then, the circuit separates into isolated subsystems. We denote the maximal size of a subsystem as $ l$.
(d) Noisy circuit: we model the dynamics of noisy quantum circuits by applying probabilistic single-qubit noise (e.g. depolarizing or amplitude damping) channels  to all qubits, after each layer of unitary evolution.
}
\label{fig:intro_alg}
\end{figure}

\section{Spoofing algorithms}\label{sec:intro_alg}

We now describe an efficient classical algorithm $C$ that, in a wide range of physically relevant situations, produces a probability distribution with XEB values larger or comparable to that of an $\epsilon$-noisy circuit, at least on average. In such situations, the existence of our algorithm suggests XEB on its own is not a good benchmark for certifying quantum advantage.

We first describe our algorithm at a high level, deferring its detailed analysis and discussion to Sec.~\ref{ssec:intro_alg_summary}. The intuition behind our algorithm borrows ideas from the following observation on noisy simulation of quantum circuits. In a quantum simulation, the presence of noise can remove entanglement and other correlations (either quantum or classical) within the system. Namely, different parts of the system are approximately decoupled. In our algorithm, we compete with an $\epsilon$-noisy quantum simulation by trying to ``rearrange'' the same amount of total noise in the most favorable way to reduce the computational complexity. 
This will also allow us to obtain relatively high XEB values owing to its vulnerabilities explained in Sec.~\ref{ssec:vulnerability}.
Specifically, we do so by dividing the quantum circuit into isolated subsystems that can be each simulated independently at much lower cost; see Fig.~\ref{fig:intro_alg} for an example of a 1D circuit, and Fig.~\ref{fig:intro_partition} for a 2D circuit). Intuitively, using a similar amount of noise ``budget'' guarantees that our algorithm achieves a better XEB value comparable to the noisy quantum simulation, while (classically) simulating the smaller isolated subsystems will be exponentially faster than simulating the original circuit. 
The above explanation is very qualitative and glosses over some important aspects.
In Sec.~\ref{ssec:stat_DR_dynamics}, we give a more quantitative analysis to motivate our algorithm based on the mapping to the diffusion-reaction model.

\subsection{Basic algorithm}\label{ssec:intro_alg}
We now describe our classical algorithm. For concreteness, we illustrate our algorithm using 1D quantum circuits although it is straightforward to generalize it to other circuit architectures. Let $N$ be the total number of qubits, $d$ be the depth, and $ l$ be
the maximum size of subsystems [see Fig.~\ref{fig:intro_alg}(a) for an example with $N=12$, $d=7$, and $ l=4$]. We start by partitioning the $N$ qubits into subsystems of size at most $ l$ by
omitting any gates acting across two different subsystems [see Fig.~\ref{fig:intro_alg}(c)].
We then simulate each subsystem separately. 
Using brute-force methods, simulating a subsystem of $ l$ qubits takes at most $2^{O(l)}d$ time. There are $\lceil N/ l\rceil$ subsystems and hence the total running time of our algorithm is at most $\frac{2^{O( l)}}{ l} Nd.$ In particular, if $ l$ is fixed and does not scale with the total system size $N$ or depth $d$, the time complexity is linear in the circuit size $Nd$.
We claim that the bitstring  distribution induced by the factorizable wavefunction obtained from our algorithm achieves relatively high XEB values.

\subsection{Improving the algorithm}
\label{ssec:algorithm_improve}

While our basic algorithm is simple and relatively straightforward to implement, it already
has significant consequences for the computational hardness of obtaining high XEB values.
Moreover, its practical performance can be 
further improved via the following modifications. 

{\bf Top-$k$ post-processing method.} Given the output distribution $q_C(U)$ produced by our algorithm $C$,  which is correlated with the ideal distribution $p_U(x)$, it is possible to amplify such correlations by using the so-called \emph{top-$k$ post-processing heuristic}.
In this method, one modifies the bitstring distribution $q_C(x)$ by ordering the bitstrings $x_i \in \{x\}$ from largest $q_C(x_i)$ to the smallest, selecting first $k$ of them (or equivalently setting the probability of the others to 0),
\begin{equation}
    q_C(x_i)\to\tilde{q}_C(x_i)=\begin{cases} 0 &{\rm if }~ i \leq k\\
    1/k & {\rm if }~ i>k
    \end{cases}.
\end{equation}
Since we can efficiently compute the probability distribution $q_C(x)$ produced by the original algorithm, we can also efficiently compute the amplified probability distribution.

The intuition behind this heuristic can be understood as follows.
The XEB is equivalent to evaluating the average of $p_U(x)$ weighted by $q(x)$ up to an unimportant scaling factor $2^N$, and a constant $-1$. 
If $q(x)$ is modified such that $q(x)$ is increased (decreased) for bitstrings $x$ with relatively large (small) values of $p_U(x)$, then the weighted average will increase.
Given that $q(x)$ and $p_U(x)$ are already positively correlated, such behavior is naturally expected for our top-$k$ post-processing heuristic, at least on average.

In fact, we can prove that the top-$k$ method increases the XEB if its value is positive and the STD over circuit realizations is not too large. The second requirement is necessary to avoid the situation where some occasional $x$ with small $p_x$ but large $q_x$ will be amplified (in another words, ``over-fitting").
Unfortunately, this second criterion is not satisfied by our basic algorithm where we simply omit gates.
This issue, however, can be straightforwardly addressed using the following method.

{\bf Self-averaging algorithm.} 
In order to decrease the STD, we make a small modification to our basic algorithm: instead of omitting gates, we insert maximal depolarizing noise or equivalently take average over different realizations of our basic algorithm with random single qubit unitary at the position of omission. This \emph{self-averaging algorithm} guarantees the positivity and small  STD conditions. However, the computational resources required are larger since we need to simulate mixed state evolution.
Interestingly, for a certain class of entangling gates (including the one used in recent experiments~\cite{arute2019quantum,USTC,zhu2021quantum}) that exhibit the ``maximal scrambling speed'' and that hinders the application of our basic algorithm, one can substantially reduce the computational resources needed for such mixed-state simulation.
This is possible because for that class of entangling gates the effect of depolarizing noise can be propagated efficiently [see SM~\cite{SM} for more detail].

{\bf Combining algorithmic  improvements.} In Figure~\ref{fig:intro_advantage_regime}, we present the increase of the XEB for the modified version of Google's gate set ensemble by several orders of magnitude after the application of the top-$k$ method on the self-averaging algorithm.
While the discussion above is mostly focused on the mean value of the XEB, it is important to show that our result also holds for typical, individual instances of quantum circuits with a high probability. In the SM~\cite{SM}, we show that the self-averaging algorithm offers a much better control over the STD, and guarantees the benefit of using the top-$k$ method. Additionally, we show evidence that the STD of the top-$k$ method decreases as $1/\sqrt{k}$.

\subsection{Performance and implications}
\label{ssec:intro_alg_summary}
Now, we present a comprehensive analysis of the performance of our algorithms and its implications.
We consider algorithms both with and without the top-$k$ post-processing heuristic introduced in the previous section.
For practical relevance, we focus on 1D and 2D circuit architectures.
For 1D circuits, we theoretically and numerically show that our basic algorithm can achieve, in linear time, a higher average XEB value than noisy quantum systems. More specifically, we show that setting subsystem size to be constant ($ l=O(1)$) is sufficient for our algorithm to obtain a higher XEB value than that of $\epsilon$-noisy quantum simulations, for every constant $\epsilon>0$, for sufficiently large $N$.
This is due to the distinct scaling behavior of the XEB value for noisy circuits and our algorithm; we discuss in detail the origin of this difference in the scaling behavior in Sec.~\ref{ssec:stat_Ising}.

\begin{figure}
\includegraphics[width=1.0\linewidth]{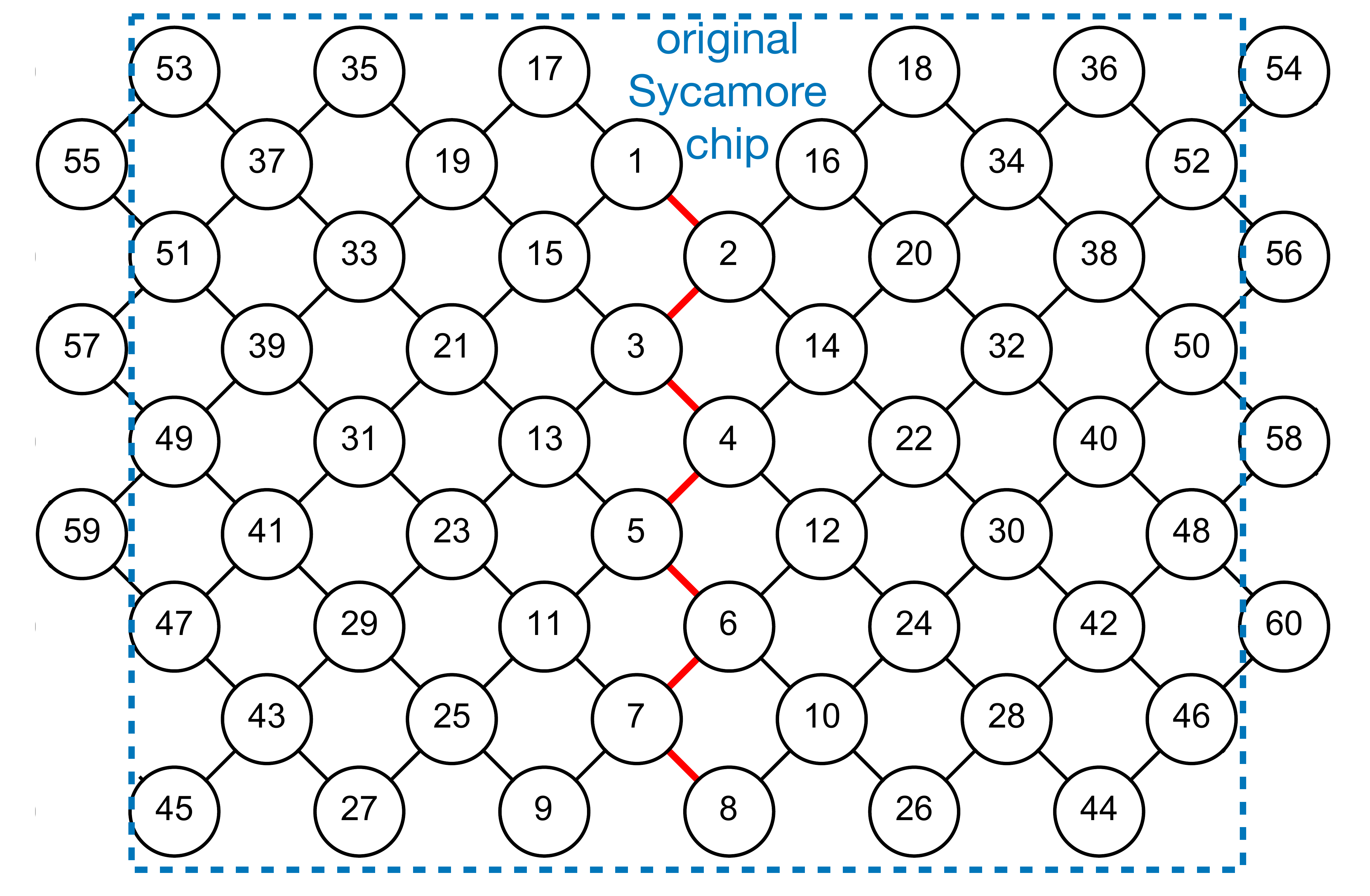}
\caption{
Sycamore circuit architecture from Ref.~\cite{arute2019quantum} and its horizontal extension. The gates marked with red lines are omitted in our algorithm. The Zuchongzhi architecture is very similar; see Ref.~\cite{USTC,zhu2021quantum} for more detail.
}
\label{fig:intro_partition}
\end{figure}

For 2D circuits, we consider Google's Sycamore architecture, which has $N=53$ qubits~\cite{arute2019quantum}, and we choose $ l\approx\left\lceil{N/2}\right\rceil = 27$ (Fig.~\ref{fig:intro_partition}). We also consider USTC's Zuchongzhi architectures which have 56 qubits and 60 qubits respectively, and we choose $l\approx28$ for both cases (with some qubits being omitted).
A subsystem of this size can be simulated by one NVIDIA Tesla V100 GPU with 32GB memory in about 1 second~\cite{julia1,julia2}. 
We analyze the performance of our algorithms on circuits constructed from  the following different quantum-gate ensembles:
\begin{description}
\item[CZ ensemble] Each random two-qubit gate is composed of the control-Z gate surrounded by four independent single-qubit Haar random gates [see Fig.~\ref{fig:intro_alg}(b)].
\item [Haar ensemble] Each random two-qubit gate is a two-qubit Haar random gate.
\item[fSim ensemble] Similar to CZ ensemble, but replacing the control-Z gate by the fSim gate, which is defined as
\begin{equation}
\label{eq:fSim_formula}
\text{fSim}_{\theta,\phi}=
\begin{pmatrix}
1 & 0 & 0 & 0 \\
0 & \cos(\theta) & -i\sin(\theta) & 0 \\
0  & -i\sin(\theta)  & \cos(\theta) & 0  \\
0 & 0 & 0 & e^{-i\phi} 
\end{pmatrix},
\end{equation}
with parameters $\theta=90^{\circ},\phi=60^{\circ}$~\cite{arute2019quantum} (denoted as fSim); we also define a new gate fSim$^*$ which has $\theta=90^{\circ},\phi=0^{\circ}$.

\item[fSim with discrete 1-qubit ensemble]
Similar to fSim ensemble, but replacing the 1-qubit Haar random gate by $Z(\theta_1)VZ(\theta_2)$
where $V$ is chosen randomly from $\{\sqrt X,\sqrt Y, \sqrt W\}$ ($W=(X+Y)/\sqrt2$) but the two $V$s between two successive layers on the same qubit should be different; and $Z(\theta_i)$ is chosen randomly from $[0,2\pi)$. 
\end{description}
The last ensemble is closely modelled after quantum circuits used in recent experiments~\cite{arute2019quantum,USTC,zhu2021quantum}.
The only modification is that, in experiments, the single qubit rotation angles $\theta_i$' are not actively controlled, but rather determined by the specific ordering of quantum gates and the qubit specification at hardware level.
We expect that this difference does not influence the performance of our algorithm significantly, because we also consider the case where $\theta_i$ is chosen randomly from either 0 or $\pi$ (which corresponds to $I$ or $Z$ operator, respectively). The numerical result shows that the average XEB values for the top-1 method in the two cases are similar: $0.00018$ ($\theta_i \in [0, 2\pi)$) and $0.0004$ ($\theta_i \in \{0, \pi\}$), respectively, for the Sycamore architecture (53 qubits, 20 depth). Therefore, we argue that the $z$-rotation part does not influence the XEB value too much.

\subsubsection{Implications for 1D quantum circuits}\label{ssec:1Dqc}

We start by discussing the performance of our algorithm on 1D circuits with gates drawn from the Haar ensemble.
For the purpose of this section, $C$ denotes either the algorithm introduced in Sec.~\ref{ssec:intro_alg} or its self-averaging version described in detail in the SM~\cite{SM}.  The self-averaging version has the same average XEB but a smaller STD, at the cost of requiring more computational power.
However, we consider constant subsystem size $l=O(1)$; thus, even the self-averaging algorithm runs in the time linear in $Nd$.

\begin{result}{(1D circuits with Haar ensemble)}
For 1D random quantum circuits with gates drawn from the Haar ensemble, 
\begin{itemize}
\item
for any constant $\epsilon>0$ and large enough $N$ (roughly $N\epsilon>1$), we have
\begin{equation}\label{eq:1D exp}
\ex{\chi_U(C)}{U} \ge \ex{\chi_U(\mathcal{N}_\epsilon)}{U}
\end{equation}
for both the basic and the self-averaging algorithms.
\item
we conjecture that
\begin{equation}\label{eq:1D var}
\sqrt{\textsf{Var}(\chi_U(C))_U} \approx \ex{\chi_U(\mathcal{N}_\epsilon)}{U} \,
\end{equation}
for the self-averaging algorithm (see the SM~\cite{SM}), which is suggested by numerical simulations. Namely, the standard deviation of $\chi_U(C)$ is comparable to its expectation value $\ex{\chi_U(C)}{U}$.
\end{itemize}
Combined, this yields a linear-time classical algorithm that spoofs XEB for any noisy quantum simulation of 1D circuits with the Haar gate ensemble, when the number of qubits is large enough.
\end{result}
Eq.~\eqref{eq:1D exp} states that the average XEB of our algorithm is at least as large as that of any noisy circuit with a constant noise level $\epsilon>0$. As mentioned previously, in practice, we would like the conclusion of Eq.~\eqref{eq:1D exp} to generalize to typical circuits $U$ (not only on average) --- this can be guaranteed by showing that the variance of the XEB value is small. This notion is expressed in Eq.~\eqref{eq:1D var}, which says that the variance is comparable to the expectation value, and hence our algorithm works for typical instances with large probability. Notice that, in the large depth limit, we expect this to hold only for the self-averaging algorithm. When discussing 1D circuits, where the purpose is to provide complexity-theoretic implications, the analysis of the STD concerns only the self-averaging algorithm.
See the SM~\cite{SM} for more detailed discussion.

From a technical point of view, our results are derived by showing that the following quantities decay exponentially with the depth of the circuit
\begin{eqnarray}
\nonumber \ex{\chi_U(C)}{U}&=&O(e^{-\Delta_1 d}), \\
\nonumber  \ex{\chi_U(\mathcal{N}_\epsilon)}{U}|_{\epsilon\rightarrow0\text{ while } N\epsilon>1} &=&O(e^{-\Delta_3 d}).
\end{eqnarray}
Additionally, numerical simulations support the scaling of the STD as
\begin{equation*}
   \sqrt{\ex{\chi^2_U(C)}{U}-\ex{\chi_U(C)}{U}^2}=O(e^{-\Delta_2 d}) 
\end{equation*}
for some constants $\Delta_1,\Delta_2>0$ that depend on the subsystem size $l$ and $\Delta_3>0$ that depends on the noise level $\epsilon$. 

\begin{figure}
\includegraphics[width=\linewidth]{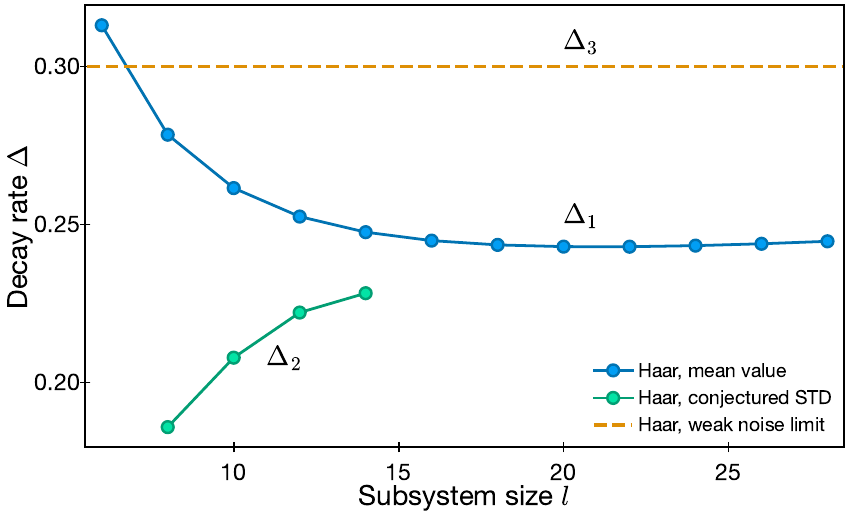}
\caption{Exponential decay rates in 1D circuits with the Haar gate ensemble. The mean value (blue) and the standard deviation (green) of the XEB obtained by our algorithm. The horizontal (dashed orange) line is the mean XEB value of the noisy circuit in the weak-noise limit. Intuitively, a smaller $\Delta$ corresponds to a larger XEB value. The STD is estimated by an approximate method~\cite{SM}, since the direct calculation is not practical. In the SM~\cite{SM}, 
 we give a strong numerical evidence  that this approximation is in fact a conservative estimation, i.e., the true STD should be even smaller ($\Delta_2$ should be larger).
}
\label{fig:intro_1D_gaps}
\end{figure}

We emphasize that this scaling is unexpected: the decay rate of the expected XEB value achieved by our algorithm does not depend on the system size but only depends on the depth of the circuit.
Numerically, we show in Fig.~\ref{fig:intro_1D_gaps} an estimate on $\Delta_1,\Delta_2$ and $\Delta_3$ with
$\epsilon\to0$ while keeping the system size large enough; i.e., $N\epsilon>1$. For the Haar ensemble,
our numerical results show $\Delta_1<\Delta_3$, where a larger $\Delta$ implies a smaller corresponding quantity in the deep-circuit limit. The numerical calculations suggest that $\Delta_1\approx \Delta_2$: around $l=14$, the gap between the two is very small and $\Delta_2$ (green curve) seems to increase continuously. The green curve is expected to be only a conservative estimation, as explained in the SM~\cite{SM}.

\subsubsection{Implications for quantum circuits in 2D experimentally relevant architectures}\label{ssec:sycamore}
\begin{figure}
\includegraphics[width=\linewidth]{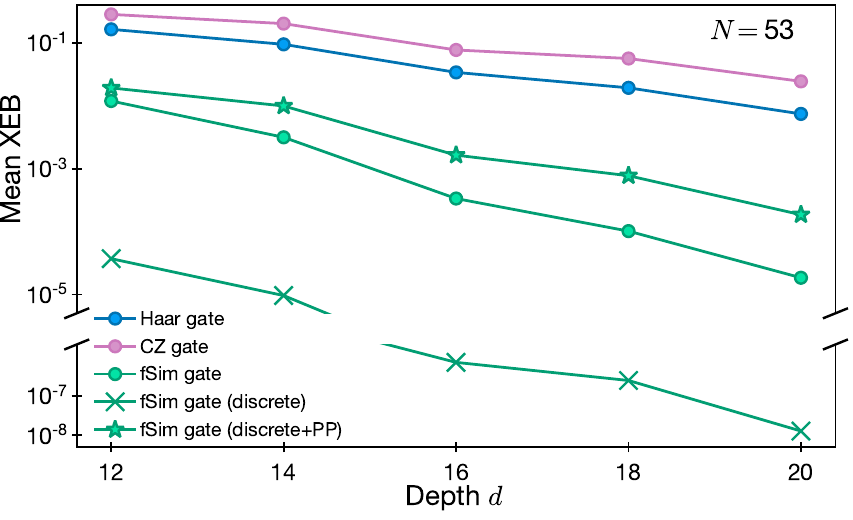}
\caption{Mean XEB obtained by our algorithm for different two-qubit gate ensembles, on Google's circuit geometry. Circles denote the Haar single-qubit gate set, while the green crosses (stars) correspond to the more experimentally relevant discrete set (with amplification using the \emph{top-$k$} method).}
\label{fig:intro_advantage_regime}
\end{figure}

Next, we consider 2D quantum circuits in the Sycamore and Zuchongzhi architectures in two different settings. First, we focus on the role of the two-qubit gate, and we analyze the performance of our algorithm for three different two-qubit gate ensembles: Haar, CZ, and fSim. For the single-qubit gate we choose either independent Haar-random gates which allows for efficient analysis using the diffusion-reaction model or the more experimentally-relevant discrete gate set. Second, we compare our algorithm against the experimental results of Refs.~\cite{arute2019quantum,USTC,zhu2021quantum}. There, we focus on the fSim gate, and we assume the experimentally relevant discrete single-qubit gate set.
These analyses lead to two main results, summarized in Fig.~\ref{fig:intro_advantage_regime} and Table~\ref{tab:running_time}.
For numerical calculations, we used a single GPU machine (32GB NVIDIA Tesla V100). 

\begin{result}{(Different gate ensembles)}
In the Sycamore architecture with $N=53$, $d=20$ with Haar-random single-qubit gates, our algorithm (using the partition in Fig.~\ref{fig:intro_partition}) has the following properties:
\begin{itemize} 
\item the algorithm achieves significant average XEB value for all depths shown in Fig.~\ref{fig:intro_advantage_regime}. As a reference, the expected XEB value of a noisy quantum device with depth 20 and error rate $\epsilon\approx0.5\%$ is $\approx0.002$;
\item the choice of the two-qubit gate affects the value of XEB, which can be understood in terms of the diffusion-reaction model~\ref{sec:DR};
\item the discrete single-qubit ensemble results in much lower XEB values (green crosses in Fig.~\ref{fig:intro_advantage_regime}), which is caused by the faster scrambling time;

\item the running time (computing the vector of output probabilities) is only 4-8 seconds;
\end{itemize}
\end{result}

\begin{table*}
\begin{center}
\begin{tabular}{|c||c|c|c|}
\hline
   & Google~\cite{arute2019quantum} & USTC-1~\cite{USTC} & USTC-2~\cite{zhu2021quantum}
\\
\hline
\hline
system size & 53 qubits, 20 depth & 56 qubits, 20 depth & 60 qubits, 24 depth
\\
\hline
claimed running time on supercomputer~\cite{zhu2021quantum} & 15.9d & 8.2yr & $4.8\times10^{4}$yr
\\
\hline
running time on quantum processor & 600s & 1.2h & 4.2h
\\
\hline
experimental XEB   & $2.24\times10^{-3}$  & $6.62\times10^{-4}$  & $3.66\times10^{-4}$
\\
\hline
\hline
running time of our algorithm (1 GPU$^{(a,b)}$) & 0.6s & 0.6s &  1.5s
\\
\hline
XEB of our algorithm$^{(b)}$ & $1.85\times10^{-4}$ & $8.18\times10^{-5}$ & $7.75\times10^{-6}$  \\%&
\hline
ratio of ours to experimental XEB  & $8.26\%$ & $12.4\%$ & $2.12\%$\\ 
\hline
\end{tabular}
\end{center}
\caption{
The comparison of XEB values (using the top-$k$ post-processing) and running times in the quantum advantage regime.
We find that the average XEB values from our algorithm is largely independent of the choice $k\lesssim 10^4$ (corresponding to more than $k^2\sim 10^8$ distinct bitstrings for two subsystems), above which they slowly decrease. See SM~\cite{SM} for the $k$-dependence as well as the estimated STD of XEB values.
(a) The running time is measured on a device using 1 GPU (NVIDIA Tesla V100).
 (b) The performance of our algorithm (XEB value and running time) listed here are measured for the partitions in SM~\cite{SM} which are not optimized and are chosen for 1 GPU simulation with bounded memory (32GB for our device). In this SM~\cite{SM}, we also discuss some other ways to make the simulation more efficient.
 The tensor network algorithm is based on Ref.~\cite{kalachev2021recursive} and implemented by a Julia package OMEinsum.jl~\cite{julia2}.
 }
\label{tab:running_time}
\end{table*}

\begin{result}{(Comparison with experimental results)}
For the experimentally relevant gate set (fSim + discrete single-qubit gates) the performance of our algorithm can be summarized (see also Table~\ref{tab:running_time}) as follows
\begin{itemize} 
\item using the top-$k$ post-processing method, the algorithm achieves average XEB values comparable ($\approx 2\%\sim 12\%$) to recent experiments up to depth 20 and 24, respectively.;
\item the running times (computing the vector of output probabilities and choosing the top-$k$ bitstrings) are on the order of one second.
\item the STD is conjectured to be comparable to the mean value for large enough $k$ but without decreasing XEB too much; this is supported numerically for Google's Sycamore architecture [see SM~\cite{SM}]. 
\end{itemize}
\end{result}
In summary, our numerical simulations show that our algorithm achieves XEB values comparable to Google's and USTC's circuits in the quantum advantage regime with the experimentally-relevant gate set.
While our basic algorithm is simple and efficient, there are ways to achieve higher XEB values by adding more sophisticated algorithmic ingredients. For example, we show that after adding a simple post-processing step (the top-$k$ method), our algorithm can achieve much higher XEB values; e.g., compare green crosses and stars in  Fig.~\ref{fig:intro_advantage_regime}. In fact, we only considered here the most straightforward way to determine the locations of omitted gates (or maximal depolarization noise), which may not be optimal. By generalizing our method, e.g., making the locations of omitted gates (maximal depolarization noise) time/depth dependent, we expect an improved version of our algorithm may produce higher XEB without substantially increasing the computational resources.
In addition, it is an interesting future direction to explore further algorithmic improvements (e.g., adding a modest amount of entanglement).

\subsection{Comparison to prior work}\label{ssec:related}

Now, we make a few remarks and compare our algorithm to several  previously introduced algorithms that challenged the XEB-based quantum advantage, which utilizes noisy-circuit experiments.
First, Ref.~\cite{zhou2020limits} proposed an MPS-based approach, which introduces effective ``noise" by greedily truncating the entanglement in the system.
In that work, the authors consider the CZ gate ensemble and achieve the average XEB value of $0.02$, while our approach achieves similar XEB of $0.024$ by simply removing the entanglement between two properly chosen subsystems. To achieve the $0.02$ XEB value, the algorithm of Ref.~\cite{zhou2020limits} requires a run-time of several hours, while our algorithm completes in only 5 minutes under the same computational resources (1 CPU with 4.5 GB memory). Moreover, Ref.~\cite{zhou2020limits} only discusses the effective fidelity and shows that the XEB generally overestimates the fidelity by roughly 10 times. In the present work, we provide a deeper understanding of the connection between the XEB and the fidelity. 

Another approach is based on tensor network contraction~\cite{gray2020hyper,kalachev2021recursive},
which explicitly computes $p_x$, represented by a tensor network, for as many bitstrings $x$ as possible, and then picks out those $x$ with large values of $p_x$. Mixing these specially chosen bitstrings with randomly chosen bitstrings forms a set of millions of bitstrings that can spoof the XEB test.
Several very recent advances~\cite{huang2020classical,pan,fu2021closing,refute1,refute2} are based on a similar idea.  In their approach, the computational resources required for simulating only Google's Sycamore chip (53 qubits, depth 20) are either based on super-computer / massive computer clusters or dozens of GPUs with dozens of hours or days. Because tensor contraction algorithms are inherently exponential in the system size, and hence do not scale to larger systems, spoofing USTC's second experiment (60 qubits, depth 24) is already beyond the scope of the above approach.
In contrast, although the XEB values we obtain are not strictly larger than those from experiments, our algorithm only requires 1 GPU with few seconds, and scales better than experiment when increasing system size. 

Next, our result for 1D circuits refutes the Linear Cross-Entropy Quantum Threshold Assumption (XQUATH)~\cite{aaronson2019classical}, at least for one of its reasonable modifications, which is a conjecture about the hardness of an approximate counting problem and the hardness of the corresponding XEB-based sampling problem can be reduced to it. In the SM~\cite{SM}, we extend the refutation of XQUATH even for 2D circuits. Concretely, XQUATH states that there is no polynomial time classical algorithm to get an estimation $q_U(0^N)$ of $p_U(0^N)$ (the probability of getting $0^N$ from the ideal circuit given a circuit $U$) up to a precision $\sim2^{-N}$ (see SM~\cite{SM} for more detail) which is slightly better than randomly guessing. In the SM~\cite{SM}, we prove that this precision is exactly the average XEB, $\ex{\chi_U(C)}{U}$. Thus if the precision $2^{-N}$ can be modified to $e^{-\Delta d}$ for some constant $\Delta $ (where $\Delta \sim\Delta_1$ for 1D circuit), then our algorithm, which runs in linear time, could achieve this approximation. We argue that the modification is reasonable because in order to get a chaotic circuit, $d\sim N$ for 1D and $d\sim\sqrt N\ll N$ for 2D~\cite{boixo2018characterizing,harrow2018approximate}. The original motivation of this conjecture was to establish a connection between the hardness of the sampling problem and the hardness of a direct simulation of quantum circuit. Since our algorithm is far from direct simulating a quantum circuit, our result implies that the precision required in XQUATH, is not accurate enough in order to capture the hardness of direct simulation; however, our result for 1D noisy circuit shows that, more accurate precision is even not reasonable to a quantum device without fault-tolerance. In the SM~\cite{SM}, we also show that a similar (although slightly weaker) refuting statement also holds for 2D or even more general circuit architectures.

Finally,
we remark that, our algorithm is not trying to simulate noisy circuits like the one in Ref.~\cite{gao2018efficient}. Instead, the only objective of our algorithm is to get high XEB value, but the associated fidelity might be very low (even much lower than what a noisy circuit could have). Conceptually, our algorithm is a generalization of the one in Ref.~\cite{lightcone} beyond shallow circuits. The present results constitute substantial improvements and extensions of this algorithm, with a thorough theoretical analysis and detailed numerical simulations.

\section{Understanding XEB and fidelity via classical statistical mechanics}
\label{sec:DR}

In this section, we assume the single qubit gate is haar random and present an analytic framework to understand the relation between the XEB and the fidelity under various conditions, including different quantum circuit architectures and the presence of noise or omitted gates.
We will find that, in these settings, both the XEB and the fidelity, averaged over an ensemble of unitary circuits, can be efficiently estimated by mapping the quantum dynamics to classical statistical mechanics models, such as the diffusion-reaction model.
This mapping to the diffusion-reaction model was previously developed in Ref.~\cite{mi2021information} for the purpose of studying quantum information scrambling under random circuit dynamics.
Here we use a similar method to study behavior of the XEB and fidelity in random circuits with various entangling gates.
In the special case of 1D circuits, the effective model can be further simplified to a ferromagnetic Ising spin model in two dimensions, allowing us to obtain the scaling behavior analytically. 

\subsection{Overall methodology}
We first outline how quantum dynamics can be mapped to a classical statistical mechanics model. The XEB and the fidelity can be written as
\begin{align}
    \label{eqn:chi_dup_hilbert}
    \chi_U+1 &= \sum_{x} \bra{x} U \rho_0 U^\dagger\ket{x} \bra{x} \mathcal{M}_U^{(a)} [\rho_0 ] \ket{x} 2^N,\\
    \label{eqn:f_dup_hilbert}
    F_U &= \sum_{x,x'} \bra{x} U\rho_0U^\dagger\ket{x'} \bra{x'} \mathcal{M}_U^{(a)} [\rho_0 ] \ket{x},
\end{align}
where $\rho_0 = \ket{0^N}\bra{0^N}$ is the initial state of the system, and $\mathcal{M}_U^{(a)}[\cdot]$ is a quantum channel associated with the ideal unitary evolution ($a$=ideal), noisy quantum dynamics ($a$=noisy), or our classical algorithm with omitted gates ($a$=algo).
For a different choice of $a=\{\textrm{ideal}, \textrm{noisy},\textrm{algo}\}$, Eqs.~\eqref{eqn:chi_dup_hilbert} and~\eqref{eqn:f_dup_hilbert} become the XEB and the fidelity of the corresponding case, respectively. The sum over $x,x'$ represents the summation over all possible $N$-qubit configurations (bitstrings). 

\begin{figure*}[tbp]
\includegraphics[width=\linewidth]{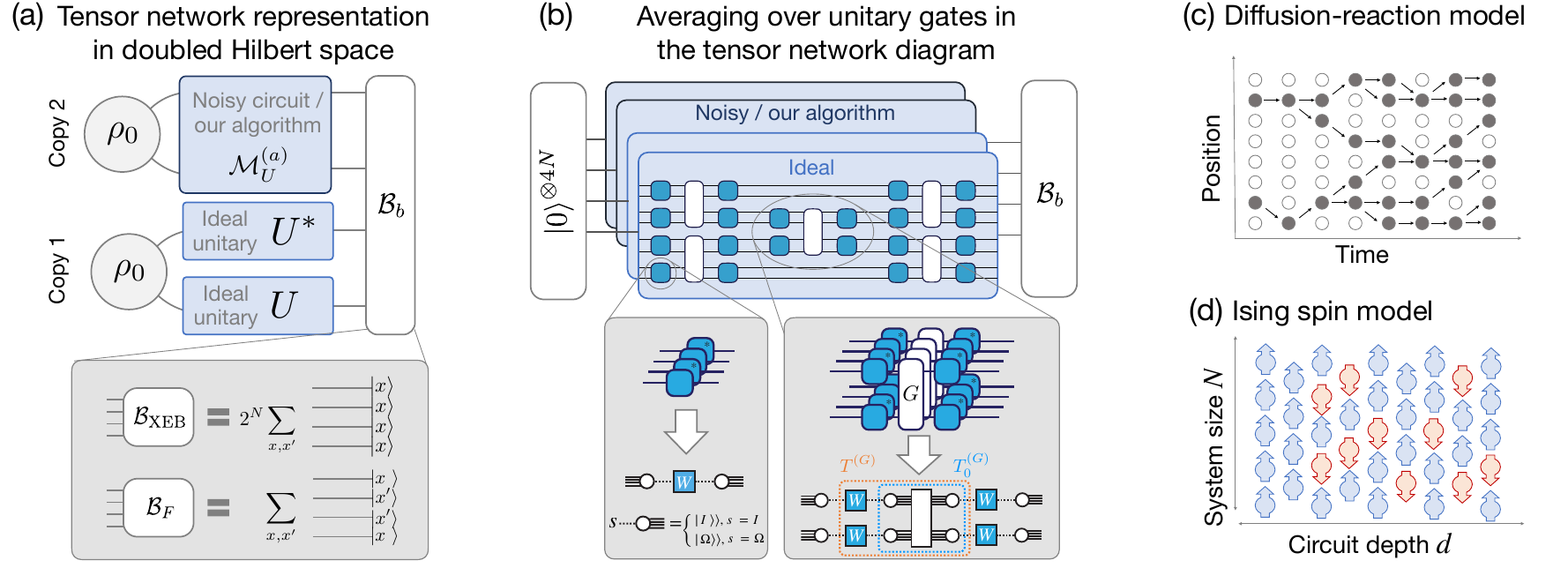}
\caption{
Mapping quantum circuits to statistical mechanics models.
(a) Both XEB and fidelity can be written as observables $\mathcal{B}_b$ with $b=\textrm{XEB},F$ in a duplicated Hilbert space by using tensor network representations.
The duplicated Hilbert space consists of the tensor product of Copy 1, representing an ideal circuit evolution, and Copy 2, representing the dynamics of either noisy circuit or our algorithm with omitted gates.
(b) For the tensor network diagrams representing XEB or fidelity, each random unitary gate (blue boxes) and its complex conjugate (blue boxes with asterisks) appear twice: in Copy 1 and in Copy 2.  One can perform averaging over an ensemble of unitary gates without explicitly evaluating the tensor network diagram, which gives rise to a simpler tensor network diagram with new classical variables, $s$, associated with each averaged single-qubit unitary gate (bottom left).
Entangling unitary gates $G$ dictate the dynamics of variables $s$, which is encapsulated in the transfer matrices of the classical statistical mechanics model (bottom right).
(c) Schematic diagram for the diffusion-reaction model. Each site can be occupied by a particle (filled) or remain unoccupied (empty). In every discrete time step, each particle may either stay on the same site, move to a neighboring site (diffusion), or duplicate itself to a neighboring site (reaction). Finally, a pair of particles located on neighboring sites may recombine into a single particle (reaction). Each of these processes has a specific probability that depends on the underlying gate ensemble. 
(d) Quantum circuits in 1D can be mapped to the classical Ising spin model in 2D.
}
\label{fig:stat_mapping_outline}
\end{figure*}

The key idea is to realize that both the XEB and the fidelity can be expressed as the expectation values of observables in an extended Hilbert space. 
More explicitly, we envision having two identical copies of the  Hilbert space: one representing the ideal circuit dynamics, and the other representing the dynamics in either the ideal circuit, noisy circuit, or our algorithm [see Fig.~\ref{fig:stat_mapping_outline}(a)].
Then, we have 
\begin{align}
    \chi_U+1 &= \Tr{ \mathcal{B}_\textrm{XEB} \left( U\rho_0U^\dagger \otimes\mathcal{M}_U^{(a)}[\rho_0] \right) },\label{eq:chi_Bchi}\\ %\mathcal{M}_a[\rho_0] \right) }\\
    F_U &= \Tr{ \mathcal{B}_F \left( U\rho_0U^\dagger \otimes \mathcal{M}_U^{(a)}[\rho_0] \right) },\label{eq:f_Bf}
    %\mathcal{M}_a[\rho_0] \right) },
\end{align}
where $\mathcal{B}_\textrm{XEB} = 2^N \sum_x \ket{x}\bra{x}\otimes \ket{x}\bra{x} $ and $\mathcal{B}_F = \sum_{x,x'} \ket{x}\bra{x'}\otimes \ket{x'}\bra{x} $ are Hermitian observables defined in the enlarged space. 
In the following, we simply use $\mathcal{B}_b$ with $b\in\{{\rm XEB},F\}$.

A convenient way to study the type of operators in Eqs.~\eqref{eq:chi_Bchi}-\eqref{eq:f_Bf} is to represent them as tensor networks whose contraction results in $\chi_U+1$ or $F_U$,  as shown in Fig.~\ref{fig:stat_mapping_outline}(a,b).
In general, the contraction of these tensor network diagrams for any given $U$ would be computationally difficult, as it is equivalent to evaluating the corresponding quantum circuit.
However, we are mostly interested in the average-case behavior of a class of random quantum circuits with gates drawn from specific gate ensembles. In this case, we can perform the averaging over the gate ensemble before contracting the network. Crucially, we find that the averaging process allows us to re-express the tensor network as a summation over exponentially many simple diagrams enumerated by different configurations of classical variables $s$ [see Fig.~\ref{fig:stat_mapping_outline}(b)].

This emergent mathematical structure---namely the summation over all possible configurations of classical variables---is similar to the path integral formulation of a classical Markov process, or a partition function in statistical mechanics models~\cite{huang1963statistical}.
Indeed, we will show that $\chi_U+1$ and $F_U$, averaged over an ensemble of unitary gates, are \emph{exactly} described by a diffusion-reaction model or a classical Ising spin model.

\subsection{The emergent diffusion-reaction model}\label{ssec:stat_DR_mapping}
We now describe the exact mapping from random unitary circuits to the diffusion-reaction model.
To derive this mapping, we will first consider the bulk of the tensor network in the absence of any noise or omitted gates, i.e., $\mathcal{M}_U^{\textrm{ideal}}[\rho_0] = U \rho_0 U^\dagger$. We will follow with the analysis of the boundaries at $t=0$ (initial state) and at $t=d$ (contraction with the observable $\mathcal{B}_b$).
Finally, we will consider how the presence of noise or omitted gates influences the system.

\emph{Bulk of the ideal circuit.}---
The central ingredient of the mapping to statistical mechanics models is the averaging over an ensemble of unitary gates~\cite{dankert2009exact}. In our case, we consider a single-qubit unitary $u \in SU(2)$ averaged over the Haar ensemble (or any other ensemble that forms a unitary 2-design). As depicted in Fig.~\ref{fig:stat_mapping_outline}(b), every random unitary $u$ appears exactly 4 times: a pair of $u$ and $u^\dagger$ for the ideal dynamics and another pair for the quantum channel $\mathcal{M}_{U}^{(a)}$. Since these sets of 4 random gates are independent, we can average them locally within the circuit using the 2-design property~\cite{dankert2009exact},
\begin{align}\label{eq:2design}
    \mathbb E_{u}[ u\otimes u^* \otimes u \otimes u^*] = \kket{I}\bbra{I} + \frac{1}{3} \kket{\Omega} \bbra \Omega, 
\end{align}
where $\kket{I}$ and $\kket{\Omega}$ are mutually orthogonal operators in the duplicated Hilbert space defined as
\begin{align}\label{eq:I_Omega}
  \nonumber  \bbra{a,b,c,d} I\rangle \rangle &= \frac{1}{2} \delta_{ab} \delta_{cd},\\
    \bbra{a,b,c,d} \Omega \rangle \rangle &= \frac{1}{2} \sum_{\mu=\text{x,y,z}} \sigma^{\mu}_{ab} \sigma^{\mu}_{cd},
\end{align}
with Pauli matrices $\sigma^\mu$, and $a,b,c,d\in\{0,1\}$.
We note that by using this notation, we are implicitly utilizing the channel-state duality (also known as  the Choi–Jamiołkowski isomorphism~\cite{choi1975completely}), where operators such as density matrices are vectorized: $\rho = \sum_{ij} \rho_{ij} \ket{i}\bra{j}\rightarrow \kket\rho =\sum_{ij}\rho_{ij}\ket i\ket j$.
Intuitively, $\bbra{I}$ and $\bbra{\Omega}$ represent the normalization and the total polarization correlation between the two copies, respectively; see the SM~\cite{SM} for the detailed derivation of these properties.

Notice that Eq.~\eqref{eq:2design} is a sum of two projectors, up to normalization factors.
Therefore, by applying Eq.~\eqref{eq:2design} to every quadruple of single-qubit unitary gates, the tensor network diagram factorizes into smaller parts, which are enumerated by different assignments of classical variables $s\in\{I,\Omega\}$ associated with every independent single-qubit unitary gate.
We interpret the classical variable $s$ at a certain site in space-time as if that site is in a vacuum state ($s=I$) or occupied by a particle ($s=\Omega$).
In this picture, the particle configuration at a specific time step is given by the assignment of $I$ or $\Omega$ values to $s$ variables within that time slice. Then, the tensor network describes how the particle configuration is advanced in every time step, which is captured by the transfer matrix $\mathcal{T}$.

The transfer matrix between two time steps is determined by the product of local transfer matrices $\mathcal T = \prod_G T^{(G)}$.
In turn, a local transfer matrix $T^{(G)}$ is given by the combination of the prefactor $1/3$, originating from Eq.~\eqref{eq:2design}, and a non-trivial contribution $T_0^{(G)}$  associated with a single two-qubit gate $G$, as shown in Fig.~\ref{fig:stat_mapping_outline}(b). 
We evaluate $T_0^{(G)}$ explicitly by contracting (four copies of) a two-qubit gate $G$ with four vectors $\kket{s}$, where $s=I,\Omega$, arising from four single-qubit random gates before and after $G$ [see Fig.~\ref{fig:intro_alg}(b)]:
\begin{align}
    T^{(G)}_{0;s_1 s_2 s_3 s_4} = \bbra{s_1}\bbra{s_2} G\otimes G^* \otimes G \otimes G^* \kket{s_3}\kket{s_4}.
\end{align}
Explicit calculations lead to the general form of the $T$-matrix
\begin{equation}\label{eq:transfer_general}
T^{(G)}=\begin{pmatrix}
1 & 0 & 0 & 0\\
0 & 1-D & D-R & R/\eta\\
0 & D-R & 1-D & R/\eta\\
0 & R & R & 1-2R/\eta
\end{pmatrix},
\end{equation}
written in the basis $\{II,I\Omega,\Omega I,\Omega\Omega\}$. 
This formula has been first derived in Ref.~\cite{mi2021information} for studying quantum scrambling. And we apply it to study  vulnerabilities of the XEB.
Here, $D\ge0$ and $R\ge0$ are parameters that depend on the specific choice of the entangling unitary gate $G$ (the gate ensemble), while $\eta=3$ for any 2-qubit gate. We call $D$, $R$ and $\eta$, the diffusion rate, reaction rate, and reaction ratio, respectively, and summarize their values for a few common entangling gates in Table~\ref{tab:DR}. 

\begin{table}
\begin{center}
\begin{tabular}{|c|c|c|c|c|}
\hline
 & CZ & Haar & fSim & fSim$^*$\\
\hline
diffusion rate $D$ & 2/3 & 4/5 & 1 & 1 \\
\hline
reaction rate $R$ & 2/3 & 3/5 & $1/3+\sqrt3/6$ & 2/3\\
\hline
\end{tabular}
\end{center}
\caption{Values of the diffusion rate $D$ and the reaction rate $R$ for a few different entangling gates.}
\label{tab:DR}
\end{table}
 
We note that each column of $T$ is normalized to unity, implying that the matrix indeed describes a transfer matrix for a stochastic process. For example, the entry in the 2nd column and the 4th row specifies the probability of the two sites going from $I\Omega$ to $\Omega\Omega$---this is an example of the ``reaction'' process. Other transitions are given in the following, with probabilities written on top of the arrows,
\begin{eqnarray}
\nonumber \text{vacuum:} && \quad II\xrightarrow{1} II\\
\nonumber \text{stay:} &&  \quad I\Omega\xrightarrow{1-D} I\Omega , \Omega I\xrightarrow{1-D} \Omega I \\
\nonumber \text{move:} && \quad  I\Omega\xrightarrow{D-R} \Omega I ,\\
 \Omega I\xrightarrow{D-R} I\Omega  ,\nonumber
&&\quad\Omega\Omega\xrightarrow{1-2R/\eta} \Omega\Omega\\
\nonumber \text{duplication:} && \quad I\Omega,\Omega I\xrightarrow{R} \Omega\Omega \\
\nonumber \text{recombination:} && \quad  \Omega\Omega\xrightarrow{R/\eta} I\Omega,\Omega I .
\end{eqnarray}
The third process (move) is the ``diffusion'' (i.e., random walk), while the last two (duplication and recombination) are reaction processes, i.e., particle creation and annihilation. Notice that a particle cannot be created from the vacuum or annihilated into the vacuum without interacting with another particle.

\emph{Boundary conditions at the initial state and at the final time.}---
Next, we turn to the boundaries of our tensor network diagram.
First, we contract the input state $\rho_0 \otimes \rho_0$, denoted as $\kket{0^{\otimes 4}}^{\otimes N}$, with tensors associated with all $2^N$ possible particle configurations.
This leads to the vector ${\mathbf u}^{\otimes N}$, where
\begin{equation}\label{eq:ini_vec}
{\mathbf u}=
\begin{pmatrix}
\langle\innerp{I}{0^{\otimes4}}\rangle \\ \langle\innerp{\Omega}{0^{\otimes4}}\rangle
\end{pmatrix}
=
\begin{pmatrix}
1/2 \\ 1/2
\end{pmatrix},
\end{equation}
which follows directly from Eq.~\eqref{eq:I_Omega}.
This vector describes the initial distribution of particles: every site is occupied by a particle or remains empty with probabilities 1/2.

Similarly, at the final layer, we contract the $\mathcal{B}_b$ observables with tensors associated with all $2^N$ possible particle configurations, leading to dual vectors
$\mathbf v_\text{\tiny XEB}^{\top\otimes N}$ and $\mathbf v_F^{\top\otimes N}$ for the XEB and the fidelity, respectively, where
\begin{eqnarray}\label{eq:ini_vec 2}
\nonumber \mathbf{v}_\text{\tiny XEB}^\top&=&\begin{pmatrix}
\langle\innerp{I}{\beta_\textrm{XEB}}\rangle & \frac{ \langle\innerp{\Omega}{\beta_\textrm{XEB}}\rangle}{3}
\end{pmatrix} =  
\begin{pmatrix}
2 & 2/3
\end{pmatrix},
\\
\mathbf{v}_F^\top&=&\begin{pmatrix}
\langle\innerp{I}{\beta_F}\rangle & \frac{ \langle\innerp{\Omega}{\beta_F}\rangle}{3}
\end{pmatrix} =  
\begin{pmatrix}
1 & 1
\end{pmatrix}.
\end{eqnarray}
and
\begin{align}
\kket{\beta_\textrm{XEB}}&=2 \sum_{i\in\{0,1\}}\ket i\ket{i}\ket{i}\ket i,\label{eq:xeb_bc}\\
\kket{\beta_F}&= \sum_{i,i^\prime\in\{0,1\}}\ket i\ket{i^\prime}\ket{i^\prime}\ket i,\label{eq:f_bc}
\end{align}
are the single-site versions of $\mathcal{B}_b$, i.e., $\mathcal{B}_b = \beta_b^{\otimes N}$.
We find that $\mathbf{v}_\text{\tiny XEB}$ is distinguished from $\mathbf{v}_F$ by unequal weights between $I$ and $\Omega$ (by a factor of $1/3$) aside from the global normalization factor $2$.
This allows an intuitive explanation: as previously mentioned,  $\bbra{\Omega}$ represents total polarization correlation between two copies of quantum states, but XEB  depends only on correlations measured in the computational basis constituting $1/3$ of the total on average.

Combining the results from bulk transfer matrices, and initial and final boundary conditions, we obtain the expression for the ensemble-averaged XEB and fidelity:
\begin{align}
\label{eq:chi_U_avg}
\chi_\text{av} + 1&\equiv   \mathbb{E}_u [\chi_U]+1 =
    \mathbf v_\text{\tiny XEB}^{\top\otimes N}
    \left(\prod_{j=1}^d \mathcal{T}_j \right)
    \mathbf{u}^{\otimes N}\\
\label{eq:F_U_avg}
F_\text{av} &\equiv
    \mathbb{E}_u[F_U] =
    \mathbf v_F^{\top\otimes N}
    \left(\prod_{j=1}^d \mathcal{T}_j\right)
    \mathbf{u}^{\otimes N},
\end{align}
where $\mathcal{T}_j$ is the transfer matrix for $N$ particles at time-step $j$.

\emph{XEB and fidelity as statistics of a particle distribution.}---
Our results in Eqs.~\eqref{eq:chi_U_avg} and~\eqref{eq:F_U_avg} allow for an intuitive understanding of the XEB and the fidelity in terms of particle distributions in the diffusion-reaction model.
We note that these two quantities differ only by the boundary condition at the final time $t=d$, as defined in Eqs.~\eqref{eq:xeb_bc}-\eqref{eq:f_bc}.
Hence, both the XEB and the fidelity are fully determined by the probability distribution of particle configurations, $\mathbf p$, obtained by evolving the initial uniform distribution $\mathbf{u}^{\otimes N}$ for $d$ time steps:
\begin{equation}\label{eq:Cpath}
{\mathbf p}\equiv\mathcal T_d\cdots\mathcal T_2\mathcal T_1{\mathbf u}^{\otimes N}.
\end{equation}
From this distribution, the XEB and the fidelity can be evaluated by simply contracting either $\mathbf v^{\top\otimes N}_{\rm XEB}$ or $\mathbf v^{\top\otimes N}_{F}$, which corresponds to computing certain statistics of the particle distribution.
For instance, all entries in $\mathbf v_F^{\top\otimes N}$ are unities, implying that $\mathbf v^{\top\otimes N}_F \mathbf{p}$ is equal to the summation over all probabilities:
\begin{align}
    F_\text{av}=\mathbf v^{\top\otimes N}_F \mathbf{p} = \mathbb E_\mathbf{p} [1],\label{eq:XEBFideal01}
\end{align}
where $\mathbb E_\mathbf{p}[\cdot]$ denotes the averaging over the distribution $\mathbf{p}$.
In the absence of any noise or omitted gates, the transfer matrix in Eq.~\eqref{eq:transfer_general} preserves the total probability, leading to $F_\text{av} = \mathbb E_\mathbf{p} [1] = 1$.
This result is trivially expected in the quantum circuit picture --- in the absence of any noise or omitted gates, the fidelity must always be unity.
We will soon see how this picture is modified when we introduce noise or omit  gates.

Similarly, the average XEB is
\begin{eqnarray}
\nonumber \chi_\text{av}+1&=&{\mathbf v^{\top\otimes N}_\text{\tiny XEB}}{\mathbf p}=2^N\mathbb E_\mathbf{p}\left[\frac{1}{3^{\#\Omega\text{ in the last layer}}}\right],\label{eq:XEBFideal02}
\end{eqnarray}
where $\#\Omega$ denotes the total number of particles.

\emph{Effects of noise or omitted gates.}---
When unitary dynamics is interspersed by noise channels ($\mathcal{M}_U^{({\rm noisy})}$) or when some of the gates are omitted in our classical algorithms ($\mathcal{M}_U^{({\rm algo})}$), only the bulk part of the tensor network changes, leading to a modified transfer matrix.
For a noisy circuit, the new transfer matrix is
\begin{equation}
\label{eq:transfer_matrix_noisy}
    T^{(G)}_\epsilon= (I_\epsilon\otimes I_\epsilon)\, T^{(G)} \quad\text{with }I_{\epsilon}=
    \begin{pmatrix}
    1 & 0\\
    0 & 1-c\epsilon
    \end{pmatrix},
\end{equation}
where $c$ is a constant depending on the type of noise. For example, $c=4/3$ for the depolarizing noise $\mathcal N_\epsilon(\rho)=(1-\epsilon)\rho+\epsilon/3\sum_{\mu}\sigma^{\mu}\rho\sigma^\mu$, and $c=2/3$ for the amplitude damping noise. 

Unlike the transfer matrix in the ideal case, the noisy-circuit transfer matrix in Eq.~\eqref{eq:transfer_matrix_noisy} no longer describes a stochastic process. 
That is, the sum of each column in $T^{(G)}_\epsilon$ is less than unity, implying that the probability is not conserved.
Thus, the effect of noise gives rise to the ``loss of probability'' in our diffusion-reaction model. In general, this leads to an unnormalized final distribution $\mathbf{p}$ and reduced average fidelity $F_\text{av} < 1$.
Crucially, the loss of probability occurs only when a particle ($\Omega$) is present at a given space-time point. The diagonal entries in $I_\epsilon$ imply that the probability associated with a given particle configuration will be damped by a factor $(1-c\epsilon)^{\#\Omega}$ at every time step. 
Therefore, we expect an interesting interplay between the diffusion-reaction dynamics of particles and the probability loss.

For our classical algorithm, it is the omission of gates that modifies the transfer matrix.
In this case, only local transfer matrices associated with an omitted gate are affected
\begin{equation}\label{eq:TP_projector}
    T^{(G)}\rightarrow  (P_I\otimes P_I)\cdot T^{(G)}=P_I\otimes P_I
    \quad\text{with }P_I=
    \begin{pmatrix}
    1 & 0\\
    0 & 0
    \end{pmatrix}.
\end{equation}
Similarly to the noisy circuit case, the omission of gates also causes the loss of probabilities; thus, the fidelity becomes smaller than 1.
More specifically, Eq.~\eqref{eq:TP_projector} implies that, at any given time, the probability weights associated with particle configurations containing at least one particle at the site of omitted gates must vanish; such configurations do not contribute to the average XEB or fidelity.
Thus, the only nonvanishing contributions arise from diffusion-reaction processes in which not a single particle ever appears at the sites of omitted gates throughout the entire dynamics.
The average fidelity will then be the total probability of such diffusion-reaction processes, and the average XEB is determined by the resultant unnormalized distribution $\mathbf{p}$.

We remark that the deterministic loss of probability at the positions of omitted gates leads to the factorization of the transfer matrix in Eq.~\eqref{eq:TP_projector} (as a product of two projectors).
Due to this factorization, $\mathbf{p}$ for the whole system also factorizes into independent probability vectors for two isolated subsystems.
This feature allows the numerical calculation of the average XEB for system sizes up to the quantum advantage regime (60 qubits, depth 24).

\subsection{Dynamics of the XEB and fidelity}\label{ssec:stat_DR_dynamics}
Having introduced the mapping of random unitary circuits to the diffusion-reaction model in the previous section, we now leverage this formalism to understand the quantitative behavior of the XEB and the fidelity under various conditions. In particular, we explain the key concepts used to obtain results presented in Section~\ref{sec:intro_alg}.

\emph{Ideal circuit. ---}
In the absence of noise and omitted gates, the fidelity remains equal to unity trivially, due to the conservation of the total probability. 
It is non-trivial, however, to see how the average XEB approaches unity in the limit of deep quantum circuits~\cite{arute2019quantum}, which we now explain in terms of diffusion-reaction dynamics.
Both the XEB and the fidelity, at late times (large depths), are determined by the output vector $\bf p$. 
For the transfer matrix in Eq.~\eqref{eq:transfer_general}, this distribution converges to a fixed point in the large-depth limit.
In the current case, there are two fixed points for local transfer matrices, ${\bf u}_1=(1/4,3/4)$ and ${\bf u}_2=(1,0)$.
The former represents a nontrivial steady-state solution in which the total normalization, and three different types of correlations (along $x$, $y$, and $z$ directions) are equally distributed, while the latter represents a trivial solution where two copies are both in completely mixed states; hence, no correlation is generated during dynamics.
It can be shown that the global stationary distribution is given as a mixture of ${\bf u}_1^{\otimes N}$ and ${\bf u}_2^{\otimes N}$, whose ratio is determined by the initial condition ${\bf u}^{\otimes N}$:
\begin{align}\label{eq:limiting_distribution}
    \lim_{d\rightarrow \infty} {\bf p} =
    (1-2^{-N}){\bf u}_1^{\otimes N} + 2^{-N} {\bf u}_2^{\otimes N}+O(4^{-N})
\end{align}
The dominant contribution originates from the non-trivial equilibrium configuration $\mathbf u_1$,
whereas the ${\bf u}_2$ term constitutes a small correction.

\begin{figure}[tbp]
\includegraphics[width=0.5\textwidth]{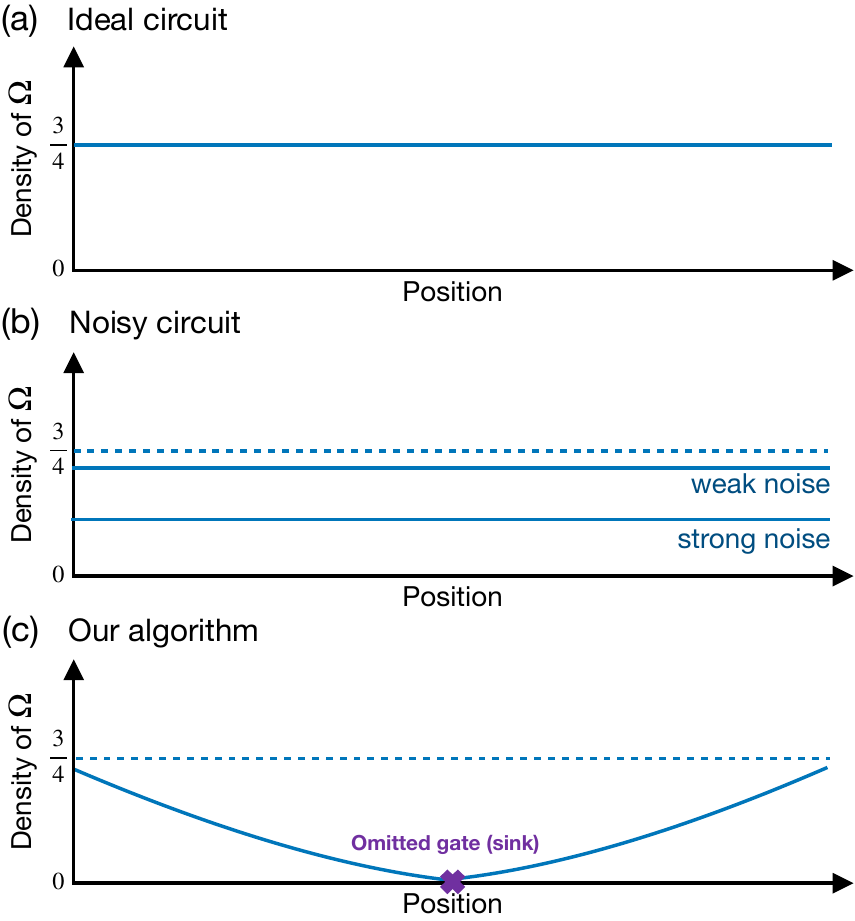}
\caption{Sketch of the particle population distribution at the last layer.
The vertical axis is the density of particles ($\Omega$) at the final layer normalized by the total probability, and the horizontal axis is the position of sites.
(a) Ideal circuits. 
(b) Noisy circuits.
The density is decreased relatively to the ideal case.
The discrepancy becomes larger for larger noise rates.
(c) Our algorithm. Close to the position of an omitted gate, or a ``sink'' (purple cross),  the density of particles is suppressed.
}
\label{fig:stat_population_last}
\end{figure}

The nontrivial term describes the homogeneous distribution of  particles with the density 3/4, as shown in Fig.~\ref{fig:stat_population_last}(a), contributing to the XEB
$$
\mathbf{v}_\text{\tiny XEB}^\top{\bf u}_1=2\left(\frac{1}{4}\cdot1+\frac{3}{4}\cdot\frac{1}{3}\right)=1.
$$
The trivial term gives $\mathbf{v}_\text{\tiny XEB}^\top{\bf u}_2=2$ per site.
Combined together with appropriate coefficients, we obtain the average XEB $\chi_\text{av}=(1-2^{-N})\approx 1$ as expected.
We note that the net contribution from the trivial solution (${\bf u}_2$ term) is always $+1$, which exactly cancels the constant term $-1$ in the definition of the XEB.

\emph{Noisy circuit.}---
If noise is introduced to the system, the total probability is no longer conserved, and ${\bf u}_1^{\otimes N}$ does not form a stationary solution. However, we can still predict the behavior of the average XEB and fidelity using our model.
We distinguish two regimes: (a) the weak noise limit where the total probability loss rate $N\epsilon$ is much smaller than the inverse equilibration time $\tau_{\rm eq}^{-1}$ of the particle distribution, $N\epsilon \ll \tau_{\rm eq}^{-1}$, and (b) strong noise limit $N\epsilon \gg \tau_{\rm eq}^{-1}$.
In terms of quantum circuit dynamics, these conditions correspond to the comparison of the total error rate to the scrambling time.

In the limit of weak noise, the steady state configuration must stay close to that of the equilibrium solution, because the system relaxes quickly before any substantial probability loss occurs.
Thus, the output probability vector at the final time is not severely affected by the probability loss during preceding times, other than a global re-scaling factor.
This leads to the (un-normalized) equilibrium state $\widetilde{\mathbf p}=\widetilde{\mathbf u}_1^{\otimes N}$, where 
\begin{equation}
\label{eq:weak_noise_solution_u}
\widetilde{\mathbf u}_1\approx \alpha\begin{pmatrix}1/4\\3/4 \beta \end{pmatrix}.
\end{equation}
Here $\alpha$ is the re-scaling factor that accounts for the probability loss (per site) during the diffusion-reaction dynamics, and it generally decreases exponentially with depth.
The parameter $\beta $ quantifies the deviation of $\widetilde{\mathbf{u}}_1$ from its equilibrium shape, and generally $\beta \approx 1$ in the weak-noise limit.
The precise value of $\beta$ depends on the strength of noise and the equilibration time.
As long as $\beta \approx 1$, $\widetilde{\mathbf p}$ is a simple re-scaling of the ideal-circuit distribution, and XEB approximates the fidelity well; both quantities are suppressed by the factor of $\alpha^N$.

In the limit of relatively strong noise (slow equilibration), the particle configuration cannot relax to its equilibrium before it is significantly affected by the probability loss.
In this limit, the deviation of $\widetilde{\mathbf{u}}_1$ from the equilibrium becomes significant, and  $\beta < 1$ decreases with the increasing strength of noise.
This is because, generically, the probability loss associated with $\Omega$ particles during dynamics results in a reduced density of particles at the last layer~[see Fig.~\ref{fig:stat_population_last}(b)].
The reduced density of particles implies that the XEB is larger than the fidelity because the boundary vector $\mathbf{v}_{\rm XEB}$ has a higher weight for the vacuum than for the particle state, whereas $\mathbf{v}_{F}$ has the same weight for both states.
Hence, the larger the noise rate, the greater the deviation of the XEB from the fidelity.
Eq.~\eqref{eq:weak_noise_solution_u} no longer holds for greater noise strengths\footnote{
In this work, we focus on the experimentally relevant regime, where the strength of noise and the depth of the circuit are not too large, such that the fidelity remains sufficiently greater than $2^{-N}$. When the fidelity is close to $2^{-N}$, the discussion in this paragraph no longer holds, as the contribution from subdominant terms in Eq.(\ref{eq:limiting_distribution}) becomes significant.}.

\begin{figure}[tbp]
\includegraphics[width=\linewidth]{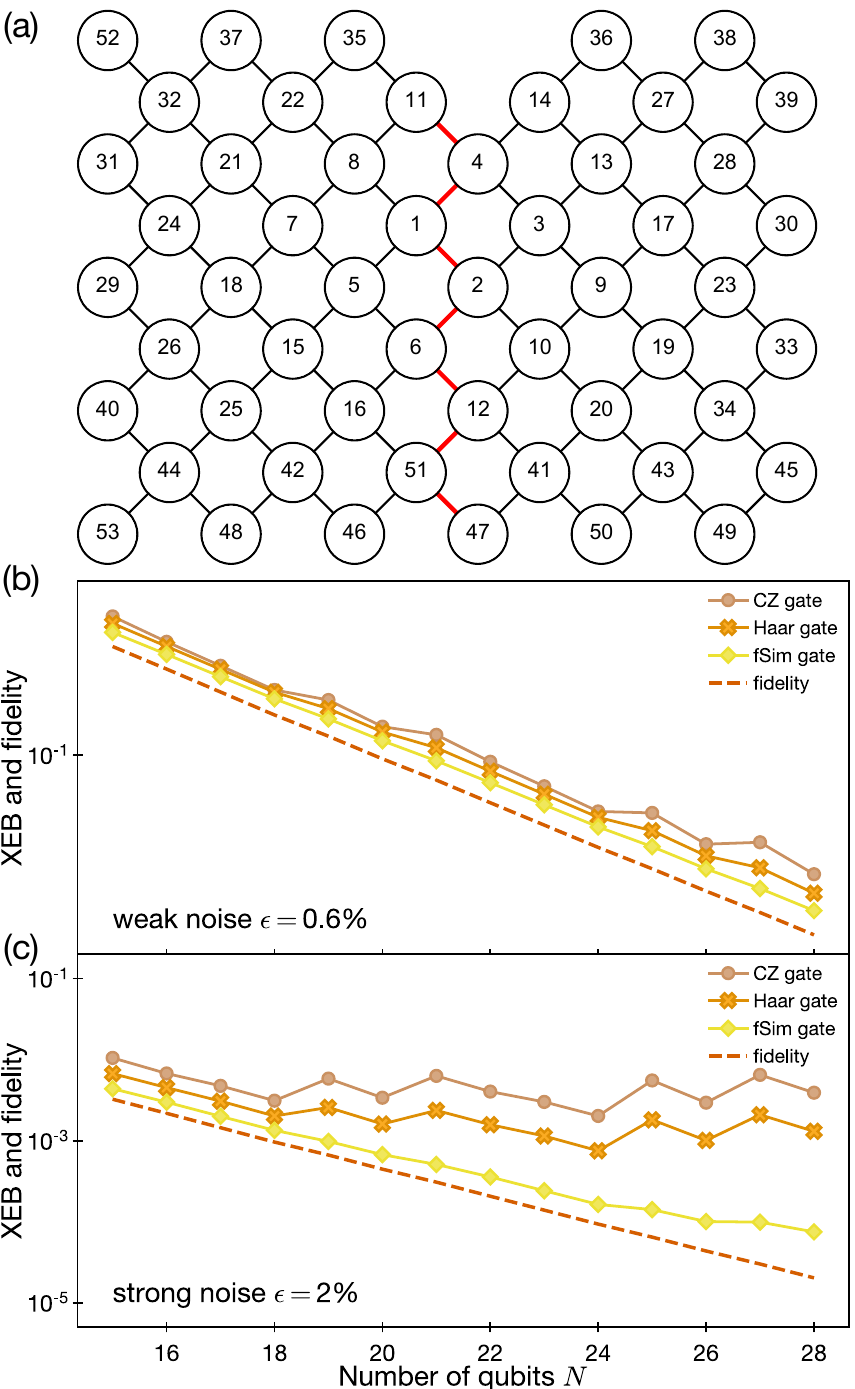}
\caption{XEB vs. fidelity in noisy circuits. The XEB always overestimates the fidelity, but the deviation depends on the gate ensemble and the strength of noise. (a) For this calculation, we use the original qubit ordering from Fig.~S25 in Ref.~\cite{arute2019quantum} (see also Fig.\ref{fig:intro_partition}). 
(b) Weak noise regime ($\epsilon=0.6$\%). The XEB approximates the fidelity well, and the fidelity values for all gate ensembles are almost the same. (c) Strong noise regime ($\epsilon=2$\%). The quality of the XEB-to-fidelity approximation strongly depends on the choice of the gate ensemble. Among the three ensembles considered here, the fSim ensemble gives the best result.}
\label{fig:stat_noise_fid_vs_xeb}
\end{figure}

\emph{Spoofing algorithm.}--- Our algorithm is designed to leverage the discrepancy between the XEB and the fidelity. 
In contrast to homogeneous errors spread over the bulk of the circuit, the errors in our algorithm are highly inhomogeneous and localized --- they appear only at specific positions where we omit gates. 
This inhomogeneity leads to a particle distribution that is far from its equilibrium counterpart.
More specifically, the position of an omitted gate behaves like a ``sink'' of probabilities --- any configurations containing particles at sink sites, at any time, will acquire vanishing contribution to $\widetilde{\mathbf p}$.
Therefore, in any non-vanishing contribution to $\widetilde{\mathbf{p}}$, the relative density of particles with respect to the density of vacuum states  is substantially lowered near the sink [see Fig.~\ref{fig:stat_population_last}(c)].
This large imbalance (relative to the equilibrium) leads to the large XEB-to-fidelity ratio.
Thus, given the same value of fidelity, which is controlled by the total number of omitted gates, one can achieve high XEB values because vacuum state $I$ has a larger weight in the XEB than in the fidelity.

The non-equilibrium, spatially inhomogeneous dynamics of particles also leads to a distinct scaling behavior. In our algorithm, the average XEB value increases with the system size $N$, when the number of omitted gates is fixed.
This can be intuitively explained: the more space for particles to diffuse to, the less likely it is for them to hit sink sites, leading to an effectively smaller particle loss rate and reduced imbalance in the particle density, relative to the equilibrium. 

Here, we make two remarks.
First, while our analysis remained qualitative and focused on two extreme cases of error models, i.e., one with completely homogeneous noise and another fully localized errors, we emphasize that our intuitive understanding can be straightforwardly generalized to arbitrary circuit geometry with arbitrary inhomogeneous error models in both space and time.
In such cases, one can directly estimate the distribution $\widetilde{\mathbf p}$ by using conventional approaches such as Monte Carlo methods.
Second, we comment that, intuitively, larger diffusion and reaction rates imply shorter time required to reach the equilibrium distribution. In other words, given a circuit architecture, the XEB will be on average a better proxy for the fidelity in circuits consisting of faster scrambling (entangling) gates, with larger $R$ and $D$.

\emph{Numerical demonstration.}---
To corroborate our predictions based on the diffusion-reaction model, we present the results of our numerical simulations.
First, we confirm that the XEB overestimates the fidelity, and that the discrepancy is larger for higher noise rates, as shown in Fig.~\ref{fig:stat_noise_fid_vs_xeb}.
We find that the fSim ensemble has the smallest XEB-to-fidelity ratio. The reason for this is clear from the diffusion-reaction model: among the three gates we considered, their reaction rates $R$ are similar (between 0.6 and 0.67), but the fSim gate has the largest possible diffusion rate $D=1$, as shown in Table.~\ref{tab:DR}. 

We use this intuition to devise an even better gate, which we call the fSim$^*$. By fixing $D=1$, we find that the fSim$^*$  gate has a larger $R=2/3$. Moreover, these values of $R$ and $D$ are now optimal, which we prove in the SM~\cite{SM}. Thus, fSim$^*$ has the smallest possible discrepancy between the  XEB and the fidelity. We compare it to the fSim gate in the inset of Fig.~\ref{fig:intro_fid_xeb}(b).

\begin{figure}[tbp]
\includegraphics[width=0.5\textwidth]{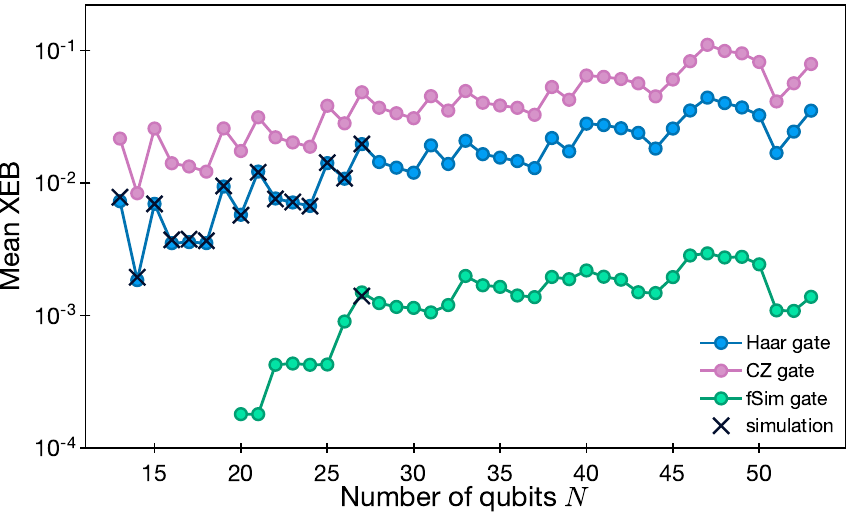}
\caption{Mean XEB value obtained by our algorithm as a function of the system size $N$, using the  ordering in Fig.~\ref{fig:stat_noise_fid_vs_xeb}(a) and the $d=14$ circuit architecture from Ref.~\cite{arute2019quantum}. The average XEB values are calculated using the diffusion-reaction model for three different gate ensembles: Haar (blue), CZ (purple), and fSim (green).
When compared with the results of the direct simulation of quantum circuits (crosses), both methods agree very well.}
\label{fig:stat_DR_mean}
\end{figure}

Next, we verify that the average XEB value of our algorithm for a specific circuit architecture (the Sycamore chip) can be very accurately predicted by our diffusion-reaction model. These results are shown in Fig.~\ref{fig:stat_DR_mean}. 
We find that our diffusion-reaction model can predict  even the fine details of the scaling with the system size $N$. For example, in Fig.~\ref{fig:stat_DR_mean}, the rise and fall in the value of XEB is caused by the lattice structure [see Fig.~\ref{fig:stat_noise_fid_vs_xeb}(a)] and its effect on the diffusion process.

\subsection{Ising model for 1D Haar ensembles}\label{ssec:stat_Ising}
The diffusion-reaction model is useful for analyzing our system qualitatively and numerically, for general circuit architectures and two qubit gate sets. However, for a certain class of systems, such as 1D circuits with Haar two-qubit gates, one can further simplify the classical statistical physics model to the 2D Ising model.
This can be understood as a  special case of the diffusion-reaction model, related to it mathematically through a basis transformation. This mapping has been studied previously in Refs.~\cite{hayden2016holographic, you2018machine, hunter2019unitary,zhou2019emergent, jian2020measurement, bao2020theory, napp2019efficient,dalzell2020random}.
The Ising model allows us to obtain more quantitative results. We find that the behavior of the XEB is related to symmetry, symmetry breaking, and magnetization.

The basis change from the diffusion-reaction model to the Ising model is 
\begin{eqnarray*}
\kket{\uparrow}&=&2\kket I,\\
\kket{\downarrow}&=&\kket I+\kket\Omega
\end{eqnarray*}
such that 
\begin{eqnarray*}
\nonumber  \bbra{a,b,c,d} \uparrow\rangle \rangle&=& \delta_{ab} \delta_{cd},\\
\bbra{a,b,c,d} \downarrow\rangle \rangle
&=&\delta_{ad} \delta_{bc},
\end{eqnarray*}
where the second equation indicates that $\kket{\downarrow}$ corresponds to a swap between indices $a$ and $c$ (or $b$ and $d$). This new basis reflects the symmetry in $u\otimes u^{*}\otimes u\otimes u^{*}$ between the two copies: the state is invariant if we exchange the positions of the two $u$s or $u^{*}$s (labeled by $a,c$ and $b,d$, respectively). 

\begin{figure}[tbp]
\includegraphics[width=\linewidth]{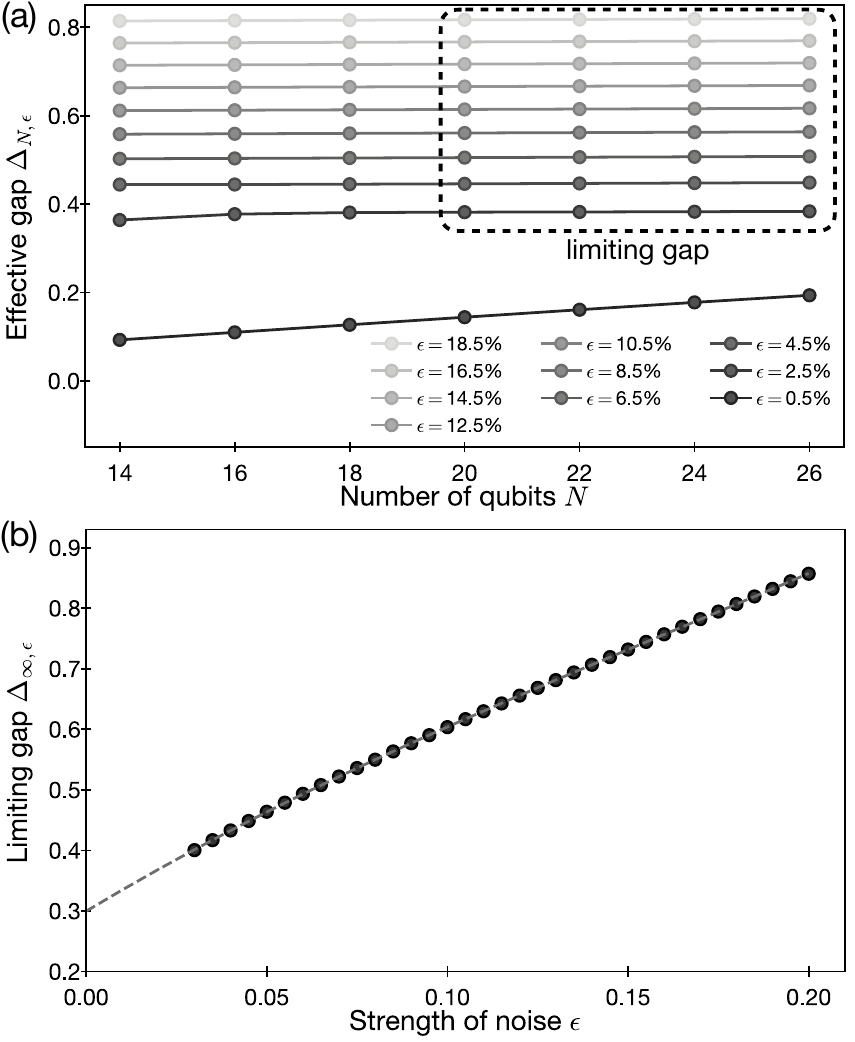}
\caption{Effective gaps of 1D noisy circuits. (a) For any noise strength $\epsilon$, the gap $\Delta_{N,\epsilon}$ saturates, for sufficiently large $N$, at the  \emph{limiting gap} value $\Delta_{\infty,\epsilon}$. (b) The limiting gap as a function of the noise strength.  Polynomial extrapolation indicates the $\epsilon\to0$ limit of the gap to be $\approx 0.03$. We define this limiting value as $\Delta_3:=\lim_{\epsilon\rightarrow0}\Delta_{\infty,\epsilon}$; in  Fig.~\ref{fig:intro_1D_gaps}, it is represented by the orange, dotted horizontal line. The subsystem considered here has only one boundary with omitted gates as the total system has open boundary condition.
}
\label{fig:1D_mean_noisy}
\end{figure} 

We regard $\uparrow$ and $\downarrow$ as the up and down spins, and the path integral of the diffusion-reaction dynamics is mapped to the partition function of the spin model [see Fig.~\ref{fig:stat_mapping_outline}(d)]. 
In the absence of noise or omitted gates, the partition function has a global $\mathbb Z_2$ Ising symmetry, such that $\kket{\uparrow}\leftrightarrow\kket{\downarrow}$ applied to all spins does not change the partition function.

After the basis change, XEB$+1$
corresponds to the partition function of the $\mathbb Z_2$-symmetric Ising spin model with identical boundary conditions at both the initial and final times. In the special case of Haar entangling gates, this model is the ordinary Ising model with 2-body interactions, which are detailed in the SM~\cite{SM}  and Refs.~\cite{hayden2016holographic, you2018machine, hunter2019unitary,zhou2019emergent, jian2020measurement, bao2020theory, napp2019efficient,dalzell2020random}. 

This mapping allows us to write the XEB for the ideal circuit in the following form:
\begin{equation}\label{eq:1D Ising partition}
\chi_\text{ideal}+1=Z=\bbra\psi \mathcal T_\text{Ising}^{(d-1)/2}\kket\psi,
\end{equation}
where $\kket\psi$ and $\bbra\psi$ are the boundary conditions, and $\mathcal T_\text{Ising}$ is the transfer matrix of the Ising model along the horizontal direction in Fig.~\ref{fig:stat_mapping_outline}(d); it is semi-definite positive and can be computed from $T_0^{(\text{Haar})}$ and Eq.~\eqref{eq:2design}. We defer the details of this calculation to the SM~\cite{SM}.  Here, we only need to know that this Ising model is in the ferromagnetic phase. Thus, the largest eigenvalue of $\mathcal T_\text{Ising}$ is doubly degenerate, which gives $Z=2$ and so  XEB$=1$ in the large-$d$ limit.

Once noise or gate defects are introduced, the Ising symmetry is violated. In the case of noisy circuits, the symmetry is violated everywhere, with each local interaction modified by the presence of effective magnetic fields with strength $\epsilon$. Then, there will be a spectral gap $\Delta_{N,\epsilon}=\lambda_1-\lambda_2$ in the modified $\mathcal T_\text{Ising}$, which we evaluate exactly.
Figure~\ref{fig:1D_mean_noisy}(a) shows the gap as a function of the system size for various error rates.
We show that in this case
\begin{equation}
   \chi_\text{noisy}=O\left(e^{-\Delta_{N,\epsilon}d}\right).
\end{equation}
If the violation is small enough ($N\epsilon\ll1$), the spectral gap is $\Delta_{N,\epsilon}\propto N\epsilon$ because the total magnetic field is only a small perturbation from the ideal (symmetric) case. However, if we consider the asymptotic behavior of noisy circuits, $\epsilon$ is assumed constant, but $N$ could be very large. In this limit, the gap will saturate to a fixed value $\Delta_{\infty,\epsilon}$, as shown in Fig.~\ref{fig:1D_mean_noisy}(a). This corresponds to the thermodynamic limit in terms of statistical physics. In this case, even if $\epsilon$ tends to 0, as long as $N\epsilon$ is still large, there is a finite gap in $\mathcal T_\text{Ising}$. This corresponds to the phenomena of spontaneous magnetization: even if the magnetic field fades away, most of the spins still point in the same direction. We numerically extrapolate the limiting gap to the vanishing noise rate $\epsilon$ and get
\begin{equation}
    \Delta_3=\lim_{\epsilon\rightarrow0}\Delta_{\infty,\epsilon}=\lim_{\epsilon\rightarrow0}\lim_{N\rightarrow+\infty}\Delta_{N,\epsilon}\approx0.3,
\end{equation}
as shown in Fig.~\ref{fig:1D_mean_noisy}(b). This corresponds to the orange dashed line in Fig.~\ref{fig:intro_1D_gaps}.

\begin{figure}[tbp]
\includegraphics[width=\linewidth]{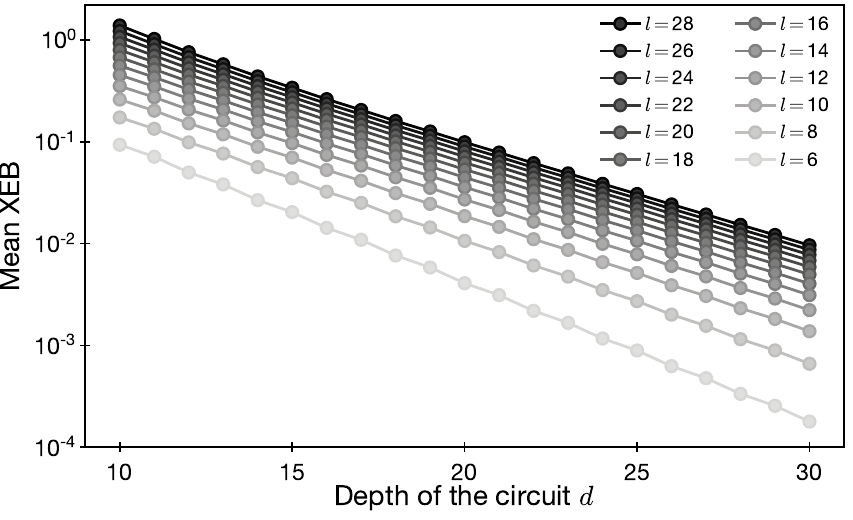}
\caption{Exponential decay of the average XEB value with increasing circuit depth $d$ for our algorithm. Through linear interpolation (on the semi-log plot), we compute the slope of the lines at each subsystem size $l$ and extract the spectral gap $\Delta_1$. The dependence of $\Delta_1$ on $l$ is shown in Fig.~\ref{fig:intro_1D_gaps} (blue solid curve).
}
\label{fig:1D_mean_our}
\end{figure}

For our algorithm, the omitted gates are mapped to a tensor product of projectors, as shown in Eq.~\eqref{eq:TP_projector}, so the partition function will also be separated into the product of partition functions of isolated subsystems
\begin{align}
\chi_\text{algo}=Z-1
&=\prod_{i=1}^{\lceil N/l\rceil} Z_l^{(i)}-1\\
&\approx \prod_{i=1}^{\lceil N/l\rceil} (e^{-\Delta_l^{(i)}d}+1) - 1\\
&\approx \sum_{i=1}^{\lceil N/l\rceil} e^{-\Delta^{(i)}_l d}\sim \frac{N}{l}e^{-\Delta_l d},\label{eq:Isingpart}
\end{align}
where $\Delta_l^{(i)}$ is the gap of the $i$-th subsystem, and $\Delta_l$ is the typical gap among these subsystems, assuming they have similar sizes. Eq.~\eqref{eq:Isingpart} shows that the XEB increases with the system size if the subsystem size $l$ is fixed. The decay rate is mainly determined by the subsystem with $\Delta_1=\min_i \Delta^{(i)}_l$. For each subsystem, the omitted gates correspond to strong magnetic fields at the bottom (or top) boundary, which have been previously identified as ``sinks'' in our diffusion-reaction model. These fields violate the $\mathbb Z_2$ symmetry, which causes the gap to open. The gap decreases if the subsystem size $l$ increases; see the discussion in the previous subsection and the  SM~\cite{SM}. We numerically compute the gap for different circuit parameters and present the results in Fig.~\ref{fig:1D_mean_our}. We find that when $l\ge15$, $\Delta_1$ approaches to a constant $\Delta_1\approx0.25$. Crucially, we see that $\Delta_1 < \Delta_3$; this means that our algorithm generates a higher XEB value, in the large-depth limit, than noisy 1D circuits --- even with arbitrarily weak noise. 

We note that many of the qualitative behaviors discussed in this section  also hold in general architectures and two qubit gate set. For example, Eq.~\eqref{eq:Isingpart} shows that the XEB obtained by our algorithm behaves more like an additive quantity, i.e., the total XEB approximately equals the sum of XEB values for each subsystems, if they are decoupled (in our algorithm) or only weakly coupled (in noisy circuits).
In contrast, fidelity exhibits multiplicative behavior, i.e., every error contributes to reducing the fidelity of the total system exponentially.

\section{Conclusion and outlook}\label{sec:outlook}

In this work, we introduced a novel framework to analyze the behavior of XEB and fidelity under random quantum circuit dynamics in arbitrary architectures. We showed that the XEB generally overestimates the fidelity, and presented an intuitive explanation for this phenomenon and more quantitative analysis using a mapping of quantum dynamics to a classical diffusion-reaction model.
Furthermore, leveraging our new framework, we designed a simple and efficient classical algorithm, which achieves XEB values comparable to, or even higher than, noisy circuit dynamics under various conditions.
We numerically demonstrated the excellent performance of our algorithm using relatively small amount of computational resources (time and memory), and showed that it achieves XEB values comparable to those obtained in experiments and by the state-of-the-art algorithms running on devices with much more computational power~\cite{huang2020classical,pan,fu2021closing,refute1,refute2}.  Our results demonstrate the shortcomings of  the XEB for estimating the circuit fidelity and certifying  quantum advantage, unless further assumptions are made.
For example, our qualitative analysis in Sec.~\ref{ssec:vulnerability} and quantitative results in Sec.~\ref{sec:DR} indicate that, for the XEB to approximate the fidelity well, the assumption of spatially homogeneous and temporally independent weak noise is crucial. The violation of any of these conditions (as in the case of our algorithm) may lead to a substantial discrepancy between the XEB and the fidelity. 

\subsection{Overcoming the vulnerabilities of linear XEB}
Our results can be used as a guideline for designing the next generation of experiments with more robust ways to certify quantum computational advantage.
Here we describe a few simple methods to circumvent the shortcomings of the XEB arising from its vulnerabilities described in Section~\ref{ssec:vulnerability}. 
First, we point out that the system-size scaling of the linear XEB in our algorithm, associated with the additive nature in Eq.~\eqref{eqn:additive_nature}, can be easily resolved by using a different, nonlinear benchmark such as the logarithm version of the XEB~\cite{arute2019quantum}.
This change, however, does not address the remaining two vulnerabilities, namely (i) the deviation of the XEB from the fidelity or from the probability of having no error in the circuit, and (ii) the amplification of correlations via post-processing in classical algorithms.

In order to address (i), one can implement quantum circuits that are  scrambling faster by choosing more optimized gate sets such as the fSim$^*$ entangling gates combined with  Google's discrete single qubit gate set. Additionally, one could design a better circuit architecture with larger depth and more non-local connectivity, in addition to improving the fidelity of individual gate operation.
Such quantum circuits make our spoofing algorithm less effective, as partitioning the circuit into subsystems  with a small number of omitted gates becomes difficult. 

The amplification vulnerability arises because the previously used benchmarks are designed to quantify the correlation between bitstring probability distributions, and they are sensitive to bitstrings occurring with high probabilities.
This aspect can be alleviated by using a different figure of merit.
In quantum information theory, the total variation distance (TVD) is used frequently~\cite{aaronson2011computational,bremner2016average,harrow2017quantum,lund2017quantum} and it seems immune to straightforward amplification methods.
Unfortunately, TVD is practically intractable to estimate and hence cannot directly serve as a new benchmark~\cite{Canonne20}.
One can argue that TVD can be instead lower bounded by the fidelity, which one can estimate using the XEB under suitable conditions.
However, this approach may not be practical either because a meaningful lower bound for the TVD is obtained only for a very high value of fidelity (close to unity), which is difficult to achieve in near-term quantum devices without quantum error correction.

\subsection{Outlook}
Our work opens up a number of other new future directions.
Besides the above discussion on the vulnerabilities in the adversarial settings, it is also interesting to explore whether XEB can be used to certify a broader class of near-term quantum devices in real-world experiments in benign settings~\cite{choi2021emergent}.
Assuming random circuit dynamics with homogeneous and independent error models, our analysis presents a systematic way to identify the regime where the XEB, fidelity, and the probability of no error are consistent with each other.
More specifically, we showed that the XEB approximates the fidelity well when $N\epsilon f(c) \ll 1$, where $\epsilon$ is the local error rate and $f(c)$ is a constant of order unity that measures a particular type of ``scrambling'' time for a given entangling gate.
%Strongly entangling gates, such as the fSim (or the more optimal fSim$^*$) with the discrete gate set introduced by Google~\cite{arute2019quantum}, are an ideal choice for minimizing $f(c)$. 
%In the experiments performed in Refs.~\cite{arute2019quantum,USTC,zhu2021quantum}, $N\approx 50\sim60$. Assuming $f(c)\approx1\sim5$, a noise rate of $\epsilon\approx0.4\%-2\%$ is sufficiently weak for the $N\epsilon f(c) \ll 1$ condition to be met.
%The error rates in the experiments ~\cite{arute2019quantum,USTC,zhu2021quantum} belong to this regime.
The mapping from quantum dynamics to the diffusion-reaction model~\cite{mi2021information}, combined with numerical algorithms such as Monte Carlo and tensor networks~\cite{gray2020hyper,kalachev2021recursive,pan,huang2020classical,zhu2021quantum,refute1,refute2},
provides a quantitative method to estimate the precise value of $\epsilon$ required. 
Even when these error rates are weak enough, the independence and the homogeneity of noise (either in space or in time) have to be unconditionally demonstrated in order to ensure a good agreement between the XEB and the fidelity.
Alternatively, it would be interesting to explore if one could either relax the requisite conditions, or develop a more generalized relation between the XEB and the fidelity, e.g., in the presence of correlated and/or strong noise. 

The mapping from quantum dynamics to classical statistical models~\cite{mi2021information} can be regarded as a de-quantization procedure of quantum circuits by randomization and averaging over an ensemble, which is similar in spirit to randomized benchmarking~\cite{emerson2005scalable,knill2008randomized,proctor2019direct,kimmel2014robust,wallman2015estimating,helsen2019new,erhard2019characterizing,harper2020efficient}. The resulting statistical model is much easier to analyze analytically and numerically. In particular, it benefits from the intuitive understanding  associated with classical models, larger number of available computational tools, and connections to well-studied machine learning models, such as probabilistic graphical models~\cite{koller2009probabilistic,niedermayer2008introduction}.
Moreover, this emergent classical model can be generalized to describe other XEB-like quantities, which are potentially useful for studying various aspects of quantum circuits. For example, by replacing ideal circuits with different quantum channels (as mentioned in the SM~\cite{SM}), the XEB can potentially detect dominant types of noise. These ideas could be explored further to design new protocols for
learning and quantifying complex quantum systems.

Our work provides strong motivation for designing new figures of merit to certify quantum advantage, which remains as an important open problem.
It would be interesting to explore other efficiently measurable benchmarks that could certify the correctness of random circuit sampling.
More broadly, the study of the sample complexity of certifying random circuit sampling is warranted.
We also notice alternative approaches to demonstrating quantum advantage.
In particular, several interactive protocols have been designed recently~\cite{kahanamoku2021classically,brakerski2020simpler,liu2021depth,hirahara2021test}, 
 where quantum features can be certified based on cryptographic and computational complexity assumptions.

\begin{acknowledgments}
We would like to thank Scott Aaronson, Ehud Altman, Yimu Bao, Edward Farhi, Jin-Guo Liu, and Pan Zhang for their insightful discussions. 
We would like to express our special thanks to Igor Aleiner, Sergio Boixo, Sergei Isakov, Hartmut Neven, Kostyantyn Kechedzhi, and Vadim Smelyanskiy for their comments and discussions, in particular pointing out that the single-qubit gate ensemble used in recent experiments leads to near optimal performance for XEB to approximate fidelity and sharing their independent analysis of XEB based on mapping to a classical statistical mechanics model.
Also, we acknowledge Jin-Guo Liu for their technical assistance regarding the use of Julia numerical package Yao and OMEinsum.
%This work is supported by NSF, CUA, ARO (W911NF1910302), ARO MURI (W911NF2010082), DARPA ONISQ program (W911NF2010021).
This work is supported by NSF, CUA, ARO, ARO MURI, DARPA ONISQ program.
XG is also
supported from Quantum Science of the Harvard-MPQ Center for Quantum Optics and the Templeton Religion Trust grant TRT 0159.
BB and CC are also supported by  NSF award CCF 1565264, and a Simons Investigator Fellowship.
\end{acknowledgments}

%\bibliographystyle{naturemag}
%\bibliography{ref}

%\newpage
%\pagebreak

%\subfile{appendix}

\end{document}

%\title{\titleofthepaper}
\date{\today}

\author{Xun Gao}
\affiliation{\HarvardPhysics}

\author{Marcin Kalinowski}
\affiliation{\HarvardPhysics}

\author{Chi-Ning Chou}
\affiliation{\HarvardSEAS}

\author{Mikhail D. Lukin}
\affiliation{\HarvardPhysics}

\author{Boaz Barak}
\affiliation{\HarvardSEAS}

\author{Soonwon Choi}

\affiliation{\MIT}

\clearpage

\onecolumngrid

\begin{center}
	\textbf{\large Supplementary Material for ``\titleofthepaper'' }\\[5pt]

\end{center}
\setcounter{equation}{0}
\setcounter{figure}{0}
\setcounter{table}{0}
\setcounter{page}{1}
\setcounter{section}{0}
\makeatletter
\renewcommand{\theequation}{S\arabic{equation}}
\renewcommand{\thefigure}{S\arabic{figure}}
\renewcommand{\thepage}{S\arabic{page}}
\renewcommand{\thetable}{S\arabic{table}}

\maketitle

\tableofcontents

%{\color{red}\bf all the equation labels should be changed when refer to main text. Fig.5 and 10 in the main text, and citations.}

\section{Effects of errors on XEB and Fidelity}

In this section, we provide a detailed discussion of the qualitative analysis presented in Sec.{\color{red}~I~A} of the main text.
More quantitative analysis, using the diffusion-reaction model, is provided in Sec.~\ref{app:DR} in this supplementary material.
In subsection \ref{sapp:fidelity_few_errors} and subsection \ref{sapp:xeb_few_errors}, we explain the effect of a few errors on the fidelity and the XEB, respectively.
Based on these analyses, in subsection~\ref{sapp:qualitative_noisy_circuit}, we derive the criterion under which the XEB approximates the fidelity well in noisy quantum circuit dynamics under certain assumptions. In subsection~\ref{sapp:add_mul}, we discuss the additive nature of the XEB, in contrast to the multiplicative nature of the fidelity, in our algorithm and hence show explicitly that the XEB can be much larger than the fidelity in an adversarial setting, even when quantum circuits are  sufficiently scrambling; i.e., the bitstring probability distribution approximately follows the Porter-Thomas distribution.

\subsection{The effect of a few errors on fidelity}\label{sapp:fidelity_few_errors}
We start by considering the effect of a single error on the fidelity in a random circuit. Suppose a Pauli error $\sigma$ occurs in the $t_1$-th layer of a quantum circuit $U=U_2U_1$ where $U_1$ is the unitary evolution up to the depth $t_1$ and $U_2$ is the evolution for the remaining part.
Then, the fidelity can be written as
\begin{equation}
F= \left| \bra{0^{N}}     U_1^\dag U_2^\dag U_2\sigma U_1 \ket{0^{N}}  \right|^2 
=\left| \bra{0^{N}}     U_1^\dag U_2^\dag U_2 U_1  \sigma(-t_1)  \ket{0^{N}} \right|^2
= \left| \bra{0^{N}}     \sigma(-t_1) \ket{0^{N}} \right|^2,
\end{equation}
where $\sigma(-t_1)=U_1^\dag \sigma U_1$ is the operator in the Heisenberg picture, representing its net effect on the initial state.

Due to the rotational symmetry associated with the ensemble of random unitary gates in $U_1$ and $U_2$, we only need to consider a Pauli error $\sigma$; any other error operator can be considered as a linear combination of Pauli errors.
In general, we can decompose $\sigma(-t_1)$ in terms of Pauli string operators:
\begin{equation}
    \sigma(-t_1)=U_1^\dag \sigma U_1\approx\sum_{i=1}^{4^s-1}\eta_i\sigma_i, \text{ where }\sigma_i\in\{I,X,Y,Z\}^{\otimes s}/I^{\otimes s},
\end{equation}
where $s=(2ct_1)^D$ is the effective light-cone size associated with the propagation of the error through $U_1$, $c<1$ is the speed of operator spreading restricted by the circuit architecture, and $D$ is the spatial dimension of the architecture. When $U_1$ is locally scrambling inside a light-cone of size $s$, the weights of the operator $\eta_i^2$ behave as pseudo random variables, approximately following a uniform  distribution~\cite{hayden2007black,roberts2017chaos} with the normalization condition $\sum_i\eta_i^2=1$, implying $\mathbb E_U[\eta_i^2] \approx 1/(4^{s}-1)\approx 4^{-s}$ and $\mathbb E_U[\eta_i\eta_j]=0$ for $i\ne j$ on average over $U_1$. This gives an approximation to the average fidelity:
\begin{eqnarray}
   \nonumber F_\text{av}&=&\mathbb E_U\left[\left| \bra{0^{N}}     \sigma(-t_1) \ket{0^{N}} \right|^2\right]=\sum_{i,j}
       \mathbb E_U[\eta_i\eta_j] \bra{0^{N}}  \sigma_i  \ket{0^{N}}
        \bra{0^{N}}  \sigma_j \ket{0^{N}}\\
       & \approx&\sum_{i}4^{-s}
      \left| \bra{0^{N}}  \sigma_i  \ket{0^{N}}\right|^2
        \approx2^{-s}
\end{eqnarray}
where the last equality is due to
\begin{equation}
     \left| \bra{0^{N}}  \sigma_i  \ket{0^{N}}\right|^2=
     \begin{cases}
        1, \text{ if }\sigma_i \in \{I,Z\}^{\otimes s};\\
        0, \text{ otherwise.}
     \end{cases}
\end{equation}
More explicitly, when $\sigma(-t_1)$ is represented in the Pauli operator basis, $\sim 2^s$ terms out of total $4^s$ terms have the initial state as an eigenstate. Therefore, the fidelity is not affected by an error with probability $2^s/4^s = 1/2^s$ on average.

Next we turn to the presence of two errors.
Consider two errors $\sigma_1$ and $\sigma_2$ in the circuit occurring at two different times. We decompose the unitary dynamics into three parts based on the timing of the error $U =U_2U_3U_1$  such that the dynamics in the presence of the errors is described by $U_2\sigma_2U_3\sigma_1U_1$. 
There are two mechanisms that can cause two errors to nontrivially influence the fidelity.
The first mechanism is the error cancellation by the back-propagation of $\sigma_2$ through $U_3$ to the layer in which $\sigma_1$ occurs. 
It is possible that the back-propagated operator $\sigma_2(-t_3)=U_3^\dag\sigma_3U_3$ has non-zero component on $\sigma_1$ and thereby the errors partially cancel one another. 
Roughly, they cancel each other with probability $4^{-(2ct_3)^D}$ (or partially cancel one another with the corresponding fraction of weights) if $U_3$ is sufficiently scrambling over a subsystem of size $2ct_3$ and the distance between these two errors is smaller than $ct_3$. 
As a specific example, when two errors occur consecutively in time just before and after a Haar-random gate, the cancellation occurs with probability $1/15$ (or the $1/15$ fraction of their weights get cancelled on average).
If the two errors occur around an fSim gate, the probability of cancellation depends on the relative position between the two errors, i.e., whether they are acting on the same qubit or not. Regardless, the cancellation probability (or the fraction of weight) is of order $1/15$.

The second mechanism concerns both errors back-propagated to the input boundary. Similar to the one-error case, the average fidelity is $2^{-s}$, where $s$ is the union of the light-cones caused by these two errors. The two extreme cases are (1) the light-cones with sizes $s_1$ and $s_2$ are disjoint, hence the fidelity is $2^{-(s_1+s_2)}$; (2) $\sigma_1$ is inside the light-cone of $\sigma_2$, hence the fidelity is $2^{-s_2}$.  In the latter case, we see that the presence of the error $\sigma_1$ does not lead to any further reduction in fidelity when the error $\sigma_2$ already exists.

Our analysis can be straightforwardly generalized to situations where there are multiple errors.
In general, multiple errors occurring closely to one another may cause  mutual cancellation of their effects. 
Also, depending on the relative location of multiple errors, the net reduction in the fidelity may be equivalent to that of only a single error.
As we elaborate below, this may give rise to a substantial correction compared to a simple expectation that the many-body fidelity equals the probability of having no error in the entire quantum circuit.

\subsection{The effect of a few errors on XEB}\label{sapp:xeb_few_errors}
Next, we consider the effect of a single error on the XEB, where the circuit $U=U_2U_1$ becomes $U_2\sigma U_1$.
\begin{align*}
\chi_\text{av}&=
2^N\mathbb E_U\left[
\sum_{\vec x}|\bra{0^N}U_2U_1\ket{\vec x} |^2
\cdot|\bra{0^N}U_2\sigma U_1\ket{\vec x} |^2
\right]-1\\\notag
&=
2^N\mathbb E_U\left[
\sum_{\vec x}|\bra{0^N}U_2U_1X^{\vec x}\ket{0^N} |^2
\cdot|\bra{0^N}U_2\sigma U_1X^{\vec x}\ket{0^N} |^2
\right]-1,
\end{align*}
where $X^{\vec x}$ denotes a product of $I$ and $X$ determined by $\vec x$ so that $X^{\vec x}\ket{0^N}=\ket{\vec x}$. 
Unlike the fidelity, the XEB is obtained from measurements in the computational basis.
Consequently, the mathematical structure of the XEB is symmetric between input and output (up to a unimportant choice of bitstring for the initial state).
Thus, one can propagate the error operator in  two different directions: back-propagation to the input boundary and forward-propagation to the output boundary. This is the most significant difference betwee the XEB and the fidelity.
Similar to the analysis of the fidelity, we will see that many terms in the forward-propagated operator will not affect the measurement outcome probability distribution, leading to no change in the XEB's value.
Before discussing which direction should be chosen to estimate the XEB, let us estimate the expected XEB value $\chi_\text{av}$ by considering, as an example, the effect of an error on the output boundary.
More explicitly, 
\begin{align*}
\chi_\text{av}&= 4^N\mathbb E_U\left[
|\bra{0^N}U_2U_1\ket{0^N} |^2
\cdot|\bra{0^N}U_2\sigma U_1\ket{0^N} |^2
\right]-1\\\notag
&= \begin{cases}
   4^N\mathbb E_U\left[
|\bra{0^N}U\ket{0^N} |^2
\cdot|\bra{0^N}U\sigma(-t_1)\ket{0^N} |^2
\right]-1\\
4^N\mathbb E_U\left[
|\bra{0^N}U\ket{0^N} |^2
\cdot|\bra{0^N}\sigma(t_2)U\ket{0^N} |^2
\right]-1 
\end{cases}
\end{align*}
where $\sigma(t_2)$ describes the effect of the error \emph{propagated forward} in time on output states.
 We denote $\ket\psi=U\ket{0^N}$ and $\rho=\ket\psi\bra\psi$, and assume it is uncorrelated with $\sigma(t_2)$. When $U$ is scrambling enough, we have
\begin{equation*}
    \rho=\gamma_0I^{\otimes N}+\sum_{i=1}^{4^N-1}\gamma_i\sigma_i 
\end{equation*}
where
\begin{eqnarray*}
\gamma^2_0=4^{-N}, \mathbb E_U[\gamma_i^2]\approx8^{-N}  \text{ if }i\ge1\text{ and }E_U[\gamma_i\gamma_j]=0 \text{ if }i\ne j
\end{eqnarray*}
where $\sigma_0=I^{\otimes N}$, $\tr\rho=1$, $\tr\rho^2=1$, and so $\gamma_i^2$ behaves as the weight of a uniform distribution.

Now, let $\sigma^\prime\in\{I,X,Y,Z\}^{\otimes s}$ be the Pauli operator that represents the effect of the propagated error $\sigma$ to a boundary (either forward or backward).
We show in the following that any term $\sigma^\prime$ of the form $\{I,Z\}^{\otimes s}$ has substantial contribution to the average XEB, implying that the XEB does not vanish even when an error occurs.
Consider the effect of a tensor product of Pauli operators $\sigma^\prime$ on the output boundary:
\begin{eqnarray}
\nonumber\chi_\text{av}&=& 4^N\mathbb E_U\left[
\bra{0^N}\rho\ket{0^N}
\bra{0^N}\sigma^\prime\rho \sigma^\prime\ket{0^N} 
\right]-1\\
\nonumber &=& 
4^N\sum_{i,j}\mathbb E_U[\gamma_i\gamma_j]\bra{0^N}\sigma_i\ket{0^N}
\bra{0^N}\sigma^\prime\sigma_j \sigma^\prime\ket{0^N}-1\\
\nonumber &=& 4^N\sum_{i}\mathbb E_U[\gamma_i^2]\bra{0^N}\sigma_i\ket{0^N}
\bra{0^N}\sigma^\prime\sigma_i \sigma^\prime\ket{0^N}-1 \\
\nonumber 
&=&
4^N\gamma_0^2-1+4^N\sum_{i\ge1}\mathbb E_U[\gamma_i^2]\bra{0^N}\sigma_i\ket{0^N}
\bra{0^N}\sigma^\prime\sigma_i \sigma^\prime\ket{0^N}\\
&=& 2^{-N}\sum_{i\ge1} \bra{0^N}\sigma_i\ket{0^N}
\bra{0^N}\sigma^\prime\sigma_i \sigma^\prime\ket{0^N}.
\end{eqnarray}
Because $\bra{0^N}\sigma_i\ket{0^N}
\bra{0^N}\sigma^\prime\sigma_i \sigma^\prime\ket{0^N}=0$ if $\sigma_i$ contains some Pauli $X,Y$ on at least 1 qubit, the only non-trivial contribution comes from $\sigma_i\in\{I,Z\}^{\otimes N}$.
There are $2^N$ such kind of terms. Then,
\begin{equation}
2^{-N}\sum_{i^\prime\ge1}^{2^N-1} \bra{0^N}\sigma_{i^\prime}\ket{0^N}
\bra{0^N}\sigma^\prime\sigma_{i^\prime} \sigma^\prime\ket{0^N}
\approx\begin{cases}
   1 \text{ if }\sigma^\prime\in\{I,Z\}^{\otimes s}\\
   0 \text{ if }\sigma^\prime \text{ contains }X\text{ or }Y 
\end{cases}
\end{equation}
where $i^\prime$ is summed over those Paulis that are of the form $\{I,Z\}^{\otimes N}$. The first case holds because $\sigma^\prime\sigma_{i^\prime}\sigma^\prime=\sigma^\prime$ in this case and $\bra{0^N}\sigma_{i^\prime}\ket{0^N}^2=1$.
The second case holds because for a fixed $\sigma^\prime$, containing at least one $X$ or $Y$ operator, roughly half of $\sigma_{i^\prime}$s commute with $\sigma^\prime$ while another half anti-commute with it, leading to the vanishing contribution, i.e., the summation has almost the same numbers of $+1$ and $-1$. 

In summary, the effect of a Pauli operator $\sigma^\prime$ on both the input and output boundary has a similar effect as in the case of the fidelity: if $\sigma^\prime\in\{I,Z\}^{\otimes s}$, the contribution is 1; otherwise, the contribution is 0. So the total contribution is $2^{-s}$.

As previously mentioned, we note that there is a freedom to choose regarding whether one should propagate an error operator either backward to input or forward to output boundaries. We argue that choosing the smaller light-cone would give a more accurate estimation on $\chi_{\text{av}}$. 
This is because, in our current qualitative analysis, we implicitly assume that the circuit is sufficiently scrambling and that the error $\sigma(t_2)$ or $\sigma(-t_1)$ is uncorrelated with $U$.
Strictly speaking, this is not the case: $\sigma(t_2)$ is determined by $U_2$ and $U=U_2U_1$ is also related to $U_2$. But if the depth of $U_2$ is sufficiently short enough whereas $U_1$ is sufficiently deep, our analysis here can be justified (and similarly for the opposite case). Indeed, from Fig.~{\color{red}1(a)} in the main text, we can see this is quite a good approximation.
The effects of two or more errors can be analyzed similarly to the case of fidelity. See also Fig.~{\color{red}1(b)} in the main text.

\subsection{Estimating the amount of corrections in noisy circuits}\label{sapp:qualitative_noisy_circuit}
It has been argued that the probability of observing no error in the entire circuit, $P_0$, equals the fidelity as well as the XEB.
In the main text, we argued that this statement is not strictly true because one needs to consider corrections to the XEB and the fidelity, away from $P_0$.
Here we will estimate the amount of total corrections to XEB or fidelity in noisy circuit dynamics. We will assume that Pauli errors occur independently from one another and are homogeneously distributed over the entire quantum circuits in space and time.
Our general strategy is to consider a few classes of positioning of single or multiple errors, estimate their contribution to the XEB or fidelity, and count the number of error configurations in each class. 

Let $\epsilon$ be the error rate (i.e., the probability of a gate having an error), the expected fidelity can be expanded as follows.
\begin{align*}
F_{\text{av}} &= \sum_{k=0}\Pr[\#\text{errors}=k]\cdot\mathbb{E}[F_{\text{av}}\, |\, \#\text{errors}=k]\\
&=\sum_{k=0}\binom{Nd}{k}\epsilon^k(1-\epsilon)^{Nd-k}\cdot\mathbb{E}[F_{\text{av}}\, |\, \#\text{errors}=k]\\
&=(1-\epsilon)^{Nd}\cdot\left(\sum_{k=0}\binom{Nd}{k}\left(\frac{\epsilon}{1-\epsilon}\right)^k\cdot\mathbb{E}[F_{\text{av}}\, |\, \#\text{errors}=k]\right) \, ,
\end{align*}
where $\Pr[\#\text{errors}=k]$ denotes the probability to have $k$ errors in a noisy circuit, and $\mathbb{E}[F_{\text{av}}\, |\, \#\text{errors}=k]$ is the fidelity of the circuit given that $k$ errors occurred in the circuit, averaged over an ensemble of unitary gates.
The combinatorial factor $\binom{Nd}{k}$ arises from different positionings of $k$ errors in quantum circuits with $Nd$ gates. 
Similar formula can be derived for the average XEB.
We will treat the average XEB and fidelity together by using the expression:
\begin{equation}
    (1-\epsilon)^{Nd}\cdot\left(
    1+C^{(b)}_1+\cdots
    \right)
    \approx
    e^{-\epsilon Nd}\cdot\left(
    \sum_{k=0}C^{(b)}_k
    \right)=P_0\cdot\left(
    \sum_{k=0}C^{(b)}_k
    \right)
\end{equation}
where $b\in\{\text{XEB},F\}$ and $C^{(b)}_k$ represents the correction caused by $k$ errors relative to the probability of no error, $P_0 \equiv \Pr[\#\text{errors}=0]$.
For the convenience of discussion, we factorize $C^{(b)}_k$ into two contributions: ``energy" and ``entropy". The ``energy" factor come from the decay caused by the presence of errors, e.g., the probability of one configuration with $k$ errors, the effect of light-cones and the probability of error cancellations. The ``entropy" factor counts the number of configurations with $k$ errors. For example, the $k$-th correction term of the expected fidelity can be decomposed as
\begin{equation*}
C^{(F)}_k = \underbrace{\binom{Nd}{k}}_{\text{entropy}}\underbrace{\left(\frac{\epsilon}{1-\epsilon}\right)^k\cdot\mathbb{E}[F_{\text{av}}\, |\, \#\text{errors}=k]}_{\text{energy}} \, .
\end{equation*}

The exact evaluation of $C_k^{(b)}$ is difficult.
Instead, we mainly focus here on the effect of the correction terms with small $k$ and argue that they are enough to capture the qualitative behavior of expected fidelity and XEB values in weakly noisy circuits.
The intuition is that if the error rate is small enough, the ``energy" terms dominate (i.e., it decays faster than the increase of entropy terms) and the correction is mainly determined by terms with small $k$.
In such cases, the correction of fidelity and XEB compared to $P_0$ will remain small.
If this is not the case, i.e., the ``entropy'' terms grows faster than the energy terms as a function of $k$, one would get a very large, non-perturbative corrections to XEB and fidelity, away from $P_0$.
This indeed occurs when the error rate is not small enough and leads to catastrophic differences between XEB, fidelity and $P_0$.

In the rest of this subsection, we show that $N\epsilon\ll1$ is a necessary and sufficient criterion for the correction terms to be perturbatively small when the circuit is $D$-dimensional and the depth is $O(N^{1/D})$  which is a reasonable requirement for a sufficiently scrambling circuit dynamics~\cite{boixo2018characterizing,harrow2018approximate} and is also relevant for recent quantum advantage experiments.

\subsubsection{A necessary condition to make the correction small}
In order to make the total correction small, at least $C^{(b)}_1$ should be small enough.
A rough estimation is given in the following:
\begin{equation}\label{Seq:correction_first_fidelity}
    C^{(F)}_1=N\widetilde\epsilon\sum_{l=1}^{d}2^{-2cl}\approx
    N\epsilon \frac{1}{2^{2c}-1},
\end{equation}
where $l$ denotes the depth of the error and $\widetilde\epsilon=\epsilon/(1-\epsilon)$. Due to the exponential decay with depth / light-cone size, only errors near the boundary have non-negligible contribution. Thus, for XEB, the correction is roughly
\begin{equation}\label{Seq:correction_first_xeb}
    C^{(\text{XEB})}_1\approx 2 C^{(F)}_1\approx
    N\epsilon \frac{2}{2^{2c}-1},
\end{equation}
where the factor of 2 arises from the fact that one needs to consider both boundaries at the input and output of  circuits.
From~\eqref{Seq:correction_first_fidelity} and~\eqref{Seq:correction_first_xeb}, in order to guarantee $C^{(b)}_1$ is small, the error rate has to satisfy 
\begin{equation}\label{Seq:nec_condition}
    N\epsilon f(c)\ll1
\end{equation}
where we used $f(c)$ to denote a decreasing function of $c$ of order unity.

\subsubsection{A sufficient condition to make the correction small}
Here we try to give an upper bound for the correction terms of fidelity and XEB compared to $P_0$. To get a tighter estimation, we further divide each correction term $C^{(b)}_k$ into two categories: $C^{(b)}_{k,\text{cancellation}}$ and $C^{(b)}_{k,\text{light-cones}}$. Concretely,  $C^{(b)}_{k,\text{cancellation}}$ contains the contribution such that all the errors cancel out with each other while $C^{(b)}_{k,\text{light-cones}}$ contains the contribution coming from the light-cone propagation of errors.
For convenience, we denote $C^{(b)}_\text{cancellation}=\sum_{k=1}C^{(b)}_{k,\text{cancellation}}$ and $C^{(b)}_\text{light-cones}=\sum_{k=1}C^{(b)}_{k,\text{light-cones}}$. Now, we can upper bound the expected fidelity and XEB as follows.
\begin{eqnarray}
F_\text{av}&\le& \left(1+ C^{(F)}_\text{cancellation} \right)\left(1+ C^{(F)}_\text{light-cones} \right)\\
\chi_\text{av}&\le&\left(1+ C^{(\text{XEB})}_\text{cancellation} \right)\left(1+ C^{(\text{XEB})}_\text{light-cones} \right)\lesssim \left(1+ C^{(F)}_\text{cancellation} \right)
\left(1+ C^{(F)}_\text{light-cones} \right)^2
\end{eqnarray}
where we include the mixture of error cancellation and light-cones by expanding the product. Note that  $C^{(F)}_\text{cancellation}=C^{(\text{XEB})}_\text{cancellation}$. $\left(1+ C^{(\text{XEB})}_\text{light-cones} \right) \lesssim \left(1+ C^{(F)}_\text{light-cones} \right)^2$ because there are two boundaries for the XEB and because, when the correction is small, only the errors near the boundary have non-negligible contribution.
Thus, to show that the total correction is small for both the XEB and the fidelity, it is sufficient to show that $C^{(F)}_\text{cancellation}$ and $C^{(F)}_\text{light-cones}$ are small. For simplicity, we consider the case where $d\sim N^{1/D}$ as we mentioned previously.

\begin{figure}[tbp]
    \centering
    \includegraphics[width=\textwidth]{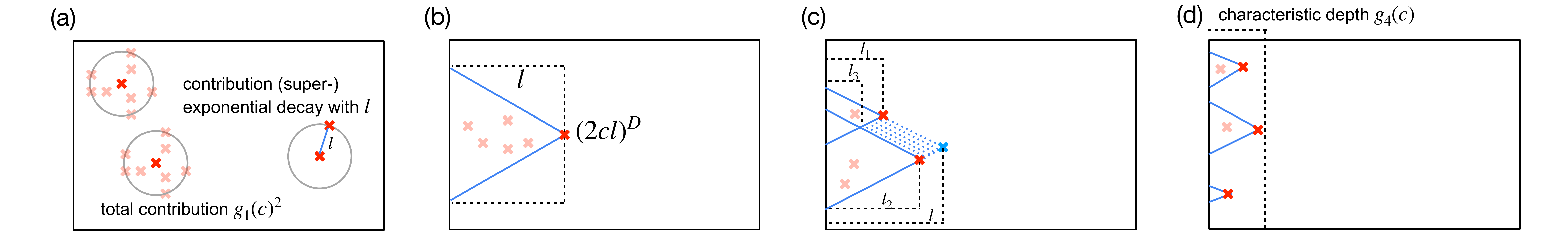}
    \caption{Different types of error contributions. (a) Error cancellations. (b) One large light-cone. (c) Overlap of two light-cones. (d) Multiple small light-cones.}
    \label{fig:app_bound_errors}
\end{figure}

First, we consider $C^{(F)}_\text{cancellation}$. Recall that in subsection~\ref{sapp:fidelity_few_errors}, we show that the probability of a pair of errors adjacent to a 2-qubit gate is around $1/15$. Similarly, if the pair of errors has distance $l$ from each other, the probability of cancellation is exponentially small with $l$. Meanwhile, the number of possible pairing of two errors is only polynomially increasing ($\propto l^D$ for fixed dimension $D$). See Fig.~\ref{fig:app_bound_errors}(a).
As a consequence, one can define an effective radius that depends on the interplay of the exponential decay of the distance to pair up errors and the polynomial increase (for fixed dimension $D$) of the number of positionings.
Only errors within this radius have non-negligible contributions.
The total contribution is denoted as $g_1(c)^2$, which is a function of $c$ . Intuitively, if $\epsilon^2Nd\ll1$, the contribution due to error cancellation will be small because this condition implies that errors are ``dilute'' in the circuit of size $Nd$; thus, it is difficult to pair them together. More concretely,
suppose we have $k$ errors, then the contribution to the correction due to error cancellation is
\begin{equation}
    C^{(F)}_{k,\text{cancellation}}={\widetilde\epsilon}^k\left(g_1(c)^2\right)^{k/2}\binom{Nd}{k/2}
    \lesssim (\epsilon g_1(c))^k (Nd)^{k/2}=\left(\sqrt{Nd}\epsilon g_1(c)\right)^k.
\end{equation}
Thus, if
\begin{equation}\label{Seq:b1}
    \sqrt{Nd}\epsilon g_1(c)\ll 1,
\end{equation}
$C^{(F)}_{\text{cancellation}}$ will be small. This is consistent with our intuition.

Second, we consider $C^{(F)}_\text{light-cones}$. For simplicity, we begin with the case where we only have one large light-cone, with additional possible errors inside of it, as shown in Fig.~\ref{fig:app_bound_errors}(b).
Next, we define the coherence length as $l_0=\ln2/\epsilon$, which characterizes ``the length of mean free path", namely, the typical length of an error can propagate without hitting another error. Intuitively, if $l_0<d$, the light-cone should only contain few errors such that the ``energy" terms dominate. Moreover, due to the decay caused by light-cone size, this condition implies that only small-size light-cones have non-negligible contributions. More concretely,
denote the depth of the light-cone as $l$, then the light-cone size will be $(2cl)^D$ for a $D$-dimensional circuit. Then the ``volume'' of the light-cone is $l(2cl)^D/D$. The total correction terms coming from error configurations with a depth-$l$ light-cone can be estimated in the following:
\begin{equation}\label{seq:1_light_cone}
    N\widetilde\epsilon\sum_{k=0}^{Nd} {\widetilde\epsilon}^k\binom{l(2cl)^D/D}{k} 2^{-(2cl)^D}
    \approx N\epsilon \sum_{k=0}\frac{\left(\epsilon l\cdot(2cl)^D/D\right)^k}{k!}2^{-(2cl)^D}
    =N\epsilon e^{(2cl)^D(\epsilon l/D-\ln 2)}.
\end{equation}
The factor $N$ is due to $N$ different positions at depth $l$.
Since $l\le d$, if
\begin{equation}\label{Seq:b2}
    d\epsilon\cdot\frac{1}{\ln2 D}\ll1,
\end{equation}
only terms with $l=1$ have non-negligible contribution, which is roughly bounded by $
    N\epsilon e^{(2c)^D(\epsilon d/D-\ln 2)}\sim N\epsilon g_2(c)$
where $g_2(c)\sim 2e^{-(2c)^D}$. Thus,
\begin{equation}\label{Seq:b3}
    N\epsilon g_2(c) \ll 1
\end{equation}
guarantees that the total contribution caused by one light-cone is small. In the following, we give a sufficient condition for the correction from multiple light-cones to also be small.

For overlapping light-cones, we focus on the case of two light-cones in a 1D circuit. It is not difficult to generalize our argument to multiple light-cone overlaps, as well as higher-dimensional cases. Let $l_1$ and $l_2$ denote the depth of the two light-cones, and let $l_3$ denote the depth of their intersecting gate; see Fig.~\ref{fig:app_bound_errors}(c). Let $l=l_1+l_2-l_3$, the two errors together span a light-cone of size $cl$ and the contribution is roughly
\begin{equation}
    \sum_{l_1,l_2=1}^{l}N\epsilon^2 e^{c\epsilon(l_1^2+l_2^2-l_3^2)-2c(l_1+l_2-l_3)\ln2}=N\epsilon e^{cl(\epsilon l-\ln2)} \cdot
    \sum_{l_1,l_2=1}^{l}\epsilon e^{-\epsilon A(l_1,l_2)}\sim
    N\epsilon e^{cl(\epsilon l-\ln2)}\cdot \epsilon l\cdot g_3(c)\label{Seq:correction_lightcones_overlapping}
\end{equation}
based on a similar estimation as in the single light-cone case, where $A(l_1,l_2)$ is the area of the dotted area shown in Fig.~\ref{fig:app_bound_errors}(c) and $g_3(c)$ is a function of $c$. The summation in the second term is proportional to $l$ because, intuitively, only small areas have non-negligible contribution due to the exponential decay. There are at most $O(l)$ such kinds of areas. To make sure the contribution of the overlap is smaller than the case with only one light-cone, we require
\begin{equation}\label{Seq:b4}
    d\epsilon g_3(c)\ll1.
\end{equation}
Note that~\eqref{Seq:b4} implies $\epsilon l< \ln 2$ and hence~\eqref{Seq:correction_lightcones_overlapping} is small compared to the single light-cone cases. Thus, we only need to focus on Eq.~(\ref{seq:1_light_cone}).

Finally, we consider the case with multiple disjoint light-cones. If the above condition is satisfied, then only small light-cones have non-negligible contribution as shown in Fig.~\ref{fig:app_bound_errors}(d). Let us denote the characteristic depth as $g_4(c)$, then the contribution with $k$ light-cones is bounded as
\begin{equation}
    {\widetilde\epsilon}^k2^{-k(2c)^D}\binom{Ng_4(c)}{k}\lesssim \left(N\epsilon g_5(c)\right)^k.
\end{equation}
Thus, if
\begin{equation}\label{Seq:b5}
    N\epsilon g_5(c)\ll1,
\end{equation}
only a small number of non-overlapping small light-cones has non-negligible contribution and hence $C^{(F)}_{\text{light-cones}}$ is small compared to $P_0$.

In summary, combining our results in Eqs.~(\ref{Seq:b1},\ref{Seq:b2},\ref{Seq:b3},\ref{Seq:b4},\ref{Seq:b5}) under the depth condition $d\sim N^{1/D}$, the estimated sufficient condition to guarantee both $C^{(F)}_{\text{cancellation}}$ and $C^{(F)}_{\text{light-cones}}$ to be small is
\begin{equation}
    N\epsilon g(c)\ll1,
\end{equation}
where $g(c)$ is a function of $c$ of order unity. This sufficient condition has the same scaling behavior as the necessary condition in Eq.~(\ref{Seq:nec_condition}).

\subsection{Additivity of XEB vs Multiplicativity of fidelity}\label{sapp:add_mul}
Here we construct a specific error distribution (in the adversarial setting) for which the difference between the fidelity and the XEB becomes very large.
We will see that our original algorithm in the main text directly utilizes this discrepancy.
%can be largely amplified. Moreover, we will use this example to illustrate different scaling behaviors between fidelity and XEB.

%This construction is basically what we use in our algorithm.

\begin{figure}[tbp]
    \centering
    \includegraphics[width=\textwidth]{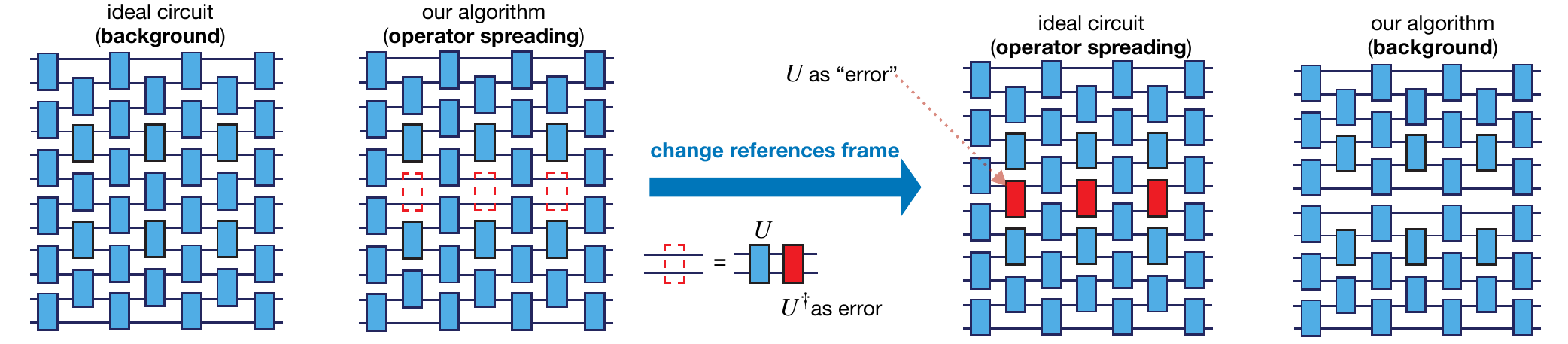}
    \caption{Additivity of the XEB vs. multiplicativity of the fidelity. After changing the reference frame, we consider two isolated quantum circuits as the ``ideal'' reference point. 
    }
    \label{fig:app_add_vs_mul}
\end{figure}

\begin{figure}[tbp]
    \centering
    \includegraphics[width=\textwidth]{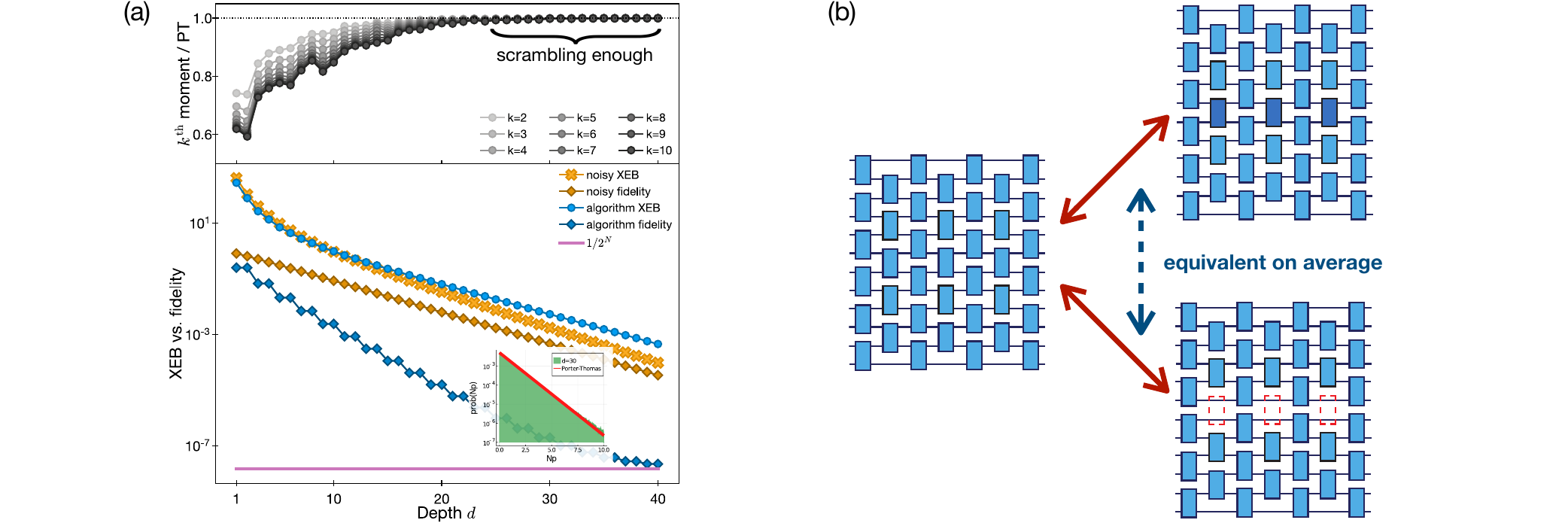}
    \caption{
    A large discrepancy between the XEB and the fidelity for deep quantum circuits. (a) Top. We use the $k$-th moments of the bitstring output distribution as an indicator of the circuit being in the scrambling regime.
    Concretely, we plot the ratio between the $k$-th moments of the output distribution and that of the Porter-Thomas (PT) distribution for circuits with 24 qubits and the 2-qubit Haar random gates. 
    The approximate PT distribution can be seen explicitly (inset of Bottom). 
    The numerical result suggests that the circuit is sufficiently scrambling when the depth is greater than 20, indicated by ``scrambling enough''.
    Bottom. XEB and fidelity as a function of circuit depth $d$ for noisy circuit and for our algorithm.
    In both cases, the XEB and the fidelity decay exponentially in depth.
    For noisy circuits, the XEB and fidelity values stay reasonably close to one another.
    For our algorithm, however, the XEB and the fidelity deviate from one another as they exponentially decay  with different rates.
(b) The average XEB and fidelity of circuits with omitted gates in the middle (i.e., our algorithm) are the same as that of circuits with independent random gates. 
    }
    \label{fig:app_scrambling}
\end{figure}

As shown in Fig.~\ref{fig:app_add_vs_mul}(left), we consider an ideal quantum circuit in 1D architecture decomposed into two parts.
In our algorithm, this is done by omitting all entangling gates acting across two subsystems.
Naively, we may regard the omission of gates as ``errors'' occurring in an ideal circuit.
From this point of view, however, it is difficult to analyze the behavior of the fidelity and XEB because the errors (omitted gates) are strongly correlated with the choice of unitary gates in the  circuit.
This issue can be resolved by changing our reference frame: we consider the quantum circuit with omitted gates as an ``ideal reference'' whereas the full quantum circuit as the ``quantum circuits with errors'' where randomly chosen entangling gates are applied across two otherwise disjoint circuits.
This change of a reference frame is possible because the formulae for both the XEB and the fidelity are symmetric under the exchange of quantum states obtained from the ideal quantum circuit, or from our classical spoofing algorithm (or an experiment).
In fact, one can consider another situation, where one compares two unitary circuits that have statistically independent random unitaries for all entangling gates acting across two subsystems, and otherwise identical [see Fig.~\ref{fig:app_scrambling}(b)].
Then, one can show that the XEB and the fidelity averaged over the ensemble of Haar-random unitary gates are identical in all three cases.
Note that in the case of statistically independent random unitary gates, each quantum circuit is entangling across two subsystems and, for a sufficiently deep circuit, the output distribution from each may approximately follow the Porter-Thomas distribution.

In all three cases, one can check that the averaged fidelity and XEB can be written in the following form:
\begin{eqnarray}
F&=&F_1\cdot F_2\\
\chi&=&(1+\chi_1)\cdot(1+\chi_2)-1\approx \chi_1+\chi_2,
\end{eqnarray}
where we assumed that the XEB values are small, i.e., $\chi_i\ll1$, corresponding to the relevant regime in recent experiments.
The fidelity factorizes into the product of fidelities for two subsystems because averaging over a random unitary gate acting across two subsystems gives rise to a depolarization channel and disconnects two subsystems.
This statement can be derived more rigorously in subsection~\ref{sapp:DR} by using properties of random unitary gates forming unitary designs.
The derivation of the second line is presented in the main text.

For two equal-size partitions of a system, $F=F_1^2$ and $\chi=2\chi_1$.
If we assume that the XEB closely approximates the fidelity for each subsystem, i.e., $F_1 \approx \chi_1$, then it leads to $F\approx F_1^2 \approx \chi_1^2 \ll \chi$.
Hence, we already see that the global fidelity and the XEB cannot be close to one another.
Conversely, if the global fideliety and the XEB agree with one another, the subsystem XEBs and fidelities cannot agree with one another.

Figure~\ref{fig:app_scrambling} shows numerical simulation results.
We obtained exact results by averaging over random unitary gates analytically first and then evaluating the XEB and the fidelity using our mapping to the diffusion-reaction model discussed in the main text and Sec.~\ref{app:DR}.
As previously mentioned, the averaged fidelity and XEB behave identically for all three cases: circuits  with omitted gates (our original algorithm), added gates (our algorithm in a different reference frame), or statistically independent gates applied at the boundaries of two subsystems.
We find that the XEB and the fidelity both decay exponentially with different decay rates.
The decay rates differ approximately by a factor of 2, consistent with our prediction that $F \approx (\chi/2)^2$ up to an unimportant factor of order unity.

We note that the exponential decay of the fidelity and XEB continues even for relatively large circuit depths $d> 25$, for which bitstring probability distributions approximately follow the Porter-Thomas distribution [see Fig.~\ref{fig:app_scrambling}(a) Top].
Our numerical simulation results highlight that the fidelity and the XEB can be very different, even for deep entangling circuits.

\section{Detailed derivation of the diffusion-reaction model}\label{app:DR}
In this section, we assume the gate ensemble consists of single-qubit Haar random gates (or more generally, single-qubit unitaries with the 2-design property) and any 2-qubit entangling gates,
present a detailed derivation of the diffusion-reaction model and discuss some properties relevant to the results in the main text. In subsection~\ref{sapp:1qubit}, we briefly review the properties of the unitary 2-design ensemble consisting of single-qubit unitaries, and use them in subsequent subsections to derive the diffusion-reaction model. In subsection~\ref{sapp:DR_transfer}, we show that the transfer matrix $T^{(G)}$ in the resulting diffusion-reaction model has the form of Eq.~({\color{red}15}) in the main text; we also discuss the physical meaning of its parameters. Then, in subsection~\ref{sapp:DR_stationary}, we compute the stationary distribution using $T^{(G)}$  and present numerical evidence which shows that depth 20 in the Sycamore architecture is sufficiently deep to reach the equilibrium distribution.  In subsection~\ref{sapp:DR_non_ideal}, we study the effects of introducing defective gates, including noisy gates and omitted gates. We also discuss how to detect the type of noise present in the system by an algorithm similar to the one used for spoofing the XEB. Finally, in subsection~\ref{sapp:DR_alg}, we analyze the scaling behavior of our classical algorithm and its dependence on  the properties of omitted gates and the transfer matrix $T^{(G)}$.
We show that, using the language of the diffusion-reaction model, even the fine details of the $N$-dependence (e.g., ups and downs in Fig.~{\color{red}11} in the main text) can be explained in an intuitive way.

\subsection{Brief review of 2-design properties of single qubit unitaries}\label{sapp:1qubit}
We explain how the behavior of quantum circuits, averaged over an ensemble of unitary gates, can be expressed in a simple form.
In particular, we will consider averaging a single-qubit unitary gate over the Haar ensemble, which is a uniform distribution over all unitaries in $\textsf{SU}(2)$.

Consider quantum states $\otimes_{i=1}^t \rho_i$, in the $t$ copies of a Hilbert space, undergoing the same unitary evolution $u$.
We are interested in the resultant quantum state averaged over the Haar random unitary $\mathbb E_{u\in\text{Haar}}\left[\bigotimes_{i=1,\cdots, t}u\rho_i u^\dag\right]$. 
The ensemble of Haar-random unitaries is defined by the invariance of the averaged quantity by both the left and right multiplications of any unitary $v\in \textsf{SU}(d)$; i.e., for any $t\in \mathbb{Z}_{+}$,
\begin{equation}\label{Seq:HaarId}
 \mathbb E_{u\in \text{Haar}}\left[ \left(u\otimes u^*\right)^{\otimes t}\right]=\mathbb E_{u\in \text{Haar}}\left[ \left(vu\otimes v^*u^*\right)^{\otimes t}\right] = \mathbb E_{u\in \text{Haar}}\left[ \left(uv\otimes u^*v^*\right)^{\otimes t}\right].
\end{equation}
This is a natural definition for a uniform distribution: if the ensemble is uniformly distributed, any application of extra rotation by $v$ should only ``permute'' the elements of $\textsf{SU}(d)$ from $u\mapsto vu$ or $u\mapsto uv$, and the average should not be affected.
We focus solely on $d=2$ (i.e., qubits) in this work.
By considering the average behavior of the $t$-copy wavefunction under the same random unitary $u$, we can study the behavior of observables in the extended ($t$-copy) Hilbert space, which contains  observables that are nonlinear (up to power $t$) in a single-copy density matrix. 
In this work, since we focus on the expectation values of the XEB and the fidelity, it suffices to study the case where $t=1$ and $t=2$. In other words, the Haar ensemble can be replaced by any other ensemble of unitaries that behaves identically to the Haar ensemble in $t=1$- and $t=2$-copy Hilbert spaces on average, which is the defining property of the so-called \emph{unitary 2-design}.

\begin{figure*}
\includegraphics[width=0.8\textwidth]{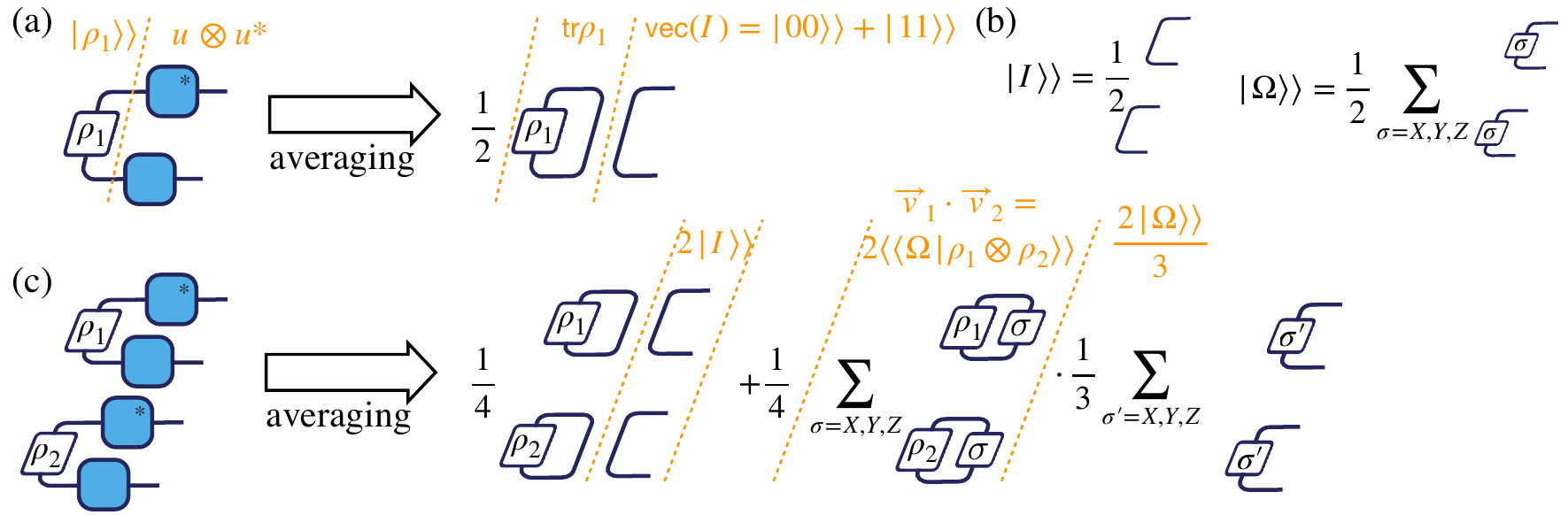}
\caption{
Diagram of averaging over Haar ensemble. The blue box represents a single qubit Haar random unitary $u$ and the blue box with a $^*$ represents $u^*$.
(a) Averaging for $t=1$ copy (i.e., unitary 1-design property). The diagram can be understood as either a tensor network representation or a circuit (non-unitary after averaging) of a state in Choi representation in which $\tr\rho_1=\langle\innerp{00}{\rho_1}\rangle+\langle\innerp{11}{\rho_1}\rangle$ . (b) Diagram of $\kket{I}$ and $\kket{\Omega}$ which are defined in Eq.~\eqref{Seq:I_Omega}. (c) Averaging for $t=2$ copies (i.e., unitary 2-design property) which is a diagram of Eq.~\eqref{eq:app_2design} by using (b).
}
\label{fig:12design}
\end{figure*}

First, we consider the case of $t=1$ to introduce some useful notations and identities, which will be helpful for the $t=2$ case. Let $u$ be sampled uniformly from $\textsf{SU}(2)$ and $\rho_1$ be any 2-qubit input state. Observe that the quantity $\mathbb E_{u\in\text{Haar}}[u\rho_1 u^\dag]$ is invariant under the action of an arbitrary $v\in\textsf{SU}(2)$, i.e., $\mathbb E_{u\in\text{Haar}}[u\rho_1 u^\dag]=\mathbb E_{u\in\text{Haar}}[uv\rho_1v^\dag u^\dag]=\mathbb E_{u\in\text{Haar}}[vu\rho_1u^\dag v^\dag]$. Thus, we have
\begin{equation}\label{eq:1-design_og}
\mathbb E_{u\in\text{Haar}}[u\rho_1 u^\dag]=(\tr\rho_1)\frac{I}{2}
\end{equation}
because only $\tr\rho_1$ and $I$ are the invariant quantities with respect to $\textsf{SU}(2)$. 
We adopt two standard representations of quantum many-body states that will be mathematically convenient for later calculation: diagram (also known as tensor network representation~\cite{bridgeman2017hand}) and Choi representation \cite{choi1975completely} (which is already used in the main text and also known as Choi-Jamio\l kowski isomorphism). The former provides intuitive graphics for gates and states while the later associates a density matrix to an ``entangled state" by the map~\cite{watrous2018theory} 
$$\text{vec:}\sum_{ij}\rho_{ij}\ket i\bra j\mapsto\sum_{ij}\rho_{ij}\kket{ij}.$$ We use $\kket\rho=\text{vec}(\rho)$ to represent this ``entangled state". Then 
\begin{equation}
\text{vec}(u\rho u^\dag)=u\otimes u^*\kket{\rho} \text{ and } \tr\rho=\sqrt2\langle\innerp{\text{Bell}}{\rho}\rangle
\end{equation}
where $\kket{\text{Bell}}=\text{vec}(I)/\sqrt{2}=(\kket{00}+\kket{11})/\sqrt2$ is the ``Bell state".
Thus Eq.~\eqref{eq:1-design_og} can be rewritten as
\begin{equation}\label{eq:1-design}
\mathbb E_{u\in\text{Haar}}[u\otimes u^*\kket{\rho_1}]=\kket{\text{Bell}}\langle\langle\text{Bell}\kket{\rho_1},
\end{equation}
the diagram of which is shown in Fig.~\ref{fig:12design}(a), where a line denotes $\text{vec}(I)$.
Focusing on the effect of unitaries averged over an ensemble, we can identify 
\begin{equation}\label{eq:1_design}
\mathbb E_{u\in\text{Haar}}[u\otimes u^*]=\kket{\text{Bell}}\bbra{\text{Bell}},
\end{equation}
which is a projector to the ``Bell state". Here we use the variation of the Dirac notation $\kket\cdot$ ($\bbra\cdot$) to represent the vector (dual vector) in the Choi representation, instead of an ordinary quantum mechanical state.
The conclusion is that, if we only have a single copy of a quantum state, all directional information on the Bloch sphere is erased after averaging over the Haar ensemble, except the normalization condition (i.e., $\tr\rho_1$ is preserved). The output state is always the maximally mixed state (or equivalently $\kket{\text{Bell}}$, in terms of the Choi representation), no matter what the initial state is.

In the case of $t=2$, as mentioned in the main text, we always have two output states after averaging, denoted as $\kket I$ and $\kket\Omega$ in the Choi representation. In particular, we use these two states as the degree of freedom in the diffusion-reaction model. The central result for a single-qubit gate is
\begin{equation}\label{eq:1qubit}
\mathbb E_{u\in\text{Haar}}[u\rho_1 u^\dag\otimes u\rho_2 u^\dag] = \frac{1}{4}\left(I\otimes I+  \frac{{\vec{\mathrm v}}_1\cdot {\vec{\mathrm v}}_2}{3}{\vec\sigma}\cdot {\vec\sigma} \right) \xrightarrow[]{\text{vec}}\frac{1}{2}\left(\kket{I}+  \frac{{\vec{\mathrm v}}_1\cdot {\vec{\mathrm v}}_2}{3}\kket{\Omega} \right)
\end{equation}
where $\vec{\mathrm v}_i = (\tr{\rho_i X},\tr{\rho_i Y},\tr{\rho_i Z})$ is the vector representation of $\rho_i$ in the Pauli-matrix basis (such that $\rho_i=(I+\vec{\mathrm v}_i\cdot \vec\sigma)/2$), ${\vec\sigma}\cdot {\vec\sigma}=\sum_{\sigma=X,Y,Z}\sigma\otimes \sigma$, $X,Y,Z$ are the 3 Pauli matrices, and 
\begin{eqnarray}\label{Seq:I_Omega}
\nonumber\kket{I}&=& \text{vec}\left(\frac{I\otimes I}{2}\right)= \kket{\text{Bell}}^{\otimes2},     \\
\kket{\Omega}&=&\text{vec}\left(\frac{{\vec\sigma}\cdot {\vec\sigma}}{2}\right)=\sum_{\sigma=X,Y,Z}[(\sigma\otimes I)\kket{\text{Bell}}]^{\otimes2} \, .
\end{eqnarray}
This can be explained according to the invariance of the expectation value under the application of an arbitrary unitary $v\otimes v$ to the density matrix  (that is two $v$s and two $v^\dag$s): (1) the output should be a linear combination of $I\otimes I$ and $\vec\sigma\cdot\vec\sigma$ because they are the only invariant 2-qubit operators (up to linear combination
); (2) by computing the expectation value of the trace with $I\otimes I$ or ${\vec\sigma}\cdot {\vec\sigma}$, the coefficients $1/4$ and $\vec{\mathrm v}_1\cdot \vec{\mathrm v}_2/(3\cdot 4)$ can be determined. The appearance of $\vec{\mathrm v}_1\cdot \vec{\mathrm v}_2$ is a consequence of this invariance, since $v$ is mapped to a rotation on the Bloch sphere while this inner product is invariant under $\text{SO}(3)$. See subsection~\ref{ssapp:Haar} for a more rigorous proof. The conclusion is that all the directional information on the Bloch sphere is deleted after the averaging process, except the normalization condition and the total polarization correlation between the two states, i.e., $\vec{\mathrm v}_1\cdot \vec{\mathrm v}_2$.

The above result can be formulated in the Choi representation (which is the same as Eq.~({\color{red}12}) in the main text):
\begin{equation}\label{eq:app_2design}
 \mathbb E_{{u\in\text{Haar}}} [u\otimes u^*\otimes u\otimes u^*]
=\kket{I}\bbra{I}+\frac{1}{3}\kket\Omega\bbra\Omega,
\end{equation}
where
$$
\langle\innerp{\Omega}{\rho_1\otimes\rho_2}\rangle=\sum_{\sigma=X,Y,Z} \bbra{\text{Bell}}\sigma \otimes I \kket{\rho_1}  \bbra{\text{Bell}}\sigma \otimes I \kket{\rho_2}=\sum_{\sigma=X,Y,Z}\frac{1}{2}\tr(\sigma\rho_1)\tr(\sigma\rho_2)=\frac{\vec{\mathrm v}_1\cdot\vec{\mathrm v}_2}{2}.
$$
The second equality is due to $\tr O=\sqrt2\langle\innerp{\text{Bell}}{O}\rangle$ for an arbitrary operator $O$ (here $O=\sigma\rho_i$ and $\kket O=\text{vec}$($O$)).
 The corresponding diagram is shown in Fig.~\ref{fig:12design}(b,c). Intuitively, $\bbra{I}$ and $\bbra{\Omega}$ encode the normalization information and the total polarization correlation information $\vec{\mathrm v}_1\cdot \vec{\mathrm v}_2$, respectively. $\kket I$ and $\kket{\Omega}/3$ represent the propagation of the corresponding information to the next time step. The $1/3$ factor in $\kket\Omega$ could be understood as the 3 polarization correlations (represented by $\text{vec}(\sigma\otimes \sigma/2)$ with $\sigma=X,Y,Z$) with equal probability being propagated. In the next subsection (subsection \ref{sapp:DR}), we will elaborate on this interpretation in terms of the diffusion-reaction model, where $I$ and $\Omega$ represent vacuum and particle states, respectively.

\subsubsection{Proof of 2-design properties}\label{ssapp:Haar}

In the following, we will prove Eq.~\eqref{eq:1-design_og} and Eq.~\eqref{eq:1qubit}. They are special cases of the Weingarten formula~\cite{weingarten1978asymptotic} for $d=2$ and $t=1,2$, respectively. In this special situation, we present simple proofs for completeness.

First, we prove Eq.~\eqref{eq:1-design_og}. The density matrix $\rho$ of a single qubit can be written in the Pauli basis as follows.
\begin{equation}\label{Seq:densitym}
    \rho=\frac{\tr(\rho) I+\vec{\mathrm v}\cdot \vec\sigma}{2},
\end{equation}
where $\vec{\mathrm v}$ is a 3-dimensional vector and the Pauli matrices $\vec\sigma=(X,Y,Z)$.
By expanding the expression of $\rho$ in this way, it suffices to understand $\mathbb{E}_{u\in\text{Haar}}[uIu^\dagger]$ and $\mathbb{E}_{u\in\text{Haar}}[u\sigma u^\dagger]$ for all $\sigma\in\{X,Y,Z\}$. First, $\mathbb E_{u\in\text{Haar}}[u\rho u^\dag]$ is straightforwardly
\begin{equation*}
   \mathbb E_{u\in\text{Haar}}[uI u^\dag]= I. 
\end{equation*}
For each $\sigma\in\{X,Y,Z\}$, we use Eq.~\eqref{Seq:HaarId} with $v=\sigma^{\prime}\in\{X,Y,Z\}\backslash\{\sigma\}$,
\begin{align}
   \mathbb E_{u\in\text{Haar}}[u \sigma u^\dag]&=\mathbb E_{u\in\text{Haar}}[u \sigma^\prime\sigma\sigma^\prime u^\dag]\nonumber \\
   &= -\mathbb E_{u\in\text{Haar}}[u \sigma u^\dag] \nonumber\\ 
   &= 0,\label{Seq:1des2} 
\end{align}
where we used the identity $\sigma^\prime\sigma\sigma^\prime= -\sigma$ for $\sigma\neq \sigma^\prime$. Putting the above results together gives
\begin{equation*}
    \mathbb E_{u\in\text{Haar}}[u \rho u^\dag] = \tr(\rho)\frac{I}{2},
\end{equation*}
which proves Eq.~\eqref{eq:1-design_og}. .

Next, we prove the 2-design property from Eq.~\eqref{eq:1qubit}. Using the parametrization from Eq.~\eqref{Seq:densitym}, the tensor product of two density matrices is
\begin{eqnarray*}
\rho_1\otimes \rho_2 &=& \frac{I\otimes I}{4}\\
&+& \frac{I\otimes \vec{\mathrm v}_2\cdot \vec\sigma}{4}+\frac{ \vec{\mathrm v}_1\cdot \vec\sigma\otimes I}{4}\\
&+& \sum_{\sigma_1\ne \sigma_2\in\{X,Y,Z\}}\frac{\mathrm v_1^{(\sigma_1)} \mathrm v_2^{(\sigma_2)}}{4}\sigma_1\otimes\sigma_2\\
&+& \sum_{\sigma\in\{X,Y,Z\}}\frac{\mathrm v_1^{(\sigma)}\mathrm v_2^{(\sigma)}}{4}\sigma\otimes \sigma.
\end{eqnarray*}
where we used the notation, where $\mathrm v_i^{(X)}$ is the $x$-th component of $\vec{\mathrm v}_i$.
The Haar-average of the first line is simply
$$
\mathbb E_{u\in\text{Haar}}[uI u^\dag \otimes uI u^\dag]=I\otimes I.
$$
The terms in the second line become zero after averaging due to the same reasoning as in Eq.~\eqref{Seq:1des2}; as an example, the first one is
\begin{equation}\label{Seq:2-design_og}
\mathbb E_{u\in\text{Haar}}[(uI u^\dag) \otimes (u \vec{\mathrm v}_2\cdot \vec\sigma u^\dag)]=I \otimes \mathbb E_{u\in\text{Haar}}[ u \vec{\mathrm v}_2\cdot \vec\sigma u^\dag]
=0.
\end{equation}
The average of the third-line term ($\sigma_1\neq \sigma_2$) also vanishes because
\begin{equation*}
\mathbb E_{u\in\text{Haar}}[u\sigma_1 u^\dag \otimes u \sigma_2 u^\dag]=E_{u\in\text{Haar}}[u(\sigma_1)^3 u^\dag \otimes u \sigma_1\sigma_2\sigma_1 u^\dag]
=-E_{u\in\text{Haar}}[u\sigma_1 u^\dag \otimes u \sigma_2 u^\dag]
= 0,
\end{equation*}
where we again used Eq.~\eqref{Seq:HaarId} with $v=\sigma_1$ and $\sigma_1\sigma_2\sigma_1= - \sigma_2$ for $\sigma_1\neq \sigma_2$.
Finally, for the fourth term, we have
\begin{equation}\label{Seq:2 design symmetry of pauli}
\mathbb E_{u\in\text{Haar}}[uX u^\dag \otimes u X u^\dag]=\mathbb E_{u\in\text{Haar}}[uY u^\dag \otimes u Y u^\dag]
=\mathbb E_{u\in\text{Haar}}[uZ u^\dag \otimes u Z u^\dag],
\end{equation}
which can be seen from the invariance under unitary rotations, namely $Y$ and $Z$ operators are related to $X$ operator by unitary transformations (e.g. Hadamard and $\pi/2$-phase gate).
Then we use the identity
\begin{equation}\label{Seq:SWAP}
    2S=I\otimes I + X\otimes X+Y\otimes Y+ Z\otimes Z,
\end{equation}
where $S$ is a SWAP operator. Crucially, $S$ commutes with any tensor product of two identical operators
\begin{equation*}
    (u\otimes u)S = S(u\otimes u),
\end{equation*}
by definition. %which can be verified by direct matrix multiplication. 
This means that $\mathbb E_{u\in\text{Haar}}[u\otimes u  S\, u^\dag \otimes u^\dag] = S$ and by Eq.~\eqref{Seq:2 design symmetry of pauli} and Eq.~\eqref{Seq:SWAP} we have
\begin{align}
    &\mathbb E_{u\in\text{Haar}}[(u\otimes u)  (X \otimes X) \, (u^\dag \otimes u^\dag)]
    =\mathbb E_{u\in\text{Haar}}[(u\otimes u)  (Y \otimes Y) \, (u^\dag \otimes u^\dag)]
    =\mathbb E_{u\in\text{Haar}}[(u\otimes u)  (Z \otimes Z) \, (u^\dag \otimes u^\dag)]\nonumber\\
    &= \mathbb E_{u\in\text{Haar}}\left[(u\otimes u)  \frac{X\otimes X + Y \otimes Y + Z\otimes Z}{3} \, (u^\dag \otimes u^\dag)\right]\nonumber\\
    &= \mathbb E_{u\in\text{Haar}}\left[(u\otimes u)  \frac{2S-I\otimes I}{3}\, (u^\dag \otimes u^\dag)\right]\nonumber\\
    &= \frac{2S-I\otimes I}{3}= \frac{X\otimes X+Y\otimes Y+ Z\otimes Z}{3}.\label{Seq:heiseninv}
\end{align}%\\&=&
We use this formula to obtain the expression for the fourth term
\begin{equation*}
    \sum_{\sigma\in\{X,Y,Z\}}  \mathrm v_1^{(\sigma)}\mathrm v_2^{(\sigma)} \mathbb E _{u\in\text{Haar}}[(u\otimes u) \sigma\otimes \sigma (u^\dag \otimes u^\dag)] 
    = \vec{\mathrm v}_1\cdot \vec{\mathrm v}_2\frac{\vec\sigma\cdot\vec\sigma }{3}.
\end{equation*}
Putting all these results together, we
proved Eq.~\eqref{eq:1qubit}.

\begin{figure*}[tbp]
\includegraphics[width=1\textwidth]{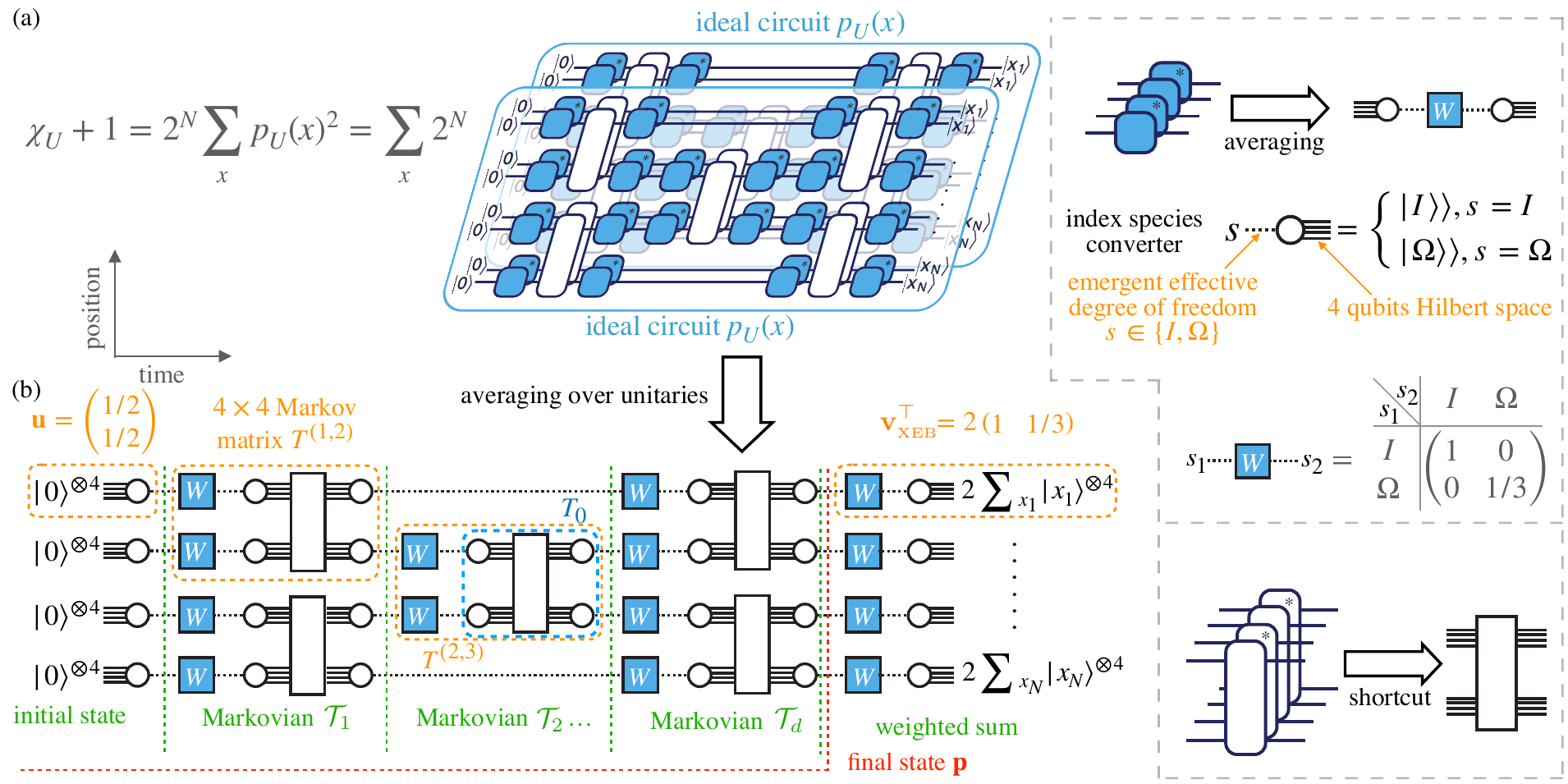}
\caption{
Illustration of the mapping from the average XEB of the ideal circuit to the diffusion-reaction model. (a) The XEB of ideal circuit can be computed by considering two copies of a state evolving under the same random quantum circuit.
Gray dashed boxes defines simple representations of tensors that appear in our tensor network diagrams. The first box gives a shortcut of the diagram for the average behavior of each single qubit gate as discussed in Fig.~\ref{fig:12design}(c) and Eq.~\eqref{eq:app_2design}.
A circle is labelled by a classical variable $s\in \{I, \Omega\}$ and represents the corresponding ``4-qubits states" (or vectorized density matrix in duplicated Hilbert space) $\kket I$ and $\kket\Omega$ defined in Fig.~\ref{fig:12design}(b) and Eq.~\eqref{eq:app_2design} (denoted as 4 lines). The $W$ is a diagonal matrix defined for the classical degree of freedom gives.
The second box defines a simplified diagram for the four copies of a 2-qubit entangling gate.
(b) The tensor network of the diffusion-reaction model where the horizontal direction is viewed as time evolution of a Markovian process, described by $\mathcal T_1,\cdots,\mathcal T_d$, on the classical degree of freedom.
In each $\mathcal T_i$, a matrix $T_0$ (where we omitted the gate dependence $^{(G)}$ here displayed in main text for $T_0^{(G)}$) on the classical degree of freedom is defined as the two copies of 2-qubit entangling gate combined with the white circles.
Then combining $T_0$ with $W$, we get the transfer matrix $T$ which is a Markovian (will be proved in subsection \ref{sapp:DR_transfer}).
In (a), there are two sets of independent single qubit Haar random gates on a wire between two successive entangling gates. They can be merged into one because the product of two independent Haar random unitary gates equals a single Haar random unitary gate. This is why we only have one layer of  $W$'s between two successive layers of entangling unitary gates in (b). Here we only present the tensor network diagram for XEB, fidelity only differs at the right boundary condition as discussed in Fig.{\color{red}8}(a) and (b) of the main text.
}
\label{fig:DR}
\end{figure*}

\subsection{Deriving the diffusion-reaction model}\label{sapp:DR}
In this subsection, we present a detailed derivation of the diffusion-reaction model. We consider the average XEB of an ideal circuit:
\begin{equation}
\chi_{\text{av}}=\mathbb E_{U\in\text{Haar}^{\otimes N_\text{\tiny single}}}\left[2^N\sum_xp_U(x)^2-1\right],
\end{equation}
where we use $\chi_{\text{av}}$ to denote $\mathbb E_{U}[\chi_U]$, and $N_\text{\tiny single}$ is the number of single-qubit Haar gates $u$.
The tensor network diagram representing this quantity is shown in Fig.~\ref{fig:DR}(a). By applying the 2-design properties (inserting Eq.~\eqref{eq:app_2design}) in the middle of two successive layers of entangling gates, i.e., applying the upper-right gray box in Fig.~\ref{fig:DR} to each single-qubit gate, we get a path integral of the diffusion-reaction model in terms of only $\{I,\Omega\}$ variables shown in Fig.~\ref{fig:DR}(b). The path integral turns out to be a Markovian evolution, in the sense that each 2-qubit gate is mapped to a transition matrix $T_0$ over the state space $\{I,\Omega\}^2$, and each single-qubit gate is mapped to a weighted diagonal matrix $W$ (see the gray boxes in Fig.~\ref{fig:DR}); then, we combine two $W$s and $T_0$s together and define $T$ to be the transition matrix over the state space $\{I,\Omega\}^2$ as follows:
\begin{equation}
T=T_0 (W\otimes W).
\end{equation}
We can show that this $T$ is indeed a stochastic matrix (subsection~\ref{sapp:DR_transfer}).
When these gates are applied to the $(i,j)$ qubit pair, we denote $T^{(i,j)}$ to be the corresponding transition matrix over the state space $\{I,\Omega\}^2$. Also, we let $\mathcal{T}_t=\otimes_{(i,j)\in t\text{-th layer}}T^{(i,j)}$ be the transition matrix of the $t$-th layer of the circuit.
To sum up, $\chi_{\text{av}}+1$ can be written as a Markovian evolution in terms of the transition matrices $\mathcal{T}_1,\dots,\mathcal{T}_d$ over the state space $\{I,\Omega\}^N$, with appropriate  boundary conditions as described in Eq.~({\color{red}16}). See also Eq.~({\color{red}20}) in the main text for a summary of the whole diffusion-reaction process. 

As discussed in Fig.{\color{red}8}(a) and (b) of the main text, the corresponding diffusion-reaction model is the same for the fidelity, except the right boundary condition which is given in Eq.~({\color{red}17}); see also Eq.~({\color{red}21}) in the main text.

\subsection{Properties of the transfer matrix $T^{(G)}$}\label{sapp:DR_transfer}
In this subsection, we discuss in more detail the properties of the transfer matrix $T$.
Here and below, we omit the superscript in $T^{(G)}$ in order to simplify our notations whenever doing so does not lead to ambiguity.
We will focus on the connection between $T$ and the amount of entanglement generated by an entangling gate.
\begin{figure}[tbp]
\includegraphics[width=0.8\textwidth]{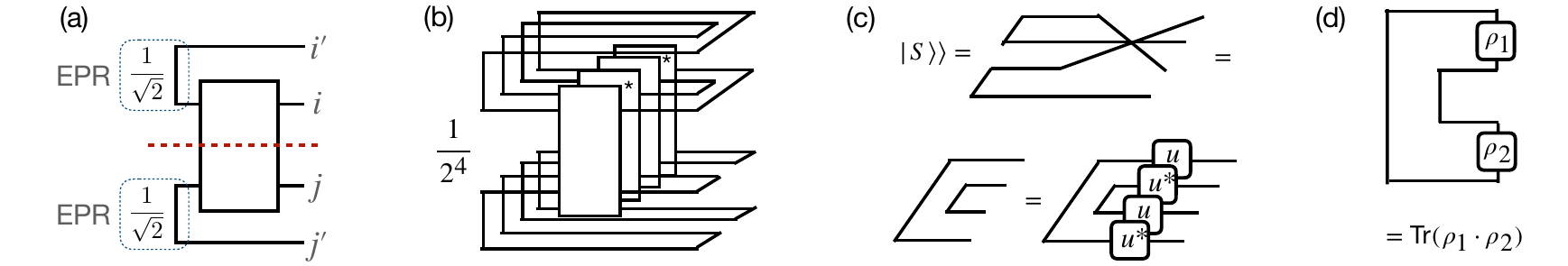}
\caption{``Apparent entanglement" and $\kket S$. (a) We consider two qubits $i$ and $j$ that are initially maximally entangled with their respective partners $i'$ and $j'$. The ``apparent entanglement'' is defined by the amount of the entanglement  (quantified by the smallness of the purity of reduced density matrices) between $ii^\prime$ and $jj^\prime$ that is generated by a unitary gate acting on $i$ and $j$.
(b) Tensor network diagram representation of  $\tr(\rho_{i^\prime i}^2)$, where $\rho_{i^\prime i}$ is the reduced density matrix of the subsystem labeled by $i$ and $i'$.
(c) Diagramatic representations of the state $\kket S$ and its invariance under the action of $u\otimes u^*\otimes u\otimes u^*$. The SWAP operator is applied to the first and third lines (could be also between the second and fourth). Here we only display the invariance under a single-qubit unitary, it is straightforward to see the invariance under 2-qubit unitaries for $\kket S^{\otimes2}$. (d) $\langle\innerp{\rho_1\otimes\rho_2}{S}\rangle=\tr{\rho_1\rho_2}$. }
\label{fig:app_D}
\end{figure} 

Recall the expression for $T$, presented in  Eq.~({\color{red}15}) of the main text, which we reproduce here for convenience
\begin{equation}\label{eq:transfer_general}
T=\begin{pmatrix}
1 & 0 & 0 & 0\\
0 & 1-D & D-R & R/\eta\\
0 & D-R & 1-D & R/\eta\\
0 & R & R & 1-2R/\eta
\end{pmatrix}.
\end{equation}
This matrix has the following properties, which we prove later in this section:
\begin{itemize}
\item The first column and the first row are all zero except the first entry. This reflects the fact that the polarization correlation can only be produced by the propagation from the polarization correlation with other qubits. Similarly, this holds for the reverse process. In terms of the diffusion-reaction model, a particle cannot be created or annihilated from vacuum. The interaction with other particles is necessary in order to change the particle number.
\item $T$ is symmetric with respect to switching the two sites (exchanging the second and third columns and rows). This reflects the fact that $T$ describes the process of entanglement changes, and that entanglement is a  concept symmetrical between the two qubits.
\item We define the quantities $R=T_{I\Omega\rightarrow \Omega\Omega}$ and $R/\eta=T_{\Omega\Omega\rightarrow I\Omega}$ to study the reaction process. We prove that the reaction ratio $\eta=3$ for every 2-qubit gate is set. This reflects the fact that there are 3 species corresponding to $\Omega$ (3 polarization directions) while there is only 1 species for $I$. In the next subsection, we will show that $\eta$ almost fully determines the stationary distribution $\mathbf p_\infty=\lim_{d\rightarrow\infty}\mathbf p$ (except some retrograde cases like $D=0$ or $R=0$) under the evolution of $\mathcal{T}_1,\cdots,\mathcal{T}_d$, when $d$ is large enough, for arbitrary circuit architectures in subsection.
\item Since $\eta=3$ is independent of the choice of the gate set, the reaction rate $R$ fully determines the reaction process. We prove that $R$ is quantitatively related to the ``2-body entanglement productivity" (also denoted as ``entanglement power"~\cite{zanardi2000entangling}) for $G^{(i,j)}\ket{\psi_i}\ket{\psi_j}$, which is defined as
\begin{equation}\label{eq:2_body_ent}
 S_2=1-\mathbb E_{|\psi_i\rangle,|\psi_j\rangle\in\text{Haar}}[\tr\rho_i^2],
\end{equation}
where $\rho_i$ is the reduced density matrix of the subsystem labeled by $i$. We have
\begin{equation}\label{eq:2_body_R}
R=3S_2.
\end{equation}
It can be shown that $0\le R\le 2/3$ (according to the result in Ref.~\cite{vatan2004optimal} that ``entanglement power" $S_2$ is at most 2/9). 
In terms of the diffusion-reaction model, $R$  characterizes the ability to change the particle number.
Generally, larger $R$ implies faster equilibration to the stationary distribution $\mathbf p_\infty$. 
\item We define the diffusion rate $D=1-T_{I\Omega\rightarrow I \Omega }$ to describe the process of particle diffusion or, equivalently, the random walk speed (by noting that the duplication process also includes a movement: e.g., $I\Omega \rightarrow \Omega\Omega$ should be viewed as the second particle is moved over 1 site and then duplicated). Intuitively, the more entanglement the 2-qubit gate can produce, the easier the polarization correlation propagates (the faster the particles move). In fact, we prove that
\begin{equation}\label{eq:apparent_D}
D=\frac{4}{3}S_\text{a} \text{ with }
S_\text{a}=(1-\tr\rho_{i^\prime i}^2),
\end{equation}
where $\rho_{i^\prime i}$ is the reduced density matrix of the subsystem labeled by $i$ and $i'$, as explained in Fig.~\ref{fig:app_D}.
Note that $0\le D\le 1$ because $0\le S_\text{a}\le 3/4$ (3/4 can be fulfilled when all the eigenvalues of the reduced density matrix are $1/4$). The quantity $S_\text{a}$ represents the ``apparent entanglement productivity'' of the state $G^{(i,j)}\ket{\text{Bell}}_{i^\prime i}\ket{\text{Bell}}_{jj^\prime}$,
as shown in Fig.~\ref{fig:app_D}(a). 
We say ``apparent'' because this quantity measures the effective bond dimension that increases when the gates is applied in the tensor network state before optimization (e.g., before truncating the bond dimension by the SVD decomposition~\cite{white1992density,vidal2003efficient,verstraete2008matrix,paeckel2019time,zhou2020limits}). However, it does not always characterize the true entanglement productivity.

We note that $D$ and $R$ characterize different aspects of an entangling gate.
For example, SWAP has $D=1$ but $R=0$, i.e., $T_{I\Omega\rightarrow \Omega I}=1$ and $T_{I\Omega\rightarrow I \Omega }=T_{I\Omega\rightarrow \Omega \Omega }=0$. 
When only SWAP gates are applied to the initial state $\ket0^{\otimes N}$, there is no entanglement produced--- no matter how many gates are applied. In terms of the  diffusion-reaction model, the particle distribution will never approach the equilibrium $\mathbf p_{\infty}$ in this case.
When $R>0$, larger $D$ implies faster equilibration time because, for example, when there are no particles around a given site, particles from other sites need to come and interact in order for the particle number to increase.
\item Finally, the quantity $T_{I\Omega\rightarrow\Omega I}=D-R$ must be non-negative, which roughly reflects the intuition that the ``apparent entanglement productivity" is larger than the ``2-body entanglement productivity", up to a rescaling factor, because the former is only ``apparent".
\end{itemize}
In summary, the reaction ratio $\eta$ is independent of the choice of a gate set. ``Reaction'', governed by the reaction rate $R>0$, is the only mechanism that changes the particle number and leads to the equilibration to $\mathbf p_\infty$. Therefore, it is necessary to produce the scrambling state. The diffusion rate $D>0$ can accelerate the equilibration if $R>0$. These quantities are essential to understand the realtion between the XEB and the fidelity for non-ideal random quantum circuits, which will be discussed in subsection~\ref{sapp:DR_non_ideal}.
In the rest of this subsection,
we prove the above properties. In the next subsection, we solve for $\mathbf p_\infty$ and explain why the average XEB $\chi_\text{av}\approx1$ for deep ideal circuits.

\subsubsection{Proofs of these properties}
In the rest of this subsection, to make equations shorter, we use $T_{s_1s_2,s_3s_4}$ to denote $T_{s_3s_4\rightarrow s_1s_2}$. This notation is consistent with the convention of using column vectors to represent a distribution (such that a Markov matrix is applied from the left).
We use the same convention for $T_0$.
First, we prove a few properties of $T_0$, defined in Eq.~({\color{red}12}) of the main text or Fig.~\ref{fig:DR}(b): (1) each entry is non-negative; (2) the entry in the first row and the first column is 1; (3) all the other entries in the first row and the first column are 0; (4) this matrix is symmetric; (5) this matrix is invariant under switching of the first site and the second site, i.e., switching the second and third rows and columns.
To prove (1), we denote $\kket{\bar\sigma}^{\otimes2}=[(\bar\sigma\otimes I)\kket{\text{Bell}}]^{\otimes2}$, where $\bar\sigma$ extends $\sigma$ by including the identity $I$,
such that 
$$\kket I=\sum_{\bar\sigma\in\{I\}}\kket{\bar\sigma}^{\otimes2}
\text{ and }\kket\Omega=\sum_{\bar\sigma\in\{X,Y,Z\}}\kket{\bar\sigma}^{\otimes2}
,$$
which could be summarized as
$$
\kket s=\sum_{\bar\sigma(s)}\kket{\bar\sigma(s)}^{\otimes2}
$$
where $s\in\{I,\Omega\}$ and the summation is over $\bar\sigma(s)\in\{I\}$ if $s=I$ and $\bar\sigma(s)\in\{X,Y,Z\}$ if $s=\Omega$. Then,
\begin{eqnarray*}
T_{0; s_1s_2,s_3s_4}&=&
\sum_{\bar\sigma(s_i)}
\bbra{\bar\sigma(s_1)}^{\otimes2} \bbra{\bar\sigma(s_2)}^{\otimes2} (G\otimes G^{*})^{\otimes 2}
\kket{\bar\sigma(s_3)}^{\otimes2} \ket{\bar\sigma(s_4)}^{\otimes2}\\
&=&\sum_{\bar\sigma(s_i)}\left(
\bbra{\bar\sigma(s_1)} \bbra{\bar\sigma(s_2)}(G\otimes G^{*})
\kket{\bar\sigma(s_3)} \kket{\bar\sigma(s_4)}\right)^2\ge0,
\end{eqnarray*}
where $G$ is the 2-qubit gate. 
To prove (2), we choose $s=I$ in the last equation, 
$$T_{0;II,II}=\left(\bbra{\text{Bell}} \bbra{\text{Bell}}(G\otimes G^{*})
\kket{\text{Bell}} \kket{\text{Bell}}\right)^2=(\tr( GG^{\dag}))^2/2^4=1.$$
To prove (3), similarly, as an example, $$T_{0;II,I\Omega}=\sum_{\sigma=X,Y,Z}(\tr(\sigma\otimes I))^2/2^4=0$$ and similarly for other entries with $II$ on the left or right. To prove (4) and (5), we need to prove the following equalities
\begin{eqnarray*}
T_{0;I\Omega,\Omega I}&=&T_{0;\Omega I,I\Omega}\\
T_{0;I\Omega, I\Omega }&=&T_{0;\Omega I,\Omega I}\\
T_{0;I\Omega, \Omega\Omega }=T_{0;\Omega I,\Omega \Omega}&=&T_{0;\Omega\Omega, I\Omega }=T_{0;\Omega \Omega,\Omega I}.
\end{eqnarray*}
As an example, we prove the first equality in detail while the others follow from the same idea. Define $\kket S=\text{vec}(S)$ where $S$ is the SWAP operators (see Fig.~\ref{fig:app_D}(c)). According to Eq.~\eqref{Seq:SWAP},
$$
\kket S=\kket I+\kket\Omega,$$
which is basically $\kket\downarrow$ as we introduced in Sec.~{\color{red}IV D} of the main text in which we discuss the mapping to the Ising spin model. Recall the symmetry of the exchange $2\kket I \leftrightarrow \kket S$ (which is equivalent to the permutation symmetry between the two $G$s or the two $G^*$ in $G\otimes G^*\otimes G\otimes G^*$):
\begin{eqnarray*}
T_{0;IS,SI}&=& T_{0;II,II}+T_{0;I\Omega,II}+T_{0;II,\Omega I}+T_{0;I\Omega,\Omega I}=T_{0;II,II}+T_{0;I\Omega,\Omega I}  \\
=T_{0;SI,IS}&=& T_{0;II,II}+T_{0;\Omega I,II}+T_{0;II,I\Omega}+T_{0;\Omega I,I\Omega}=T_{0;II,II}+T_{0;\Omega I,I\Omega}, 
\end{eqnarray*}
where the equalities in the last column are due to terms like $T_{0;II,I\Omega}=0$. Thus, we have proved all the general properties (1)-(5) of $T_0$. Then, we use them to prove properties of $T$.

\emph{Gate set independent properties of $T$.}---
Next, we prove that for all choices of gate sets, the corresponding transition matrix $T$ must have the form shown in Eq.~\eqref{eq:transfer_general}. Recall that $T=T_0\cdot W^{\otimes 2}$. The matrix $W$ does not change the first row and the first column of $T_0$, so $T$ also has the properties (1)--(3) of $T_0$. Because of the symmetry of exchanging $I\Omega\leftrightarrow \Omega I$ for $T_0$ and $W^{\otimes 2}$, $T$ also has this symmetry; this explains why $T$ is invariant under exchanging the second and third columns and rows. Then, the last row and the last column are almost the same except an extra factor $\eta=3$ due to the transpose symmetry of $T_0$ and $W$ (which caused the $1/3$ factor). Finally, we need to prove that each column is normalized. Recall that $\kket{S}=\kket{I}+\kket{\Omega}$, and in the following we associate $I$ with $0$ and $\Omega$ with $1$ for convenience of writing equations. For each $s_1,s_2\in\{I,\Omega\}$, the sum of the column indexed by $s_1s_2$ is
\begin{eqnarray*}
\sum_{s_a,s_b}T_{s_as_b,s_1s_2}&=&  T_{SS,s_1s_2}\\
&=& \frac{T_{0;SS,s_1s_2}}{3^{s_1+s_2}} \\
&=& \frac{\langle\innerp{S}{s_1}\rangle}{3^{s_1}}\cdot \frac{\langle\innerp{S}{s_2}\rangle}{3^{s_2}}\\
&=& \sum_{\bar\sigma(s_1),\bar\sigma(s_2)} \frac{\tr\left(\bar\sigma(s_1)^2\right)}{2\cdot3^{s_1}}\cdot \frac{\tr\left(\bar\sigma(s_2)^2\right)}{2\cdot3^{s_2}}\\
&=& \frac{3^{s_1}}{3^{s_1}} \cdot \frac{3^{s_2}}{3^{s_2}}=1,
\end{eqnarray*}
where the third and fourth equalities are due to Fig.~\ref{fig:app_D}(c) and (d), respectively (where the later is obtained by considering $\rho_i=\bar\sigma(s_i)$). These  fully determine the form of $T$, as shown in Eq.~\eqref{eq:transfer_general}, where $D$ and $R$ are just two parameters depending on the 2-qubit gate.

\emph{Gate set dependent properties of $T$.}---
Next, we consider gate set dependent terms by linking $D$ and $R$ to the entanglement properties of the 2-qubit entangling gate $G^{(i,j)}$. First, we consider the ``2-body entanglement productivity" $S_2$ shown in Eq.~\eqref{eq:2_body_ent},
\begin{widetext}
\begin{eqnarray*}
1-S_2=&&\mathbb E_{u_1,u_2\in\text{1-qubit Haar}}\tr(\rho_2^2)\\
=&&\mathbb E_{u_1,u_2\in\text{1-qubit Haar}}2\bbra I\bbra S(G^{(1,2)}u_1\ket0 u_2\ket0)\otimes(G^{(1,2)*}u^*_1\ket0 u^*_2\ket0)\otimes(G^{(1,2)}u_1\ket0 u_2\ket0)\otimes(G^{(1,2)*}u^*_1\ket0 u^*_2\ket0)\\
=&&\mathbb E_{u_1,u_2\in\text{1-qubit Haar}}2\bbra I\bbra S
(G^{(1,2)}\otimes G^{(1,2)*}\otimes G^{(1,2)}\otimes G^{(1,2)*})\cdot 
\left(
(u_1\otimes u_1^*\otimes u_1\otimes u_1^* \ket0^{\otimes4})(u_2\otimes u_2^*\otimes u_2\otimes u_2^* \ket0^{\otimes4})\right)\\
=&&2\bbra I\bbra S
(G^{(1,2)}\otimes G^{(1,2)*}\otimes G^{(1,2)}\otimes G^{(1,2)*})
\frac{1}{4}
\left(\kket I+\frac{1}{3}\kket\Omega \right) \left(\kket I+\frac{1}{3}\kket\Omega \right)\\
=&& \frac{1}{2}\left(T_{0;II,II}+\frac{1}{3}T_{0;I\Omega,\Omega I}+\frac{1}{3}T_{0;I\Omega,I\Omega}+\frac{1}{9}T_{0;I\Omega,\Omega \Omega}\right)\\
=&& \frac{1}{2}\left(T_{0;II,II}+\frac{1}{3}T_{0;\Omega I,I\Omega }+\frac{1}{3}T_{0;I\Omega,I\Omega}+\frac{1}{9}T_{0;\Omega \Omega,I\Omega}\right)\\
=&& \frac{1}{2}\left(T_{II,II}+T_{\Omega I,I\Omega}+T_{I\Omega,I\Omega}+\frac{1}{3}T_{\Omega\Omega,I\Omega}\right)
=\frac{1}{2}\left(
1+1-R+\frac{1}{3}R
\right)\\
=&&1-\frac{1}{3}R,
\end{eqnarray*}
where in the second line $2\kket I$ plays the role of a partial trace and $\kket S$ plays the role of matrix multiplication and then trace as shown in Fig.~\ref{fig:app_D}(d); thus it represents the purity $\tr(\rho_2^2)$ where $\rho_2$ represents the reduced density matrix of the second qubits for the output state after $G^{(1,2)}$; the fourth equality is because of $W\cdot\mathbf u$ according to Fig.~\ref{fig:DR}(b); we omitted terms which are 0 in the fifth line and notice the change of the second and fourth terms in the sixth line. Note that this proves the relation between the reaction rate $R$ and the 2-body entanglement productivity, as described in Eq.~\eqref{eq:2_body_R}.
\end{widetext}

Second, we consider the ``apparent entanglement'' $S_\text{a}$, which is defined by the state shown in Fig.~\ref{fig:app_D}(a). According to Fig.~\ref{fig:app_D}(a) and (b),
\begin{eqnarray*}
1-S_\text{a}&=&\frac{1}{4}T_{0;SI,SI}=\frac{1}{4} (T_{0;II,II}+T_{0;\Omega I,\Omega I})=\frac{1}{4} (T_{II,II}+3T_{\Omega I,\Omega I})=\frac{1+3-3D}{4}\\
&=&1-\frac{3D}{4}
\end{eqnarray*}
This proves the relation between the diffusion rate $D$ and the apparent entanglement productivity, as described in Eq.~\eqref{eq:apparent_D}.

Direct calculations could give $D$ and $R$ for arbitrary 2-qubit entangling gates like CZ, fSim and fSim$^*$. In the following, we present an example for the 2-qubit Haar ensemble.

\subsubsection{Calculation of $T_\text{Haar}$}
$T_\text{Haar}$ can be found in Ref.~\cite{dalzell2020random},  but for completeness we show the derivation here.
%Finally, we calculate $T_\text{Haar}$ by establishing some relations between $D$ and $R$ and solving the system of equations. 
Denote $\kket{\sigma}=\kket{\bar\sigma}^{\otimes2}$, then $\kket\Omega=\sum_{\sigma=X,Y,Z}\kket{\sigma}$, and consider one of the entries of $T_0$ for $G^{(i,j)}$ chosen from the Haar 2-qubit unitary ensemble
$$
T_{0;I\Omega,I\sigma}=\bbra I\bbra\Omega G^{(i,j)}\otimes G^{(i,j)*}\otimes G^{(i,j)}\otimes G^{(i,j)*}\kket I\kket\sigma
$$
. We can prove
$$
T_{0,\text{Haar};I\Omega,IX}=T_{0,\text{Haar};I\Omega,IY}=T_{0,\text{Haar};I\Omega,IZ}
$$
by using the definition of the Haar random ensemble and inserting $V=I\otimes H,I\otimes HS$ into $G^{(i,j)}$. Then,
$$
T_{\text{Haar};I\Omega,I\Omega}=\frac{1}{3}\sum_{\sigma=X,Y,Z}T_{0,\text{Haar};I\Omega,I\sigma}=T_{0,\text{Haar};I\Omega,IX}.
$$
Next, consider
$$
T_{0,\text{Haar};I\Omega,IX}=T_{0,\text{Haar};I\Omega,ZX}=T_{\text{Haar};I\Omega,\Omega\Omega},
$$
where the first equality is obtained by inserting $V=\text{CZ}$, and the last equality follows similarly with $V=H\otimes I,S\otimes I$. By inserting $V=\text{SWAP}$, we can further prove $T_{\text{Haar};I\Omega,I\Omega}=T_{\text{Haar};I\Omega,\Omega I}$. Together, these formulae show that for $T_\text{Haar}$, $1-D=D-R=R/\eta$, which implies $1-D=1/5,D-R=1/5,R=3/5$.

\subsection{Stationary distribution for ideal circuits}\label{sapp:DR_stationary}

One of the important features of the XEB is that, for ideal circuits, its value approaches 1 in the large-depth limit~\cite{arute2019quantum}. Here, we  reproduce this property using our diffusion-reaction model. This familiar example will set the stage for the more complicated cases of noisy circuits and our classical algorithms.

%Since $\chi_\text{av}+1$ is a weighted sum of the probability distribution at the last layer, as defined in Eq.~\eqref{eq:Cpath}, 
Our problem is reduced to computing the stationary distribution $ \mathbf p_\infty=\lim_{d\rightarrow\infty}\mathbf p$ under the evolution of $\mathcal{T}_1,\cdots,\mathcal{T}_d$.
We try the factorizable distribution as an ansatz first, and then show that such an ansatz is the only solution:
\begin{equation}
    \mathbf p_\infty=\bigotimes_{i=1}^N\mathbf u^{(i)}_\infty,
\end{equation}
where $\mathbf u^{(i)}_\infty$ is a proposed single-bit probability in the product distribution ansatz on the $i$-th site, and we assume that they are identical for all sites $i$. For deep-enough circuits, we expect the diffusion-reaction process to reach a fixed point, as is the case in usual Markov processes. Hence, we can write a self-consistent equation for the above ansatz
\begin{equation}
T^{(i,j)}\mathbf u^{(i)}_\infty\mathbf u^{(j)}_\infty=\mathbf u^{(i)}_\infty\mathbf u^{(j)}_\infty.
\end{equation}
This equation has two solutions that are, surprisingly, independent of both $D$ and $R$
\begin{equation}\label{eq:fixed point sol}
{\mathbf u_1}=
\begin{pmatrix}
\frac{1}{1+\eta} \\ \frac{\eta}{1+\eta}
\end{pmatrix}
=
\begin{pmatrix}
\frac{1}{4} \\ \frac{3}{4}
\end{pmatrix}
\text{ or }
{\mathbf u_2}=
\begin{pmatrix}
1\\0
\end{pmatrix}.
\end{equation}
These two vectors fully determine all the solutions of $\mathcal T_d\cdots \mathcal T_2 \mathcal T_1\mathbf p_\infty=\mathbf p_\infty$.

We note that these two vectors are the only solutions to the Markovian dynamics of the system.
This can be seen by using the Perron–Frobenius theorem, which implies that the steady state solution is unique for any Markovian process, as long as the process is ergodic, i.e., all configurations have non-zero transition probabilities to one another.
In our diffusion-reaction model, it can be easily checked that any pair of particle configurations with at least one particle in the system have non-zero transition probabilities upon the multiplication of the transfer matrix for a finite time. 
Therefore, all these configurations form an ergodic sector.
The trivial configuration with no particles in the entire system forms its own ergodic sector.  
Therefore, our Markovian process has at most two stationary solutions, corresponding to ${\mathbf u_1}^{\otimes N}-1/4^N{\mathbf u_2}^{\otimes N}$ and $\mathbf{u}_2^{\otimes N}$. Furthermore, it is not difficult to check that our Markovian dynamics is also aperiodic, i.e., for time steps larger than 2, there is always a non-zero transition probability from one configuration to itself, which avoids the problem of periodic solutions.

The probability of the all-vacuum state in the initial distribution is $1/2^N$, which results in the final probability
\begin{align}
\mathbf p_\infty &= \frac{1-1/2^N}{1-1/4^N}({\mathbf u_1}^{\otimes N}-\frac{1}{4^N}\mathbf {u_2}^{\otimes N})+\frac{1}{2^N}{\mathbf u_2}^{\otimes N}\nonumber\\
&\approx
\left(1-\frac{1}{2^N}\right){\mathbf u_1}^{\otimes N}+\frac{1}{2^N}{\mathbf u_2}^{\otimes N}+O(1/4^N).
\end{align}
Finally, we apply the appropriate boundary conditions to $\mathbf p_\infty$ at the final time (${\mathbf v^\top}_\text{\tiny XEB}^{\otimes N}$) to get the XEB for deep circuits
\begin{eqnarray}\label{eq:ideal_XEB}
\nonumber\chi_{\infty;\text{av}}&=&{\mathbf v^\top}_\text{\tiny XEB}^{\otimes N}{\mathbf p_\infty}-1\\
\nonumber&\approx& \left(1-\frac{1}{2^N}\right) \left(\mathbf v_\text{\tiny XEB}^\top\mathbf u_1 \right)^N+\frac{1}{2^N}  \left(\mathbf v_\text{\tiny XEB}^\top\mathbf u_2 \right)^N-1\\
\nonumber&\approx&  \left(\mathbf v_\text{\tiny XEB}^\top\mathbf u_1 \right)^N+\frac{1}{2^N}2^N-1\\
\nonumber&=&   \left(\mathbf v_\text{\tiny XEB}^\top\mathbf u_1 \right)^N\\
&=&\left(\frac{1}{2}+\frac{3}{2}\cdot\frac{1}{3}\right)^N= 1.
\end{eqnarray}
The small contribution $\mathbf u_2$, which represents the all-vacuum state, cancels out with the $-1$ term in the definition of XEB.

\begin{figure}[tbp]
\includegraphics[width=0.5\textwidth]{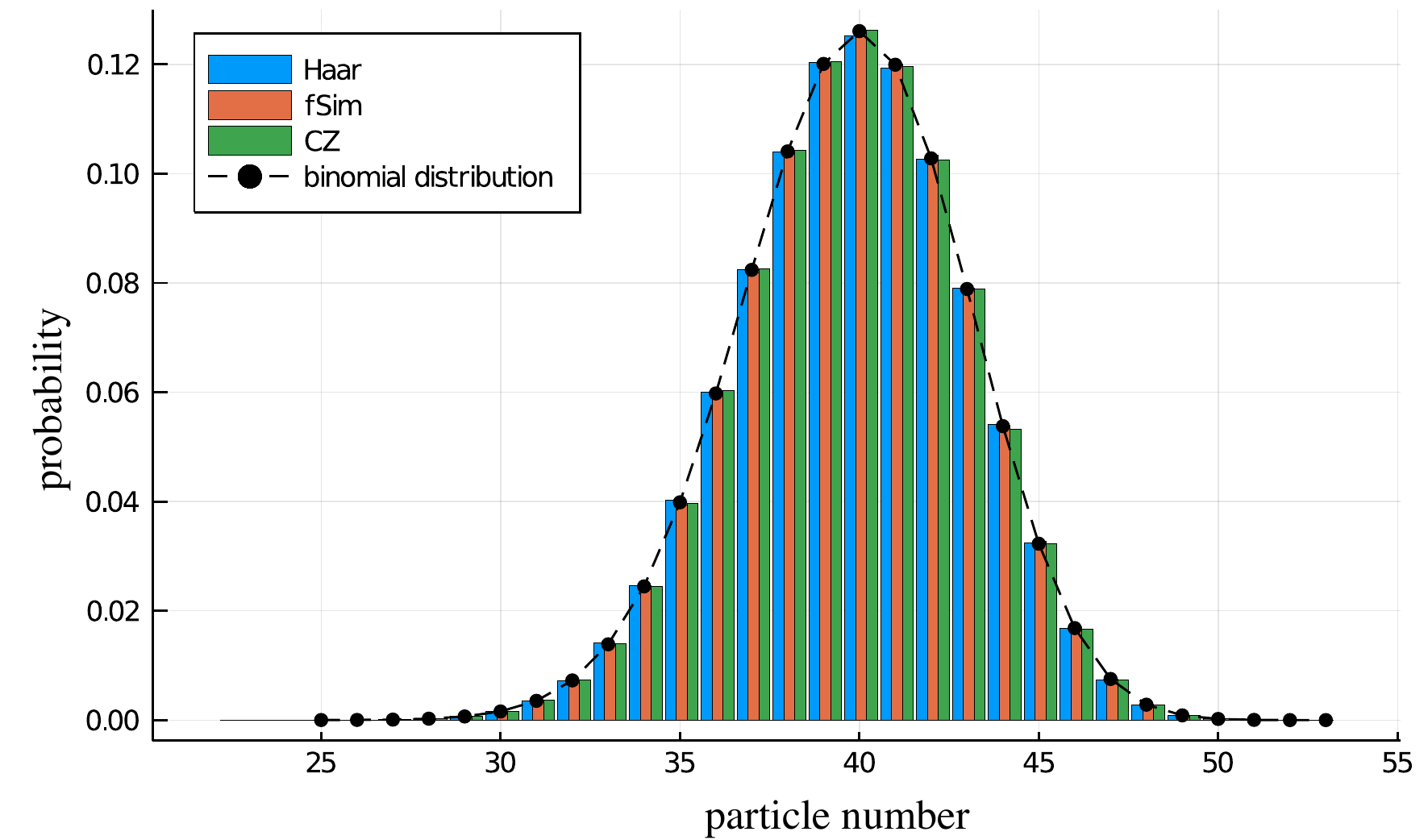}
\caption{The stationary distribution histogram fitted with a binomial distribution $B(53,3/4)$.
We use $10^7$ samples (instances of the diffusion-reaction process) to draw the normalized histogram. The circuits correspond to the Sycamore architecture, with $N=53$ and $d=20$.
}
\label{fig:stationary}
\end{figure} 

Finally, we study the distribution induced by the finite-depth random circuits. We have shown that the stationary distribution is $\mathbf p=(1/4,3/4)^{\otimes N}$ in Eq.~\eqref{eq:fixed point sol} up to an exponentially small correction, thus the particle number distribution obeys the binomial distribution $B(N,3/4)$ and this can be used to diagnose whether the depth is large enough for the equilibration to occur (strictly speaking, this is necessary but not sufficient). This can be achieved by simulating the diffusion-reaction model, which is a classical stochastic process and thus much easier to simulate (e.g., using Monte Carlo sampling). For the Sycamore architecture with 53 qubits and depth 20, we numerically test corresponding diffusion-reaction models and find that the particle number distribution can not be distinguished from $B(N,3/4)$, as shown in Fig.~\ref{fig:stationary}, which gives strong evidence that this class of circuits with $d=20$ is deep enough.

\subsection{The effects of defective gates}\label{sapp:DR_non_ideal}
For ideal circuits, the corresponding probability distribution $\mathbf p$ of a pure-state ensemble is normalized to 1 since its average fidelity is 1. However, for mixed states or in the presence of correlations between two non-identical pure states, this is not always the case. Concretely, let $\rho_1$ and $\rho_2$ be two distinct density matrices. We consider the corresponding distribution $\mathbf p$ defined as
\begin{equation*}
p(s_1\cdots s_N)=\langle\innerp{s_1\cdots s_N}{\rho_1\otimes \rho_2}\rangle \text{ such that } \sum_{s_1\cdots s_N}\langle\innerp{s_1\cdots s_N}{\rho_1\otimes \rho_2}\rangle=\langle\innerp{S^{\otimes N}}{\rho_1\otimes \rho_2}\rangle=\tr\rho_1\rho_2.
\end{equation*}
Note that $\mathbf p$ is not always normalized to 1 because the fidelity of $\tr\rho_1\rho_2$ is less than 1 in general. In this subsection, we consider two situations: (1) noisy circuits, i.e., $\rho_1=\ket{\psi}\bra{\psi}$ and $\rho_2$ is its noisy version; (2) our algorithm: $\rho_1=\ket{\psi_1}\bra{\psi_1}$ and $\rho_2=\ket{\psi_2}\bra{\psi_2}$, where $\psi_2$  is a related, but not identical, pure state. In the following, before presenting the results of noisy circuits and our algorithm respectively, we discuss single-qubit examples first, in order to build intuition.

\subsubsection{Noisy gates}

\begin{figure}[tbp]
\includegraphics[width=0.8\textwidth]{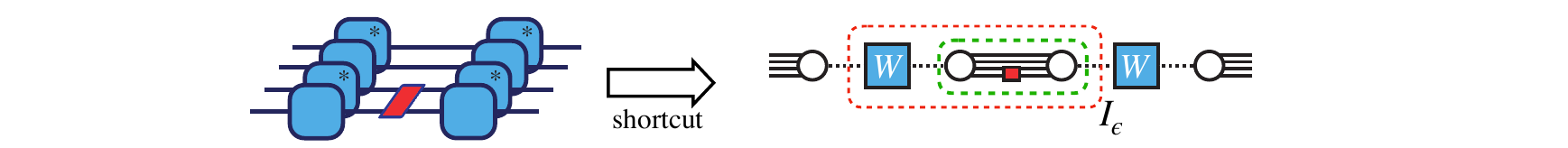}
\caption{The effect of noise. The diagram on the right represents a simplified picture of the left after averaging over two single-qubit Haar random unitaries. The matrix $I_\epsilon$ (appearing in Eq.~({\color{red}24}) in the main text) is defined as the matrix in the space of classical degrees of freedom bounded by the red, dashed box. The part inside the green box is computed as $\bbra{s_1}\hat I\otimes\hat\Phi_\epsilon\kket{s_2}$.}
\label{fig:app_noise}
\end{figure} 

Recall that in Eq.~\eqref{eq:1qubit}, the only non-trivial part is $\vec{\mathrm v}_1\cdot\vec{\mathrm v}_2$, which is the coefficient of  $\kket{\Omega}$. For two identical pure states, this inner product is equal to $|\vec{\mathrm v}_1|^2=1$. However, for noisy circuits, e.g. with depolarizing noise, $\vec{\mathrm v}_2=(1-\epsilon)\vec{\mathrm v}_1$, which follows from the density matrix represention in Eq.~\eqref{Seq:densitym}; thus, the inner product is equal to $1-\epsilon$. Intuitively, a small amount of polarization correlation is lost.
In terms of the diffusion-reaction model, the noise introduces the reduction of the probability by a factor of $1-\epsilon$ if there is a particle at the given site, which is a probability loss process. Furthermore, the picture of probability loss also works for any other type of noise. For example, for a coherent noise, we have $\vec{\mathrm v}_2\sim (1-\epsilon)\vec{\mathrm v}_1+\sqrt{2\epsilon} \vec{\mathrm v}_1^\perp$ and so the inner product is also $1-\epsilon$. The amplitude damping noise is similar: $\vec{\mathrm v}_2$ is a combination of the displacement, rotation and possibly shrinkage of $\vec{\mathrm v}_1$ by a total amount $1-\epsilon$. In summary, any type of uncorrelated noise appears in the same way in  the diffusion-reaction model.

Below, we provide a more quantitative analysis of the effect of noise.
We denote $\Phi_\epsilon$ as the quantum channel of the noise, and denote its Choi representation as $\hat\Phi_\epsilon$. Suppose a quantum channel in Pauli basis is given by
$$
\Phi_\epsilon(\rho)=\frac{1}{2}\sum_{\sigma_1,\sigma_2\in\{I,X,Y,Z\}}c_{\sigma_1,\sigma_2}\tr(\rho \sigma_1)\sigma_2,
$$
which is basically the Pauli-Liouville representation for quantum channel (see e.g., Ref.~\cite{greenbaum2015introduction}) such that $c_{I,I}=1$ and $c_{\sigma,\sigma}=1-O(\epsilon)$. Similar to $T$ for entangling gate, we could also compute the corresponding matrix element ($I_\epsilon$ in Eq.~({\color{red}24}) of the main text) for $\hat I\otimes \hat \Phi_\epsilon$ (we use $\hat I$ to denote the Choi representation of the identity operation for ideal circuits) in the $I,\Omega$ basis, explicitly:
\begin{eqnarray}\label{Seq:n}
\nonumber\bbra I  \hat I\otimes \hat\Phi_\epsilon \kket I &=& \frac{\tr(I\Phi_\epsilon(I)) \tr(I\Phi_\epsilon(I))}{4}=1\\
\frac{\bbra I  \hat I\otimes \hat\Phi_\epsilon \kket\Omega}{3}&=&\frac{\sum_{\sigma=X,Y,Z}\tr(I\sigma) \tr(\Phi_\epsilon(I)\sigma)}{3\cdot4}=0\\\nonumber
\frac{\bbra \Omega  \hat I\otimes \hat\Phi_\epsilon \kket I}{3}&=&\frac{\sum_{\sigma=X,Y,Z}\tr(\sigma I) \tr(\Phi_\epsilon(\sigma)I)}{3\cdot4}=0\\\nonumber
\frac{\bbra \Omega  \hat I\otimes \hat\Phi_\epsilon \kket\Omega}{3}&=&\frac{\sum_{\sigma,\sigma^\prime=X,Y,Z}\tr(\sigma^\prime\sigma) \tr(\sigma^\prime\Phi_\epsilon(\sigma))}{3\cdot4}=\frac{\sum_{\sigma=X,Y,Z} \tr(\sigma\Phi_\epsilon(\sigma))}{3\cdot2}\\\nonumber
&=&\frac{\sum_{\sigma\in\{X,Y,Z\}}c_{\sigma,\sigma}}{3}=1-O(\epsilon),
\end{eqnarray}
where the $1/3$ factor comes from $W$ [see Fig.~\ref{fig:app_noise}]. 

As examples, we consider depolarizing noise (parameterized by $\mathcal N_\epsilon(\rho)=(1-\epsilon)\rho+\epsilon/3\sum_{\sigma=X,Y,Z}\sigma\rho\sigma$) and amplitude damping noise. For the depolarizing noise, $c_{X,X}=c_{Y,Y}=c_{Z,Z}=1-4\epsilon/3$. For the amplitude damping noise, $c_{X,X}=c_{Y,Y}=\sqrt{1-\epsilon}$ and $c_{Z,Z}=1-\epsilon$, so the second diagonal element is roughly $1-2\epsilon/3$.

\subsubsection{Omitting gates}
Now, we consider the effect of omitting gates. Recall that there are two distinct density matrices $\rho_1$ and $\rho_2$  involved in the definition of $\mathbf{p}$, where the former corresponds to the ideal circuit and the latter would correspond to our algorithm. In $\rho_2$, the single-qubit Haar gates applied to $\rho_1$ are omitted for $\rho_2$ if the corresponding 2-qubit gates in the circuit are removed as part of the algorithm. We recall that each entangling gate is accompanied by 4 single-qubit Haar unitaries, as shown in Fig.{\color{red}4}(b) in the main text. 
Therefore, omitting a 2-qubit gate implies that we omit not only the entangling gate but also four single-qubit Haar random gates associated with it. Upon averaging, this belongs to the $t=1$ case in subsection~\ref{sapp:DR} and so only normalization survives the process. This effectively corresponds to a maximally depolarizing noise: any directional information is deleted. Another way to view this is to think of the remaining $u$ on $\rho_1$ as an extra unitary which is effectively a rotation (denoted as $\hat R$) on $\vec{\mathrm v}_2$. Then, the inner product between this vector, which is denoted as $\vec{\mathrm v}_1=\hat R\vec{\mathrm v}_2$, and $\vec{\mathrm v}_2$ is $\langle\hat R\vec{\mathrm v}_1,\vec{\mathrm v}_1\rangle$. Its average value is clearly 0. In terms of the diffusion-reaction model, this corresponds to a strong probability loss at the position of omitted gates: once a particle hits this region, the probability density of this particle configuration of $I$ and $\Omega$ over the entire space-time is set to 0. Namely, in the diffusion-reaction model, only configurations  without any probability loss would contribute to $\mathbf p$ at the last layer.

Formalizing the above discussion, $I_\epsilon$ appearing in Eq.~({\color{red}24}) of the main text, will be replaced by the projector $P_I$, since this corresponds to the situation of a noisy circuit with maximal depolarizing noise (such that $c\epsilon=1$), according to the 1-design property.

\subsubsection{Detecting noise type by generalizing XEB}
 As a remark, we note that  if one replaces the ideal circuit with a non-trivial quantum channel (as a reference state), our previous results can be used to extract information about the noise type. Mathematically, this changes $\vec{\mathrm v}_1$ (ideal circuit) to $\vec{\mathrm v}_1'$ (non-trivial quantum channel). Then, the inner product between $\vec{\mathrm v}_1'$ and $\vec{\mathrm v}_2$ (noisy circuit whose properties we want to detect) can reveal the information about the noise type. For example, if we introduce a noisy circuit with amplitude damping noise $\Phi_{\epsilon_0}$ as the reference circuit (instead of the ideal circuit), one of the entries in $I_\epsilon$ becomes $\bbra{I}\hat{\Phi}_{\epsilon_0}\otimes\hat{\Phi}_{\epsilon}\kket{\Omega}/3=\epsilon_0\epsilon/12\ne0$, which is different from Eq.~\eqref{Seq:n}.

\subsection{Analysis of the scaling behavior of our algorithm through the diffusion-reaction model}\label{app:our_algorithm}\label{sapp:DR_alg}

Applying the diffusion-reaction model we developed, it is intuitive to understand the scaling behavior of our algorithm, as discussed in the main text. Increasing $N$ while keeping the number of omitted gates fixed usually increases XEB for our circuit but the opposite is expected from noisy circuits. This can be explained through the diffusion-reaction model. For our algorithm, larger space for particles undergoing random walk will decrease the probability loss rate, because the particles that are far away from the boundary are less likely to hit the loss region (positions of omitted gates).

\begin{figure}[tbp]
\includegraphics[width=0.5\textwidth]{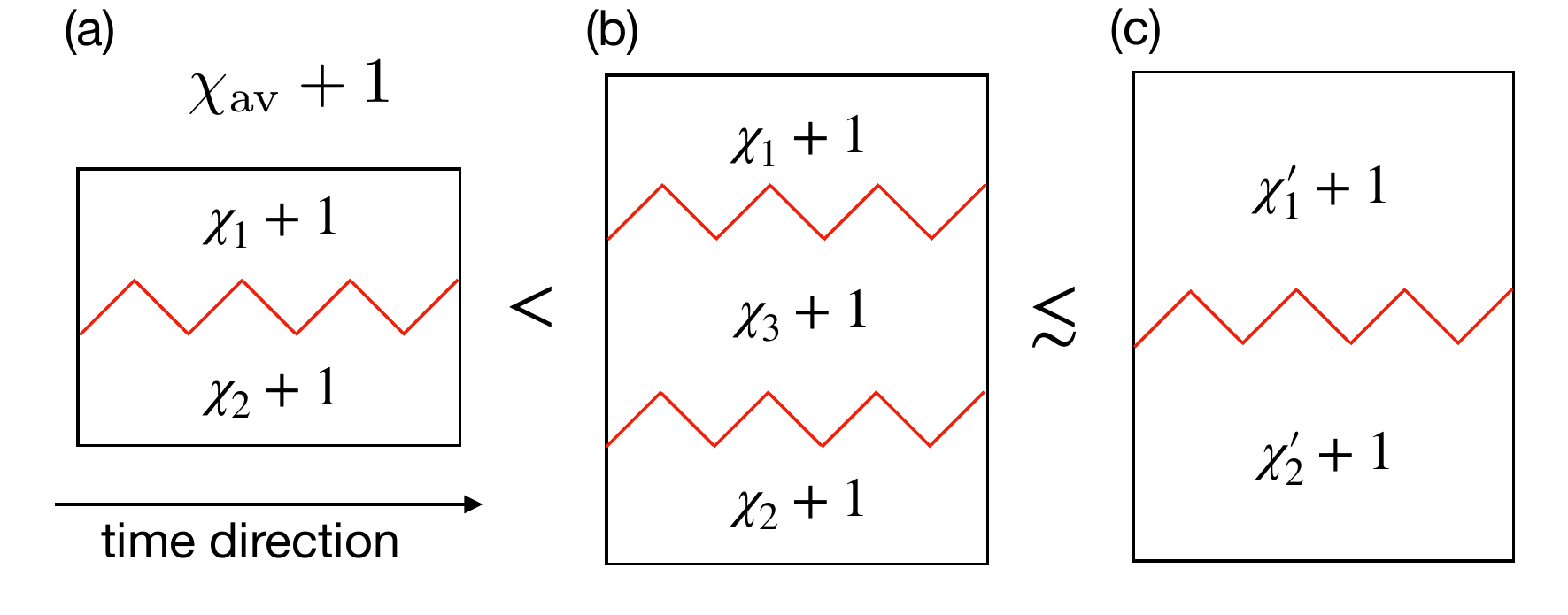}
\caption{Scaling of the XEB when the number of omitted gates is fixed. (a) The behavior of the XEB in our algorithm with two disconnected subsystems. The total $\chi_\text{av}+1$ can be written as $(\chi_1+1)(\chi_2+1) = 1+\chi_1 + \chi_2 + \chi_1 \chi_2$ owing to the decoupling of the diffusion-reaction model by omitting gates.
This quantity is larger than $1+\chi_1$ or $1+\chi_2$.
In fact, the total XEB is approximately additive, $\chi_\text{av}\approx \chi_1 + \chi_2$ when $\chi_1\chi_2$ is small.
(b) Introducing more subsystems only increases XEB, as long as $\chi_3>0$.
(c) It is highly likely that by reducing the number of omitted gates, the XEB can be further increased.}
\label{fig:app_scaling}
\end{figure}

Another way to describe the same phenomena is based on the following observation (also mentioned in Sec.~{\color{red}I D} in the main text): XEB behaves more like an additive (rather than multiplicative) quantity; thus, we should consider the average loss over particles.
In contrast, fidelity behaves multiplicatively, such that we should consider the total loss over particles, thus increasing the system size will decrease fidelity. In Fig.~\ref{fig:app_scaling}, we present a step-by-step argument regarding the scaling behavior of the XEB for our algorithm.

Fig.~{\color{red}11} in the main text shows that the scaling behavior with the system size is quantitatively similar for different gate sets. We observe that the CZ ensemble has much larger XEB than the fSim ensemble. This is because the diffusion rate of the fSim ensemble is the largest (see Table {\color{red}II} in the main text), which means that particles require the least amount of time to hit the loss region. At the first sight, it might seem this result suggests that the smaller XEB for the fSim ensemble is due to the fact that each fSim gate produces larger (at least ``apparent'') entanglement, or larger bond dimension in the tensor network representation--- thus, it is more dangerous to omit fSim gates. However, this is not correct: if all the omitted gates were fSim gates, but the remaining gates would belong to the CZ ensemble, the mean value of the XEB would be the same as in the case where all the omitted gates were CZ gates. In short, the XEB value does not depend on the properties of the omitted gates, but rather on those in the rest of the circuit.

\subsubsection{Fine structure of the scaling in the Sycamore architecture}

The intuitive picture based on the random walk (diffusion) can explain even finer details of the scaling with the system size $N$. For example, in Fig.~{\color{red}11} in the main text, the rise and fall in the value of the XEB is caused by the lattice structure [see Fig.~{\color{red}10}(a) in the main text] and its effect on the diffusion process.
For examples, the large fall occurring at $N=51$ is caused by adding 2 omitted gates that enlarge the loss region and connect qubit 16 and 46 closer to the loss region.
Other examples include adding qubits 14, 16, 20: they shorten the distance for particles at position 13, 15, 19 from the loss region. The qubits that have only a single connection to the rest of the system (before adding subsequent qubits), such as qubits 15, 27, 47, 52, 53, for example, contribute a lot to increasing the XEB value since particles at these positions have only one way out and are kept away from the loss region. This is reflected in the sudden rises in the XEB value when those qubits are included.

\section{Detailed analysis of 1D circuits through the Ising spin model}\label{app:1D}
\begin{figure*}
\includegraphics[width=0.8\textwidth]{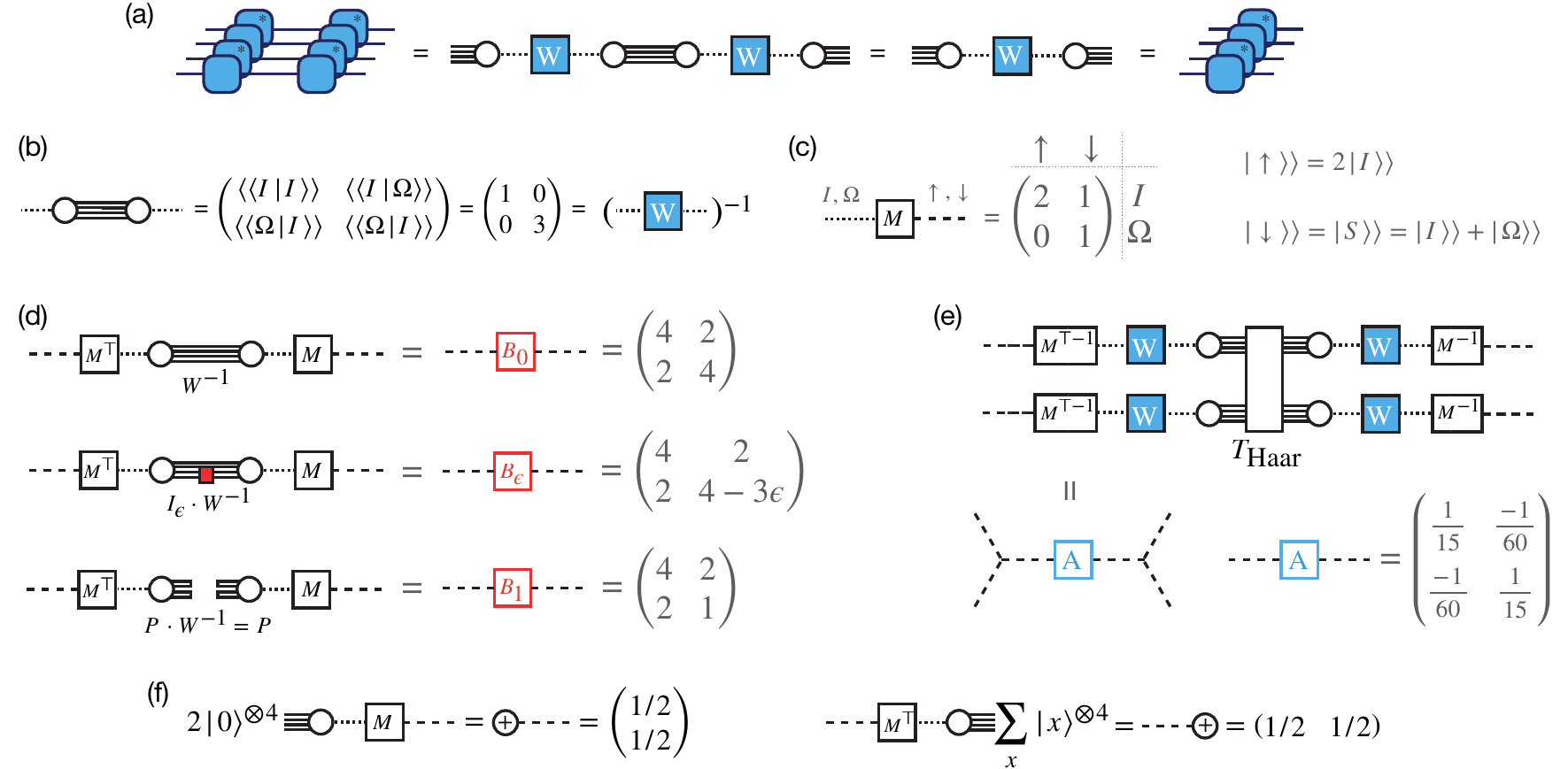}
\caption{Diagramatic representation illustrating the mapping from quantum circuits to the Ising model. This mapping can be understood as a basis change from the diffusion-reaction model. (a) The rule $\mathbb E_\text{Haar}[u\otimes u^*\otimes u\otimes u^*]\cdot\mathbb E_\text{Haar}[v\otimes v^*\otimes v\otimes v^*]=\mathbb E_\text{Haar}[u\otimes u^*\otimes u\otimes u^*]$, which has been used explicitly in Fig.~\ref{fig:DR}. (b) Inverse of the matrix $W$.
(c) A matrix describing the basis change from species $I,\Omega$ in the diffusion-reaction model (dotted line) to spin variables $\uparrow,\downarrow$ in the Ising model (dashed line). (d) The Ising interaction between two classical spin variables associated with different gates (successive in time). The matrices describe Boltzmann weights $e^{a s_is_j+b s_i+b s_j+c}$. 
(e) The Ising interaction between two spins belonging to one gate. This also derives the Weingarten formula~\cite{weingarten1978asymptotic,collins2003moments,collins2006integration} in the case of $t=2,d=2,N=2$. 
(f) The boundary conditions for the XEB, which are both equal-weight summation of $\uparrow,\downarrow$. A new notation, a circle with a plus inside, is introduced and used in Fig.~\ref{fig:partition_function}.
}
\label{fig:ising}
\end{figure*} 

In this section, we present the detailed derivation of the Ising spin model, obtained from the mapping of 1D Haar-random circuits, directly  from the diffusion-reaction model in subsection~\ref{sapp:Ising_model}. 
In particular, we show that the effective Ising model for the Haar-random ensemble is in the ferromagnetic phase.
Then, in subsection~\ref{sapp:1D_fSim}, we present numerical results for STD.

\subsection{Deriving the Ising model from the diffusion-reaction model}\label{sapp:Ising_model}

\begin{figure}
\includegraphics[width=0.45\textwidth]{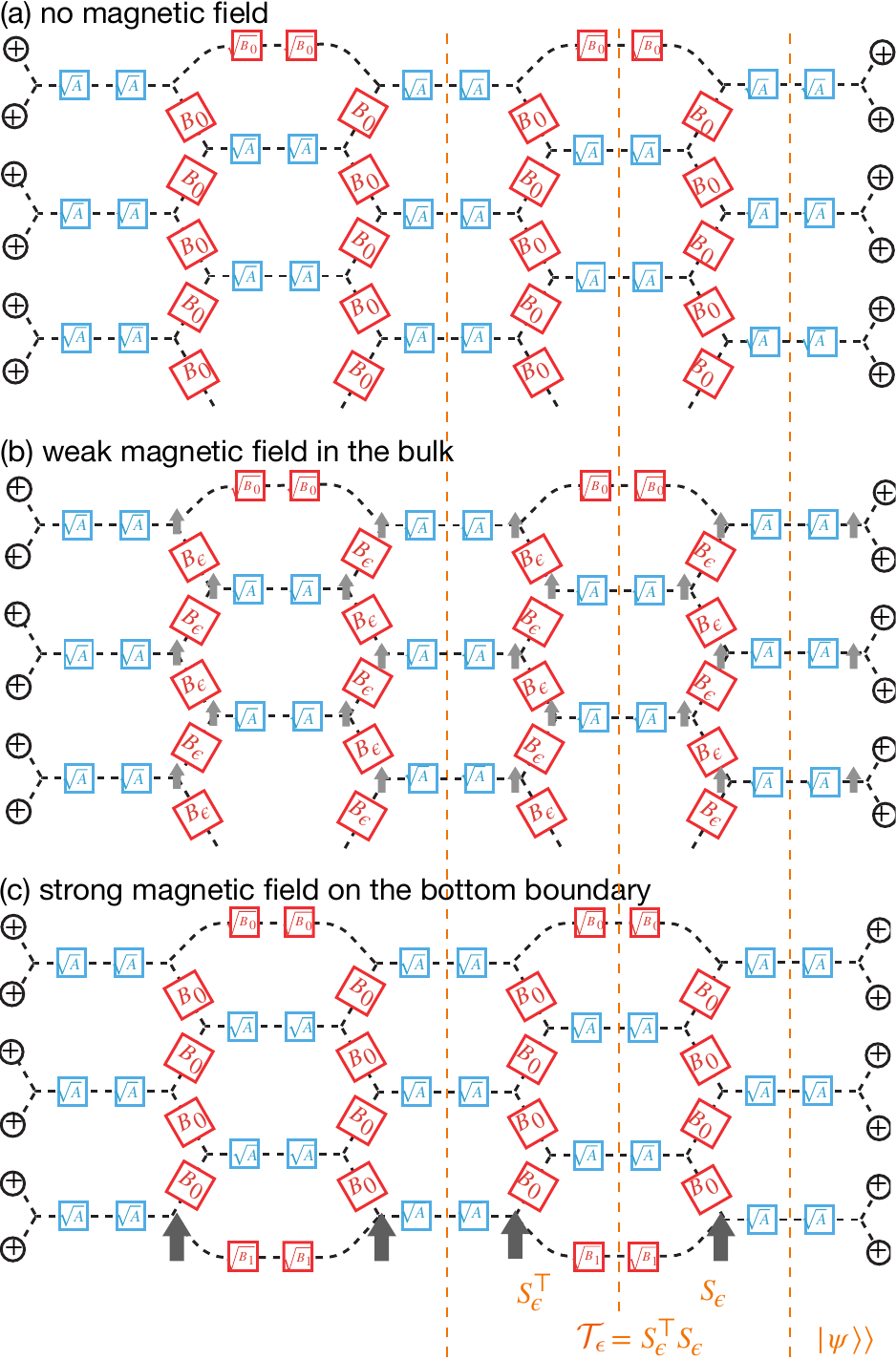}
\caption{$\chi_\text{av}+1$ as a partition function of the Ising spin model. Here we only draw the top part instead of the whole circuit. (a) Ideal circuits. There is a global $\mathbb Z_2$ Ising symmetry.
(b) Noisy circuits. There are weak magnetic fields over the entire bulk with strength $\sim \epsilon$, which breaks the Ising symmetry.
(c) Our algorithm. There are strong magnetic fields on the boundary (at the positions of omitted gates).  The bottom is the transfer matrix view of the partition function. Because $\mathcal T_\epsilon=S^\top_\epsilon S_\epsilon$, the transfer matrix is positive semi-definite.
}
\label{fig:partition_function}
\end{figure}

The mapping from the XEB to the partition function of an Ising model is diagramatically illustrated in Fig.~\ref{fig:ising}.
This mapping is obtained simply by a basis change from the diffusion-reaction model.
The new basis is motivated by the symmetry in the Choi representation of unitary operation $u\otimes u^{*}\otimes u\otimes u^{*}$ on two copies of the states when we exchange the position of the two $u$s or $u^{*}$s (labeled by $1,2$ or $\bar 1,\bar 2$ respectively). This symmetry is hidden in the basis $\kket I$ and $\kket\Omega$, but explicitly exhibited by $\kket I$ and $\kket S$ (which is a SWAP operator acting on the first and the second line of $2\kket I$, as shown in Fig.~\ref{fig:app_D}(c)). This inspires us to try the following basis transformation 
\begin{eqnarray*}
\kket{\uparrow}&=&2\kket I,\\
\kket{\downarrow}&=& \kket S=\kket I+\kket\Omega.
\end{eqnarray*}
Then, the corresponding transfer matrices for ideal circuits, as shown in Fig.~\ref{fig:ising}(d) and (e), are unchanged by exchanging $\kket{\uparrow}\leftrightarrow\kket{\downarrow}$. We regard these two variables as the spins in an Ising model. Concretely, the mapping works in the following way:
(i) turn each local $2$-qubit gate into two spins and a blue box $A$, (ii) turn each wire connecting two 2-qubit gates into a red box $B_\epsilon$ ($\epsilon=0$ and $\epsilon=1$ correspond to ideal gates and omitted gates, respectively; other non-trivial $\epsilon$ values correspond to the noise strength in noisy circuits), (iii) the boundary condition for input and output states are simply equal-weight summations over all possible spin configurations, (iv) the presence of noise or gate defects corresponds to magnetic field towards $\uparrow$ directions with non-zero $\epsilon$. See Fig.~\ref{fig:ising} and Fig.~\ref{fig:partition_function} for a pictorial explanation.

Next, after the basis change, $\chi_\text{av}+1$ from Fig.~\ref{fig:DR}, as well as its versions for noisy circuits (by inserting $I_\epsilon$) and our algorithm (by replacing the omitted gates by projectors $P\otimes P$), correspond to the partition functions of the respective Ising spin models shown in Fig.~\ref{fig:partition_function}. 

We note that there are negative Boltzmann weights in the matrix $A$, which is represented by blue boxes in Fig.~\ref{fig:ising}. Na\"ively, it seems to indicate that we have a non-classical Ising model. However, this model can be transformed into a spin problem with non-negative Boltzmann weights by either integrating out some of the spins~\cite{hunter2019unitary} (also known as the ``star-triangle transformation''), or via the Kramers-Wannier duality~\cite{baxter2016exactly}. Ref.~\cite{hunter2019unitary} makes use of the first approach and shows that this model is in the ferromagnetic phase by counting the domain walls in the model after the transformation.

We start with the XEB of ideal circuits. Let $Z$ be the partition function of an ideal circuit. The key observation is that $Z$ (i.e., $\chi_\text{av}+1$) can be written in the following form ,when the depth $d$ is odd,
\begin{equation}\label{eq:1D Ising partition}
\chi_\text{av}+1=Z=\bbra\psi \mathcal T_0^{(d-1)/2}\kket\psi
\end{equation}
where $\kket\psi$ and $\mathcal T_0$ are shown in the bottom of Fig.~\ref{fig:partition_function} for the case of $\epsilon=0$. The transfer matrix $\mathcal T_0$ is  positive semi-definite and, hence, it has an eigen-decomposition $\mathcal T_0=\sum_i \lambda_i \kket i\bbra i$ with $\lambda_0\geq\lambda_1\geq\cdots\geq0$ and $\langle\innerp{i}{j}\rangle=\delta_{ij}$. Thus, the partition function can be further written as follows,
\begin{widetext}
\begin{eqnarray}\label{eq:Z}
Z&=&\lambda_0^{(d-1)/2}|\langle\innerp{0}{\psi}\rangle|^2
\left(1+\sum_{i>0}\left(\frac{\lambda_i}{\lambda_0}\right)^{(d-1)/2}\frac{|\langle\innerp{i}{\psi}\rangle|^2}{|\langle\innerp{0}{\psi}\rangle|^2}
\right)\nonumber\\
&\xrightarrow{\text{large }d}&\lambda_0^{(d-1)/2}|\langle\innerp{0}{\psi}\rangle|^2\left(
1+\left(\frac{\lambda_1}{\lambda_0}\right)^{(d-1)/2}\frac{|\langle\innerp{1}{\psi}\rangle|^2}{|\langle\innerp{0}{\psi}\rangle|^2}\right)\nonumber\\
&=&
1+\lambda_1^{(d-1)/2}|\langle\innerp{1}{\psi}\rangle|^2.\nonumber
\end{eqnarray}
\end{widetext}
As discussed in the section about the diffusion-reaction model, we know that when $d\rightarrow +\infty$, $Z\rightarrow 2$. Thus, $\lambda_0=\lambda_1=1$ and $|\langle\innerp{0}{\psi}\rangle |=|\langle\innerp{1}{\psi}\rangle |=1$.  This coincides with the $\mathbb Z_2$ symmetry in the ferromagnetic phase.
For noisy circuits or our algorithm, $\lambda_0$ and $\lambda_1$ are no longer equal because of the violation of the Ising symmetry caused by the presence of magnetic fields, as shown in Fig.~\ref{fig:partition_function}(b,c).
We denote $\Delta=(\lambda_0-\lambda_1)/2=(1-\lambda_1)/2$ as the gap and obtain
\begin{equation}
\chi_\text{av}=O\left(e^{-\Delta d}\right),
\end{equation}
where $\lambda_0=1$ and $|\innerp{0}{\psi}|=1$ because $Z\rightarrow 1$ when $d\rightarrow +\infty$. Note that for ideal circuits, we have $\Delta=0$.

\subsection{The numerical results for STD}\label{sapp:1D_fSim}

In the main text, we mainly focus on the mean value of the XEB. To complete our understanding, we also need to address the fluctuations of the XEB value caused by the random unitaries. If the fluctuations turned out to be much larger than the mean value, then it suggests that our result might not hold for individual instances of random circuits with a large probability. 
If the fluctuation of XEB over different instances of random circuits is sufficiently small, it implies that our algorithm can spoof the XEB for any given randomly choosen instance with a large probability. 
In this section, we numerically estimate the STD of our algorithm in several settings. Additionally, we propose a variant of our algorithm to significantly decrease the STD, however, with the cost of a higher running time.

\begin{figure*}[tbp]
\includegraphics[width=1\textwidth]{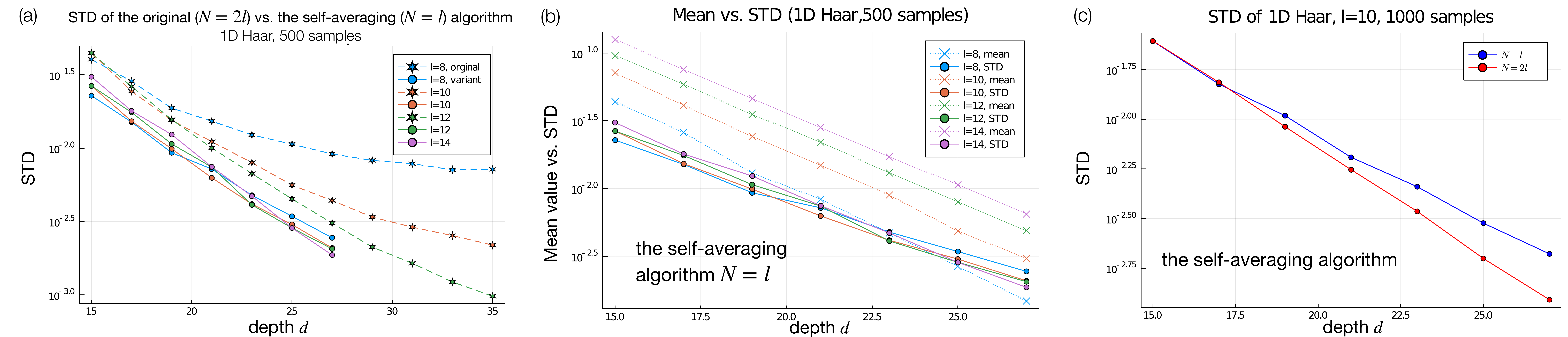}
\caption{(a) The STD of the original algorithm vs. a self-averaging version of our original algorithm (see (b) for more detail) for 1D Haar ensembles with open boundary condition. The former saturates  for a sufficiently large   depth $d$, while the latter does not and is smaller than the former even when the depth is small.
(b) Mean value vs. STD of the self-averaging algorithm (by inserting maximal depolarizing noise instead of omitting gates, see subsection \ref{app:new_algorithm}). Here, the STD is estimated for a subsystem. 
By computing the slope of the solid lines, we extract the decay rate $\Delta_3$, shown in Fig.~{\color{red}6} of the main text (green curve). 
(c) Comparison between the STD in (b), i.e., the $N=l$ case and the $N=2l$ case. This indicates that (b) actually overestimates the actual value of the STD.
}
\label{fig:1D_STD}
\end{figure*} 

It is likely that the STD of the original algorithm saturates to a depth-independent value $2^{-O(l)}$, as suggested by Fig.~\ref{fig:1D_STD}(a).
This is expected because there is no mechanism that would decrease fluctuations further below $2^{-O(l)}$ for disconnected evolution in increasing depth (limited by the Hilbert space dimension of the subsystem).
More specifically,  we note that the (sub)system size is the only characteristic length in random circuits~\cite{brandao2016local,harrow2018approximate} (because the subsystems decouple with each other thus one of them are not influenced by other subsystems). For a sufficiently deep circuit, this is analogous to considering a fully scrambling system. There, the variance of observables is indeed depth-independent, since the system wavefunction approaches Haar-random (or more precisely, 4-design) states within each subsystem of size $l$.

For complexity-theoretic purposes, we must consider the limit of deep circuits. Thus, in 1D systems, the original algorithm does not provide a good asymptotic scaling with $d$, because the mean XEB value will eventually drop below the STD value. However, it is still practical for finite-depth systems.

Therefore, in 1D systems, we focus only on the STD of a self-averaging version of our algorithm (see subsection \ref{app:1D_fSim_STD}) and estimate its value numerically. In this case, we expect the fluctuations to decrease with the depth of the circuit since the maximal depolarizing noise on the boundary adds entropy to the system and causes it to decay to the maximally-mixed state. Compared to the original algorithm, the numerical analysis of such mixed state evolution requires further computational resources. To reduce the amount of required computational resources, we focus on the analysis of only one subsystem.
We argue that it overestimates the STD, which means that the true magnitude of fluctuations is even smaller.
This is because the STD of a single subsystem turns out to be smaller than the STD of the joint distribution of $N/l$ identical subsystems that comprise the whole circuit. We demonstrate this numerically in Fig.~\ref{fig:1D_STD}(c) on the example of $N=2l$. Intuitively, this is because the joint system makes the scrambling more complete (for both ideal circuit and the self-averaging algorithm; the later is regarded as a fully connected system but with very strong depolarizing noise at the positions of omitted gates). Thus, the joint system experiences smaller  fluctuations around the typical cases. 

For more discussion of the self-averaging algorithm and a more efficient implementation for fSim gate, see subsection~\ref{sapp:fSim_trick}.

\section{Properties of circuits with fSim entangling gates}

In this section, we discuss several special properties of quantum circuits consisting of single qubit rotations and the fSim gate.
We will see that these properties both improve and obstruct the performance of our spoofing algorithm: on the one hand, fSim gives rise to the optimal ``scrambling speed" such that our original algorithm becomes relatively less efficient;
on the other, we can take advantage of this ``optimal scrambling'' property to design an improved, more efficient algorithm for spoofing the XEB.
To illustrate the role of this ``optimal scrambling'' property, we first study the effect of maximally depolarizing noise in fSim circuits in subsection~\ref{ssec:fSim identity}, and then analyze the effect of omitted gates on the XEB in subsection~\ref{ssec:fSim omitting gate effect}.

\subsection{Maximally depolarizing noise
in fSim circuits}\label{ssec:fSim identity}

\begin{figure}[tbp]
    \centering
    \includegraphics[width=\linewidth]{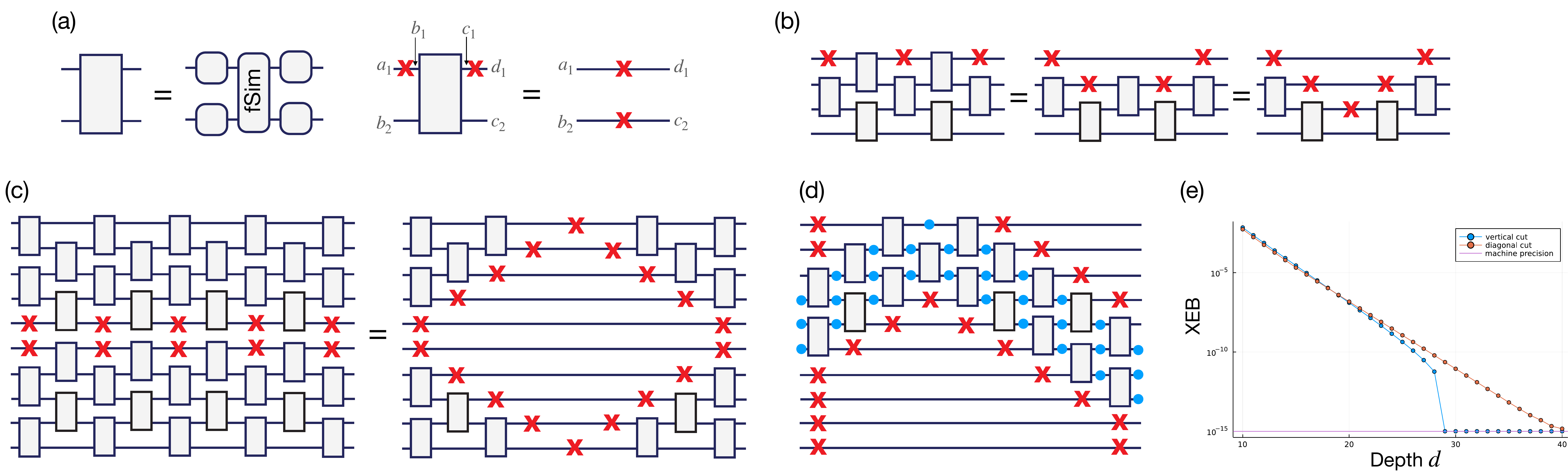}
    \caption{Properties of MDN applied to fSim circuits. The red crosses represent instances of inserting MDN. (a) When MDN is applied to the same qubit before and after the action of an fSim gate (or any gate of the form $U_1 U_2$, where $U_1$ and $U_2$ are the SWAP and controlled phase gate, respectively) while surrounded by single-qubit unitaries, the fSim gate can be effectively replaced by the MDN acting on the other qubit. (b) By repeatedly applying the identity, one can propagate the MDNs in an fSim circuits. (c) For a sufficiently deep fSim circuit, the XEB equals 0, since all the gates in the middle are removed an replaced by MDN; this effectively stop all information flow from the input to the output of the circuit. (d) Different pattern of MDN-insertions, such that the XEB is non-zero even for deep circuits. The blue dots represent particles from the diffusion-reaction model, which in the language of Sec.~\ref{app:DR} corresponds to a
    path with non-zero contribution to the XEB. (e) Numerical simulations verifying (c) and (d), where the subsystem size is 29, and we used periodic boundary conditions and  Haar-random single-qubit gates.}
    \label{fig:app_fSim_identity}
\end{figure}

Here we present a useful property of maximally depolarizing noise (MDN), when applied to fSim circuits.
Formally, MDN is defined as
\begin{align}
    \rho \mapsto \mathcal{D}[\rho] \equiv  \tr{[\rho]} \;  \mathds{1}/2,
\end{align}
where $\rho$ is the density matrix of a single qubit.
When the MDN is applied twice to a qubit: before and after an fSim gate, its effect is equivalent to removing the fSim gate and applying MDN to both qubits [see Fig.\ref{fig:app_fSim_identity}(a)].
Concretely,
\begin{align}
   \mathcal{D}_1\left[ U_\textrm{fSim}\mathcal{D}_1[\rho_{12}] U_\textrm{fSim}^\dagger\right] = \mathcal{D}_2[\mathcal{D}_1[\rho_{12}]],\label{seq:fsimmdn}
\end{align}
where $\rho_{12}$ is a two-qubit density operator, $\mathcal{D}_i$ is the MDN applied to qubit $i$, and $U_\textrm{fSim}$ is the unitary representing the fSim gate.
In fact, this relation holds much more generally for any unitary gate that is equivalent (up to single-qubit rotations) to $U =U_1 U_2$, where $U_1$ and $U_2$ represent the SWAP and the controlled-phase gate, respectively.

This identity can be understood in the following way.
The SWAP operation moves the MDN from the first qubit to the second qubit, while the controlled phase operation preserves the MDN.
More formally, we can write the action of the unitary $U$ on a density matrix in the tensor notation $U_{b_1c_1,b_2c_2}U^*_{b^\prime_1c^\prime_1,b^\prime_2c^\prime_2}$, and similarly the action of the MDN channel $\delta_{aa^\prime}\delta_{bb^\prime}/2$, where the index with $^\prime$ labels the complex conjugate part. When $U$ is of the abovementioned ``SWAP$+$control-phase gate'' form, we have  $U_{b_1c_1,b_2c_2}=\delta_{b_1c_2}\delta_{b_2c_1}e^{i\phi_{b_1b_2}}$. Then, the left-hand side of Fig.~\ref{fig:app_fSim_identity}(a) is 
\begin{eqnarray*}
  \sum_{b_1,b_1^\prime,c_1,c_1^\prime} \frac{1}{2}\delta_{a_1a_1^\prime}
   \delta_{b_1b_1^\prime}\cdot
   \delta_{b_1c_2}\delta_{b_2c_1}e^{i\phi_{b_1c_2}}
   \delta_{b^\prime_1c^\prime_2}\delta_{b^\prime_2c^\prime_1}e^{-i\phi_{b^\prime_1c^\prime_2}}
   \cdot \frac{1}{2}\delta_{c_1c_1^\prime}
   \delta_{d_1d_1^\prime}
   &=&\sum_{c_1,c_1^\prime}\frac{1}{2}\delta_{a_1a_1^\prime}\delta_{c_2c_2^\prime}\delta_{b_2c_1}\delta_{b^\prime_2c^\prime_1}e^{i\phi_{b_1c_2}}e^{-i\phi_{b_1c_2}}\cdot \frac{1}{2}\delta_{c_1c_1^\prime}
   \delta_{d_1d_1^\prime}\\
   &=& 
   \frac{1}{2}\delta_{a_1a_1^\prime}\delta_{d_1d_1^\prime}
    \frac{1}{2}\delta_{b_2b_2^\prime}\delta_{c_2c_2^\prime},
\end{eqnarray*}
where the final result corresponds exactly to the right-hand side of Fig.~\ref{fig:app_fSim_identity}(a) and the statement in Eq.~\eqref{seq:fsimmdn}

Note that this MDN property can be used to explain the propagation of noise in this system. 
For example, as shown in Fig.~\ref{fig:app_fSim_identity}(b), the MDNs applied to the first qubit (represented by the top line) can propagate to the second, and the third qubit by repeatedly applying the identity depicted pictorially in Fig.~\ref{fig:app_fSim_identity}(a).

\subsection{Limitations of our original algorithm applied to fSim circuits}\label{ssec:fSim omitting gate effect}

The MDN identity described in subsection~\ref{ssec:fSim identity} is helpful in studying the effect of omitting gates on the XEB value, in fSim circuits. This is because, once averaged over random unitary gates, omitting gates and applying MDN leads to the same fidelity and XEB in both cases.
Therefore, we consider our algorithm, in which, instead of omitting gates, we apply MDNs for every gates in the middle of the circuit, as depicted in Fig.~\ref{fig:app_fSim_identity}(c).
The MDN property tells us that the MDNs would propagate all the way to the (top and bottom) boundaries of the circuit as long as the circuit is sufficiently deep.
In this case, we find that inputs and outputs for all qubits are completely disconnected by MDNs, implying that the output of the quantum circuit is exactly the maximally mixed state [see Fig.~\ref{fig:app_fSim_identity}(c)].
In other words, when the fSim circuit is sufficiently deep, our algorithm with omitted gates in the middle of the circuit cannot produce any meaningful output bitstring distribution; i.e., the XEB value of our algorithm using this particular positioning of omitted gates will be zero.

There is a simple way to bypass this catastrophic situation by judiciously choosing the position of omitted gates.
As an example, see the ``zig-zag'' pattern in Fig.~\ref{fig:app_fSim_identity}(d).
In this case, the input and output of the circuit remains connected. 
Consequently, the XEB value of our algorithm using this particular choice of omitted gates will be positive even when the circuit is deep. For example, if the subsystem size is 4, as shown in Fig.~\ref{fig:app_fSim_identity}(d), the expected XEB value will scale at least (e.g., there exist other non-zero transition paths) as 
$$
\left(T^{(\text{fSim})}_{I\Omega\rightarrow \Omega\Omega}T^{(\text{fSim})}_{\Omega\Omega\rightarrow \Omega I}\right)^d=
[(1/3+\sqrt{3}/6)^2/3]^d\approx0.13^d,
$$
where $d$ is the depth of the circuit, and we obtained this result using a direct calculation within the diffusion-reaction model.

\section{Improved algorithm: Mixed state simulation and top-$k$ heuristics for fSim circuits}

In this section, we provide more details for our improved algorithm. This algorithm has two steps: (1) replace the omission of gates by inserting maximal depolarizing noise (MDN) and get a probability distribution $\{\widetilde q_x\}$; (2) sort $\{\widetilde q_x\}$ in the decreasing order and choose the first $k$ bistrings with the largest $\widetilde q_x$ as our samples.
In many cases, the distribution from step (1) has exactly the same expected XEB value as the original algorithm due to the 2-design property of the gate set, e.g., for all the ensembles that contain single-qubit Haar ensemble. However, for Google's gate set or its modification (fSim+discrete single-qubit gate), this is not guaranteed, and one has to rely on numerical results.

The motivation of step (1) is two-fold: (i) to get a provable positive XEB; (ii) to have a reduced STD over random circuits. Both (i) and (ii) are important for the step (2) of our improved algorithm. 
This is because as the step (2) amplifies the weight of $x$ with large $\widetilde q_x$, it may also amplify the fluctuation, i.e., the difference between $\widetilde q_x$ and true probabilities.
The two properties (i) and (ii) of the step (1) of our improved algorithm can aid successful amplification in the step (2).
To be more precise, we restate the two properties as follows. (i) $\widetilde q_x$ should have positive correlation with $p_x$; (ii) The STD should be small in order to avoid the case that some occasional $x$ with small $p_x$ but large $\widetilde q_x$ will be amplified (in another words, ``over-fitting'').

In the rest of this section, we first prove the 1-design property of Google's gate set which is the key to prove (i). Second, we prove properties (i) and (ii) of step (1). Third, we discuss in detail how to get $\widetilde q_x$ by simulating the mixed state evolution for the gate set with fSim as the entangling gate. Finally, we discuss the top-$k$ amplification method.

\subsection{1-design property of the  modified Google's single-qubit gate set}

Google's gate set has two ingredients: the single qubit random gate of the form $Z(\theta_1)VZ(\theta_2)$, and the two-qubit fSim entangling gate. For the single-qubit gate, $V$ is chosen randomly from $\{\sqrt X,\sqrt Y, \sqrt W\}$ except with a constraint that two $V$s in successive layers on the same qubit must be different; $Z(\theta_i)$ is a $z$-axis rotation on the Bloch sphere with a site-dependent angle $\theta_i$ which is not actively chosen but is known to be constant and can be potentially calibrated. For simplicity, we introduce two analogous ensembles, with small modifications to the behavior of the $Z$ gate.
To the best of our knowledge, these modifications do not lead to any significant changes in the behavior of quantum circuits.

Ensemble 1: $\theta_i$ is chosen randomly from $[0,2\pi)$. Ensemble 2: $\theta_i$ is chosen randomly from either 0 or $\pi$ (which is $I$ or $Z$ operator respectively). For these ensembles, numerical simulations show that the average XEB values using the top-1 method are $0.00018$ and $0.0004$, respectively, for the Sycamore architecture (53 qubits, 20 depth). Since these values are similar, we argue that the details of the $z$-rotation do not influence the XEB value too much at least not in orders of magnitudes. 

Next, we prove that the single qubit random gates (after the slight modification) form a 1-design ensemble, even with the constraint that two successive $V$'s on the same qubit must be different. 
To see this, we first observe that a single qubit rotation always maps computational states $\ket{0}$ and $\ket{1}$ into their equal superpositions with different relative phases:
\begin{equation*}
    V\ket0=\frac{\ket0+e^{i\phi}\ket1}{\sqrt2}\text{ and } V\ket1=\frac{\ket0-e^{-i\phi}\ket1}{\sqrt2},
\end{equation*}
where $\phi$ depends on the specific gate $V$ we are considering.
We note that this property makes the circuit scramble faster for a fixed entangling gate, compared to the Haar single-qubit gate ensemble. Finally, consider a matrix $M$ under the action of $Z(\theta_1)VZ(\theta_2)$
\begin{eqnarray*}
&&M=\begin{pmatrix}
a & b\\
c & d
\end{pmatrix}\\
\Longrightarrow &&
\mathbb E_{\theta_2}[Z(\theta_2)MZ^\dag(\theta_2)]=
\begin{pmatrix}
a & 0\\
0 & d
\end{pmatrix}\\
\Longrightarrow &&
V\mathbb E_{\theta_2}[Z(\theta_2)MZ^\dag(\theta_2)]V^\dag
=\frac{a}{2}
\begin{pmatrix}
1 & e^{-i\phi}\\
e^{i\phi} & 1
\end{pmatrix}
+
\frac{d}{2}
\begin{pmatrix}
1 & -e^{i\phi}\\
-e^{-i\phi} & 1
\end{pmatrix}\\
\Longrightarrow&&
E_{\theta_1,\theta_2}[Z(\theta_1)V\mathbb Z(\theta_2)MZ^\dag(\theta_2)V^\dag Z^\dag(\theta_1)]=
\frac{a+d}{2}
\begin{pmatrix}
1 & 0\\
0 & 1
\end{pmatrix}
=\tr{M}\cdot\frac{I}{2},
\end{eqnarray*}
where the expectation is over either the ensemble 1 or 2. This proves that these ensembles form a 1-design, no matter which $V$ is chosen.

\subsection{Self-averaging algorithm with maximally depolarizing noise}\label{app:new_algorithm}\label{app:1D_fSim_STD}

Recall that we denote the bitstring distribution from an ideal quantum circuit as $p_x$ and denote the probability distribution after inserting MDNs as $\widetilde q_x$. Since the two subsystems decouple after the insertion of MDNs, we have $\widetilde q_x=\widetilde q_{x_1}\widetilde q_{x_2}$. In the rest of this Supplementary Material, we use $\langle\cdot\rangle$ to denote expectation value instead of $\mathbb E[\cdot]$, since the Dirac notation is no longer used.

We first show property (i) for step (1) of our improved algorithm. The XEB between $p_x$ and $\widetilde q_x$ is
\begin{equation}\label{seq:pq_qq}
    2^{N}\sum_{x}\langle p_x\widetilde q_x \rangle-1
    =2^{N}\sum_{x_1,x_2}\langle p_{x_1x_2}\widetilde q_{x_1}\widetilde q_{x_2} \rangle-1
    =2^{N}\sum_{x_1,x_2}\langle \widetilde q_{x_1}\widetilde q_{x_2}\widetilde q_{x_1}\widetilde q_{x_2} \rangle-1
    =2^{N}\sum_{x}\langle \widetilde q_x^2 \rangle-1
    >
    0,
\end{equation}
where the average is taken over the ensemble of omitted gates. The second equality is due to the 1-design property of the gate set, and the last equality is due to the fact that the XEB is 0 if and only if the distribution $\widetilde q_x$ is uniform (which can be shown by applying the Cauchy-Schwartz inequality). Since $\widetilde q_x$ is non-uniform, as long as the subsystem density matrix is not maximally mixed, the last inequality is strict, leading to property (i).

Next, we move on to show property (ii) for step (1) of our improved algorithm.
Compared to our original algorithm, which directly removes gates along a cut dividing the circuit into two, it is reasonable to expect that the STD of this new algorithm is smaller since the introduction of MDNs intuitively produces averaging effects. The mean value, however,  remains the same at least in the case that the single-qubit gate set is Haar random (or has the 2-design property). Concretely, the improved algorithm can also be realized by averaging over many realizations of the original algorithm by inserting different Paulis (plus $I$) or other arbitrary $t$-design single-qubit gates with $t\ge1$ at the positions of omitted gates. 
Each realization is equivalent to the original algorithm, because the omitted gates are random as well, which would effectively introduce these extra Paulis (plus $I$) (by choosing $v$ as the corresponding Paulis in Eq.~\eqref{Seq:HaarId}). The mean value of the new algorithm is exactly the same as that of the original one because of the 1-design property of random Paulis. Meanwhile, this improved algorithm effectively averages over  different instances of the original algorithm,  hence reducing the STD.
For this reason,  we call this step ``self-averaging''.
Because the output bitstrings could be influenced after the propagation of these single qubit gates in the middle of the circuit, the STD over different circuits is also associated with the STD in the property (ii).

\begin{figure}\includegraphics[width=0.8\textwidth]{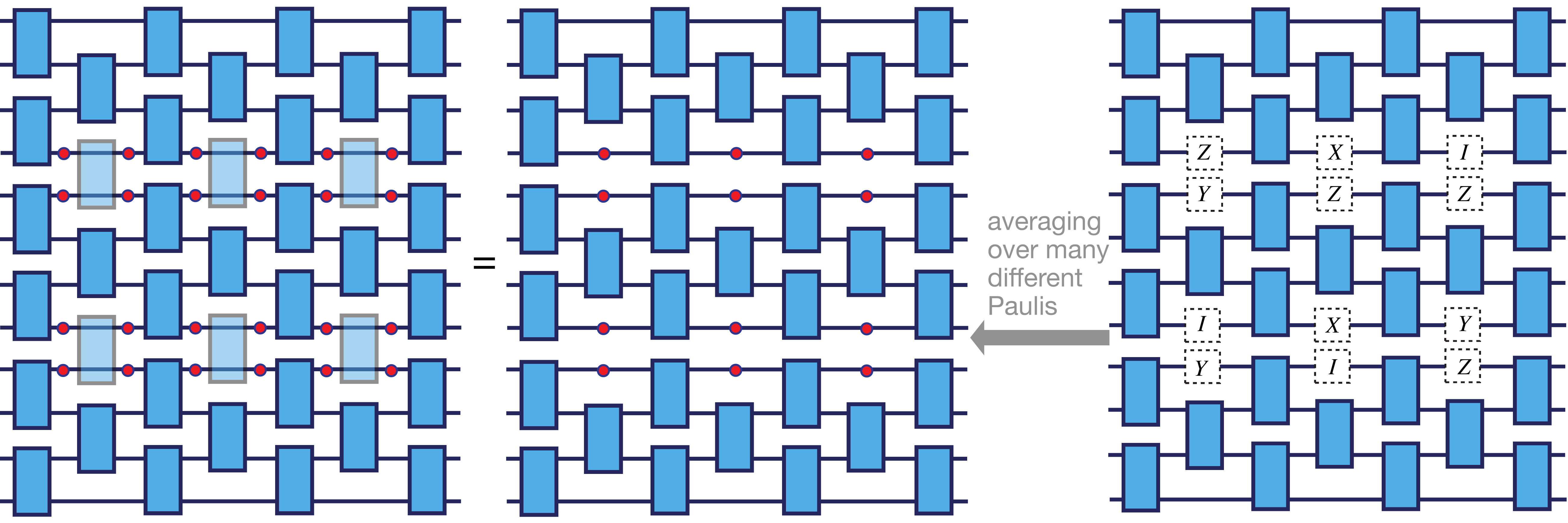}\caption{Illustration of the self-averaging algorithm: simulating the target circuit by inserting MDNs (red circles) or, equivalently, taking partial trace and preparing maximally-mixed states at the positions of light blue gates in the target circuit. In this case, the light blue gates can be omitted since the state is still maximally-mixed even after applying arbitrary unitary gates. This algorithm can also be realized by taking the average over many realizations of the original algorithm while inserting different Paulis (plus $I$) at the positions of omitted gates. }
\label{fig:algorithm2}
\end{figure}

\subsection{Numerical techniques for simulating fSim circuits with MDNs}\label{sapp:fSim_trick}
In order to exactly simulate MDN, the memory resources necessary for simulating the dynamics increase substantially as one needs to use density matrices to represent mixed states.
Naively, this is equivalent to doubling the system size.
Therefore, a direct and exact simulation of 53 qubits is no longer numerically viable with our resources.
Instead, one can sample many different realizations of the original algorithm, by applying single-qubit gates, randomly chosen from a 1-design ensemble, at the positions of omitted gates.

\begin{figure}\includegraphics[width=0.8\textwidth]{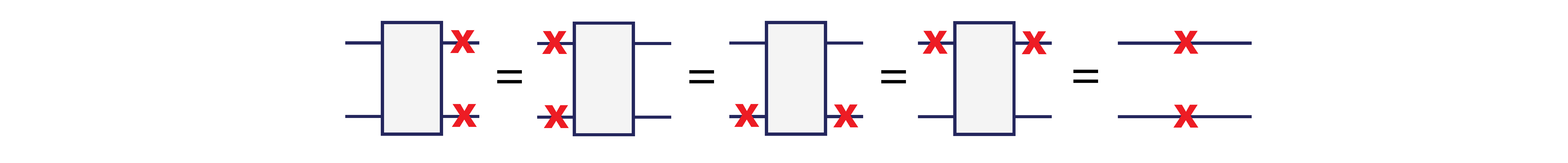}
\caption{Identities used for simplifying the tensor network representing the mixed state evolution of the fSim quantum circuits (or any other SWAP+control-phase gate).}
\label{fig:app_identities}
\end{figure}

In the special case where the entangling gate is fSim, however, we can take advantage of the property discussed in the previous section, and presented in Fig.~\ref{fig:app_fSim_identity}(a), to efficiently simulate the dynamics of mixed states. For completeness, Fig.~\ref{fig:app_identities} shows all the identities used for simplifying the tensor network that represents the mixed state evolution of the quantum circuit.
After simplifying a quantum circuit using these identities, we use a tensor network contraction algorithm based on a Julia package OMEinsum\footnote{``https://under-peter.github.io/OMEinsum.jl/dev/"} in which the contraction order is found using the algorithm in Ref.~\cite{kalachev2021recursive}. The subsystems we consider are given in Table~\ref{tab:app_subsystems}.

In this work, we only consider the most direct way to insert MDNs which is time/depth independent.
The choice of the subsystem is not optimized either.
By generalizing the way of inserting MDNs, e.g., making their locations to be time/depth dependent, the tensor network contraction algorithm can still be used straightforwardly. We expect that this type of optimization can produce higher XEB without increasing the necessary computational resources too much.

\begin{table*}
\begin{center}
\begin{tabular}{|c|c|c|}
\hline
 & subsystem 1 & subsystem 2\\
\hline
   Google: Fig.~S27 in Ref.~\cite{arute2019quantum} &
   the left part of red lines &
   the right part of red lines\\
\hline
USTC-1: Fig.~S11 in Ref.~\cite{USTC} &
   the blue part &
   the black part with 1 being excluded\\
\hline
USTC-2: Fig.~3(a) in Ref.~\cite{zhu2021quantum} & \makecell{
   1,2,3,7,8,9,10,12,13,14,15,18,19,20,\\21,24,25,26,30,31,32,36,37,42,43,48 }&
   \makecell{
   23,27,28,29,33,34,35,38,39,40,41,44,45,\\46,47,49,50,51,52,53,55,56,57,58,59}\\
\hline
\end{tabular}
\end{center}
\caption{
The subsystems in our simulation. They are not optimized and chosen for simulating the mixed state evolution using 1 GPU (NVIDIA Tesla V100).}
\label{tab:app_subsystems}
\end{table*}

\subsection{Top-$k$ method}
Suppose we replace $\widetilde q_x$ by another distribution $r_x=r_{x_1}r_{x_2}$, then
\begin{equation}
   \chi_\text{av}= 2^{N}\sum_{x}\langle p_x r_x \rangle-1
    =2^{N}\sum_{x_1,x_2}\langle p_{x_1x_2} r_{x_1} r_{x_2} \rangle-1
    =2^{N}\sum_{x_1,x_2}\langle \widetilde q_{x_1}\widetilde q_{x_2} r_{x_1}r_{x_2} \rangle-1
    =2^{N}\sum_{x}\langle \widetilde q_xr_x \rangle-1,
\end{equation}
where the expectation value is taken over the omitted gates, and the second equality follows from the 1-design property. Here, we choose $r_x$ in the following way:
\begin{equation}
    r_x=\begin{cases}
       \frac{1}{k}, \text{ if }\widetilde q_x \text{ is in the first }k\text{ largest probabilities;}\\
       0, \text{ otherwise.}
    \end{cases}
\end{equation}
This is called a ``top-$k$ method'', and it can substantially amplify the XEB. The resulting XEB is
\begin{equation}
    2^{N}\sum_{x^*}\frac{\widetilde q_{x^*}}{k}-1 \text{ where }x^*\in\{\widetilde q_{x^*}\text{ is in the first }k\text{ largest probabilities}\}.
\end{equation}

\begin{figure}\includegraphics[width=1.0\textwidth]{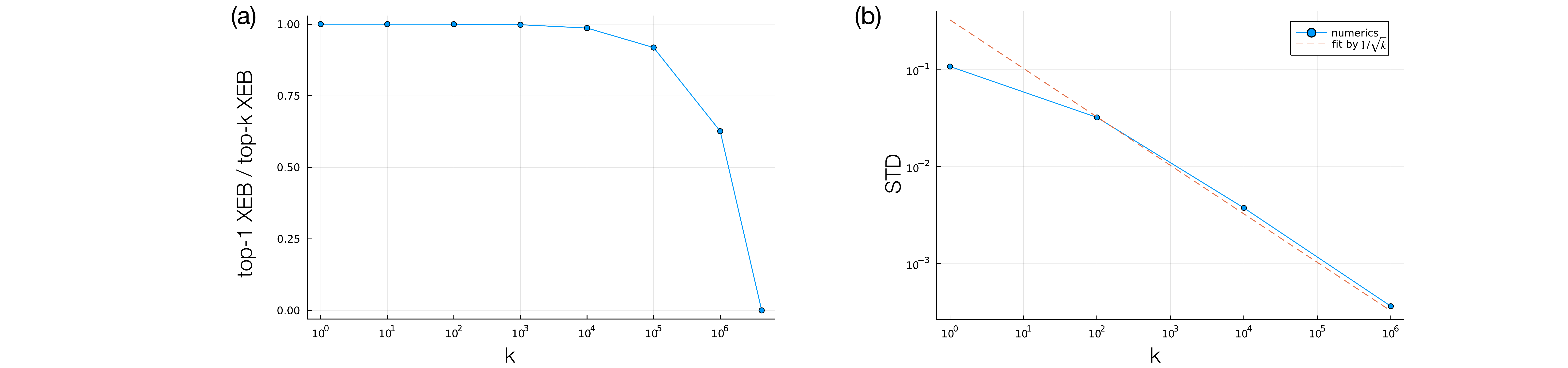}
\caption{Performance of the top-$k$ method. (a) The ratio between XEB values of top-$1$ and top-$k$ methods. (b) The STD of top-$k$ is approximately $\propto 1/\sqrt{k}$. Direct verification of the STD is difficult, since it requires simulating the ideal circuit. Here, we replace the omitted gates with the tensor product of two random, single-qubit gates in the ideal circuit as a reasonable approximation.}
\label{fig:app_top_k}
\end{figure}

In the actual simulation, we choose top-$k_1$ and top-$k_2$ bitstrings from $\widetilde q_{x_1}$ and $\widetilde q_{x_2}$ respectively. More accurately, we only choose top-$k_i$ bitstrings from the non-trivial part of $\widetilde q_{x_i}$.
Here we mention ``non-trivial" because many bitstrings share exactly the same value for $\widetilde q_{x_i}$.
This is because usually some output qubits experience MDN right before measurements, and thus the corresponding distribution is perfectly uniform, leading to degeneracies in $\widetilde q_{x_i}$.
In this case, we say the output qubits are ``trivial''.
Denoting the total number of trivial qubits as $m$, we can get $k_1k_22^m$ distinct bitstrings with $k_1 k_2$ distinct values of $\widetilde q_{x_i}$ (up to potential accidental degeneracies).
In Fig.~\ref{fig:app_top_k}, we present the performance of the top-$k$ method for the non-trivial part of subsystem 1 of Google's Sycamore architecture using the modification of their gate set. We can see that the mean value hardly decreases when increasing $k$ until $k$ becomes very large $\sim 10^5$.
However, the STD decreases $\propto 1/\sqrt k$. Intuitively, this suggests that the top-$k$ bitstrings are roughly independent due to strong scrambling.

\section{Refuting XQUATH}
The Linear Cross-Entropy Quantum Threshold Assumption (XQUATH) proposed by Aaronson and Gunn~\cite{aaronson2019classical} serves as the complexity-theoretic foundation of the XEB-based quantum computational advantage. The assumption is that there is no classical efficient algorithm to estimate the output probability of a string from a randomly sampled quantum circuit. Formally, we restate XQUATH from Ref.~\cite{aaronson2019classical} as follows.

\begin{conjecture}[Linear Cross-Entropy Quantum Threshold Assumption (XQUATH)~\cite{aaronson2019classical}]
Given a random circuit description $U$, there is no polynomial time classical algorithm to compute an estimation $q_U(0^N)$ of $p_U(0^N)$ (the probability of getting $0^N$ for the ideal quantum circuit $U$) such that
\begin{equation}\label{Seq:XQUATH}
   2^{2N}\ex{\left(p_U(0^N)-\frac{1}{2^N}\right)^2-\left(p_U(0^N)-q_U(0^N)\right)^2}{U}=\delta
\end{equation}
where $\delta=\Omega(2^{-N})$ and the expectation is over a random circuit ensemble.
\end{conjecture}

In this section, we show that our techniques can refute XQUATH, at least up to a reasonable modification from the original setup and most of the circuit architectures.

\subsection{Reduction from XQUATH to the hardness of average XEB}\label{ssec:XQUATHreduction}
Here we adopt our self-averaging algorithm shown in Fig.~\ref{fig:algorithm2} to refute XQUATH, since this algorithm has smaller STD as discussed in Sec.~\ref{app:1D_fSim_STD}. The key idea is to reduce Eq.~\eqref{Seq:XQUATH} in XQUATH to the average XEB value of our algorithm.

First, we show that the quantity on the left-hand side of Eq.~\eqref{Seq:XQUATH} is exactly the same as the average XEB if $\widetilde q_U$ is computed from the self-averaging algorithm introduced in Fig.~\ref{fig:algorithm2} and if the circuits are random enough:
\begin{eqnarray*}
2^{2N}\ex{\left(p_U(0^N)-\frac{1}{2^N}\right)^2-\left(p_U(0^N)-\widetilde q_U(0^N)\right)^2}{U}&=&2^{2N}\left(
2\ex{p_U(0^N)\widetilde q_U(0^N)}{U}-\ex{\widetilde q^2_U(0^N)}{U}
-\frac{2\ex{p_U(0^N)}{U}}{2^N}+\frac{1}{2^{2N}}\right)\\
&=&2^{2N}\left(
2\ex{p_U(0^N)\widetilde q_U(0^N)}{U}-\ex{\widetilde q^2_U(0^N)}{U}
\right)-1\\
&=&
2^{2N}\ex{p_U(0^N)\widetilde q_U(0^N)}{U}-1\\
&=&
2^{N}\ex{\sum_xp_U(x)\widetilde q_U(x)}{U}-1\\
&=&\ex{\chi_U(C)}{U}
\end{eqnarray*}
where the second line is due to $\ex{p_U(0^N)}{U}=1/2^N$; the third line is due to
$\ex{p_U(0^N)\widetilde q_U(0^N)}{U}=\ex{\widetilde q^2_U(0^N)}{U}$ (see Eq.~(\ref{seq:pq_qq})); the fourth line is due to that the circuit behaves identically for all bitstrinigs, i.e., there is nothing special about the particular choice of $0^N$, i.e., $\ex{p_U(0^N)\widetilde q_U(0^N)}{U}=\ex{p_U(x)\widetilde q_U(x)}{U}$ for every $x\in\{0,1\}^N$. 

In summary, we proved that, for our self-averaging algorithm, the $\delta$ defined in Eq.~\eqref{Seq:XQUATH} of XQUATH is exactly the same as the average XEB. In the next subsection, we discuss the value of $\delta$ obtained from our algorithm.

\subsection{Refuting XQUATH for 2D circuits}
In the main text and subsection \ref{app:1D} in this supplementary material, we have shown that, for 1D circuits, $\chi_\text{av}=\Omega(e^{-\Delta_1 d})$ where $\Delta_1\approx 0.25$. Here, we show that for a much larger family of circuit architecture, including 2D circuits, our algorithm achieves $\chi_\text{av}=\Omega(e^{-\Delta d})$ with some constant $\Delta $. As a consequence, we refute XQUATH in these circuit architectures. We have two kinds of arguments, one is the additivity of XEB and another is the path integral point of view. The former works for generic quantum circuits (such as Haar random unitiary circuits) but not work for circuits consisting of more tailored, maximally entangling gates such as the fSim due to the property shown in Fig.\ref{fig:app_fSim_identity}.
The latter provides a more flexible analysis which also works for fSim gates.

\subsubsection{Additivity of XEB}

We start with a generic analysis for non-maximal-entangling 2-qubit gates like Haar-random and CZ gates. For special gate sets, such as fSim, one needs to carefully cut the subsystem, as described in Fig.~\ref{fig:app_fSim_identity}.

The key observation is that after our algorithm breaks a circuit into several subsystems, the corresponding diffusion-reaction model (or Ising spin model for 1D) is also decoupled into $\lceil N/l\rceil$ isolated subsystems, where $l$ is the subsystem size (measured in the number of qubits). Then, similar to Eqs.~({\color{red}31-33}) in the main text, we have
$$
\chi_\text{av}=\prod_i^{\lceil N/l\rceil}(\chi^{(i)}_\text{av}+1)-1,
$$
where $\chi_\text{av}^{(i)}+1$ is equal to the partition function in the $i$-th subsystem. Then, we have
$$
\chi^{(i)}_\text{av}=c_l^{(i)}e^{-\Delta_l^{(i)}d},
$$
where $c_l^{(i)}$ and $\Delta_l^{(i)}$ are some constants that depend only on the subsystem size $l$ (up to the detailed arrangement of these $l$ qubits). Specifically, $c_l^{(i)}$ and $\Delta_l^{(i)}$ would not have any dependency on $N$ because each subsystem can only see the maximal depolarizing noise at the boundaries (positions of omitted gates) and is disconnected from any information about other subsystems. Next, we choose $l$ as a constant so that each $\chi^{(i)}_\text{av}$ has the form $\Omega(e^{-\Delta d})$. Thus,
$$
\chi_\text{av}=\prod_i^{\lceil N/l\rceil}(\chi^{(i)}_\text{av}+1)-1\approx\sum_i^{\lceil N/l\rceil}\chi^{(i)}_\text{av}
=\sum_i^{\lceil N/l\rceil}c_l^{(i)}e^{-\Delta_l^{(i)} d} = \Omega(e^{-\Delta d})
$$
for some constant $\Delta>0$, as desired. Together with the reduction in subsection~\ref{ssec:XQUATHreduction}, this shows that our algorithm refutes XQUATH for circuit architectures with non-maximal-entangling 2-qubit gates because $d\sim N^{1/D}$ is a reasonable requirement for a sufficiently scrambling circuit dynamics to demonstrate quantum computational advantage.

\subsubsection{Path integral in Pauli basis}
The proposal of XQUATH was based on the observation that sub-sampling the path integral for XEB over the computational basis only achieves $2^{-\Omega(Nd)}$ XEB value with high probability. Concretely, if we consider the path integral for XEB using the computational basis, there are roughly $2^{Nd}$ different paths and their weights (i.e., their contribution to the XEB) are roughly equal. The belief underlying XQUATH is that a polynomial time classical algorithm can only compute the contribution from polynomial number of paths and thus the attainable XEB is only $\text{poly}(Nd)2^{-Nd}$.

However, the weight of each path is far from uniform when we consider performing path integral with respect to the Pauli basis ($I$ and $X,Y,Z$ as we discussed in section~\ref{app:DR}). Here, we denote 
\begin{equation}
\widetilde\sigma_0=\frac{1}{\sqrt2}
\begin{pmatrix}
0&1\\
1&0
\end{pmatrix}, \quad
\widetilde\sigma_1=\frac{1}{\sqrt2}
\begin{pmatrix}
0&1\\
1&0
\end{pmatrix}, \quad
\widetilde\sigma_2=\frac{1}{\sqrt2}
\begin{pmatrix}
0&-i\\
i&0
\end{pmatrix},\quad
\widetilde\sigma_3=\frac{1}{\sqrt2}
\begin{pmatrix}
1&0\\
0&-1
\end{pmatrix},
\end{equation}
where the factor $1/\sqrt2$ makes them different from conventional Pauli matrices.
For any $2\times 2$ density matrix $\rho$, we have the following decomposition:
\begin{equation}\label{eq:dec_Her}
\rho=\sum_{i=0,1,2,3}(\tr{\widetilde\sigma_i\rho})\widetilde\sigma_i.
\end{equation}
This is the Pauli-Liouville representation~\cite{greenbaum2015introduction}, and since $\rho$ is Hermitian, $\tr{\widetilde\sigma_i\rho}$ is real. For a $4\times4$ matrix $\rho$
\begin{equation}
\rho=\sum_{i,j=0,1,2,3}(\tr{\widetilde\sigma_i\otimes\widetilde\sigma_j \rho})\widetilde\sigma_i\otimes\widetilde\sigma_j.
\end{equation}
The $\{\widetilde\sigma_i\}^{\otimes n}$ basis is complete for all Hermitian matrices acting on on $n$-qubits.

Similarly to inserting $I=\ket0\bra0+\ket1\bra1$ at every space-time position in the quantum circuit to get the path integral in the computational basis, we can formulate a path integral in the Pauli basis by instead inserting a different, appropriately chosen, resolution of identity. To illustrate the idea of this path integral, we use a very simple example, which  can be straightforwardly extended to most general cases. First, we consider a matrix element of a single-qubit density matrix after the application of two gates
\begin{eqnarray}\label{eq:path}
\notag  \bra x V^\dag U^\dag \rho U V \ket x&=&
\sum_{i=0,1,2,3}(\tr{\widetilde\sigma_i\rho})\bra xV^\dag U^\dag \widetilde\sigma_iUV \ket x\\
\notag&=&\sum_{i,j=0,1,2,3}(\tr{\widetilde\sigma_i\rho}) ( \tr{U^\dag \widetilde\sigma_iU\widetilde\sigma_j })
 \bra xV^\dag \widetilde\sigma_j  V \ket x\\
 &=&\sum_{i,j,k=0,1,2,3}(\tr{\widetilde\sigma_i\rho}) ( \tr{U^\dag \widetilde\sigma_iU\widetilde\sigma_j} )
(\tr{V^\dag \widetilde\sigma_j  V \widetilde\sigma_k}) (\tr{\widetilde\sigma_k\ket x\bra x}),
\end{eqnarray}
where we used Eq.~(\ref{eq:dec_Her}) iteratively.

While now there are roughly $4^{Nd}$ different paths (because we have $\propto Nd$ space-time positions and 4 basis matrices), the weights of different paths can be vastly different, as contrasted with the nearly-uniform distribution of weights in the computational-basis path integral. In particular,  
a path in the Pauli basis has a weight exponentially decaying with the number of non-trivial Paulis (i.e., other than the identity). This can be understood by the fact that the transition $I\to I$ satisfies  $|\tr{U^\dag I U I}|>|\tr{U^\dag \sigma U \sigma^\prime}|$, if any of $\sigma$,$\sigma^\prime$ is not $I$ (Clifford gates are a special case where  should use $\ge$). Thus, a transition involving a non-trivial Pauli ($X,Y,Z$) will cause a decay relative to the transition involving only $I$. The path involving only $I$s corresponds to the contribution which is cancelled by $-1$ in the definition of the XEB.
 It would seem that we could simply choose a path with a small number of non-trivial Paulis, which would have a large contribution to the XEB. However, it can be shown that the smallest number of non-trivial Paulis in a path must be $d$ (see also Ref.~\cite{gao2018efficient}), since $\tr{U^\dag I U \sigma}=0$ if $\sigma\ne I$. In order to avoid this situation, a minimal path should contain non-trivial Paulis connecting the input and output boundaries. In this case, the contribution to the XEB is at least $e^{-O(d)}$.
For example, in Fig.~\ref{fig:app_fSim_identity}(d), we show that in a 1D circuit one can find a path of Pauli strings that obeys the transition rules and has non-trivial Paulis connecting the two boundaries (within the blue dots). Note that for circuits with fixed dimension, we can always generalize the idea in Fig.~\ref{fig:app_fSim_identity}(d) and get a desired subsystem with constant width (so that it can be fully simulated in polynomial time). 
To conclude, the supporting argument for XQUATH, based on the conventional path integral formulation, does not hold if we consider a Pauli-basis path integral. This is because the contribution of each path becomes polarized (not equally weighted) and our subsystem algorithm is equivalent to outputting paths with weight $e^{-O(d)}$ in the Pauli basis.